\documentclass[twocolumn]{aastex63}

\received{Nov. 2024}
\revised{TBD}
\accepted{TBD}%\today
\pdfoutput=1 %for arXiv submission
\usepackage{amsmath,amstext}
\usepackage[T1]{fontenc}
\usepackage[figure,figure*]{hypcap}

\usepackage{graphicx}
\usepackage{graphics}
\usepackage{color}
\usepackage[caption=false]{subfig}
\definecolor{dartmouthgreen}{rgb}{0.05, 0.5, 0.06}
\usepackage{afterpage}
\usepackage{natbib}
\usepackage{makecell}

\def\arcsec{\hbox{$^{\prime\prime}$}}
\def\deg{\hbox{$^\circ$}}

\def\h{h^{-1}_{70}}
\graphicspath{ {Images/} }

\shorttitle{The Big Multi-AGN Catalog}
\shortauthors{Pfeifle et al.}
\begin{document}
\title{Super-Size Me: The Big Multi-AGN Catalog (The Big MAC)\\ Data Release 1: The Source Catalog}
%or\\
%Can I Get a \#1? The Big Multi-AGN Catalog (The Big MAC):\\
%Data Release 1, The Source Catalog}

\correspondingauthor{Ryan W. Pfeifle}
\email{ryan.w.pfeifle@nasa.gov}

\author[0000-0001-8640-8522]{Ryan W. Pfeifle}
\altaffiliation{NASA Postdoctoral Program Fellow}
\altaffiliation{Surname pronunciation: ``\textit{Fife-Lee}''}
\affiliation{X-ray Astrophysics Laboratory, NASA Goddard Space Flight Center, Code 662, Greenbelt, MD 20771, USA}
\affiliation{Oak Ridge Associated Universities, NASA NPP Program, Oak Ridge, TN 37831, USA}

\author[0009-0008-4232-486X]{Kimberly A. Weaver}
\affiliation{X-ray Astrophysics Laboratory, NASA Goddard Space Flight Center, Code 662, Greenbelt, MD 20771, USA}

\author[0000-0002-4902-8077]{Nathan J. Secrest}
\affiliation{U.S. Naval Observatory, 3450 Massachusetts Avenue NW, Washington, DC 20392, USA}

\author[0000-0003-2283-2185]{Barry Rothberg}
\affiliation{U.S. Naval Observatory, 3450 Massachusetts Avenue NW, Washington, DC 20392, USA}
\affiliation{Department of Physics and Astronomy, George Mason University, 4400 University Drive, MSN 3F3, Fairfax, VA 22030, USA}

\author[0000-0002-1871-4154]{David R. Patton}
\affiliation{Department of Physics and Astronomy, Trent University, 1600 West Bank Drive, Peterborough, ON, K9L 0G2, Canada}

\begin{abstract}
Galaxy mergers represent the most transformative and dramatic avenue for galaxy and supermassive black hole (SMBH) evolution. Multi-active galactic nuclei (multi-AGNs) are expected to ignite, grow, and evolve alongside the host galaxies, and these represent different evolutionary stages of the SMBHs over the merger sequence. However, no comprehensive census exists of observed multi-AGN systems. Here we present The Big Multi-AGN Catalog (The Big MAC), the first literature-complete catalog of all known (confirmed and candidate) multi-AGN systems, which includes dual AGNs (separations $\sim0.03-110$\,kpc), binary AGNs (gravitationally bound, $\lesssim30$\,pc), and recoiling AGNs, gleaned from hundreds of literature articles spanning the years 1970-2020. The Big MAC is the first archive to assemble all multi-AGN systems and candidates across all selection methods, redshifts, and galaxy mass ratios, and this catalog offers a solid foundation for archival and targeted multiwavelength follow-up investigations. In this work, we provide an overview of the creation of the multi-AGN literature library and the catalog itself, present definitions for different multi-AGN classes (including new definitions for dual AGNs derived from galaxy pairs in Illustris-TNG100), describe the general properties of the catalog as a function of redshift space and separation, and we provide a thorough examination of selection and confirmation method usage within the literature. We also discuss best practices for the multi-AGN literature, and we emphasize that a diverse, multiwavelength array of selection approaches is crucial for a complete understanding of multi-AGNs and -- by extension -- answering long-standing, open questions regarding the importance of AGNs and galaxy mergers.

\end{abstract}

 
\section{Introduction}
\label{sec:intro}
\subsection{Motivation}
\label{sec:motivation}

From the numerical simulation work of \citet{holmberg1940,holmberg1941} showing the clustering of gas in nebulae to the groundbreaking numerical simulation work of \citet{toomre1972} demonstrating the direct consequence of the interaction of two galaxies, galaxy mergers have long been argued to be a fundamental -- and perhaps the most transformative -- channel of evolution for a galaxy in terms of structure, mass, gas distribution, and the stellar populations \citep{barnes1991,barnes1996,mihos1994,mihos1996,cox2008,hopkins2005,capelo2015,blecha2018}. Upon a sufficiently close passage of two galaxies, complex combinations of gravitational and hydrodynamical forces conspire to torque the gas and give rise to a variety of gas-inflow mechanisms, including ram pressure stripping \citep[][]{capelo2017shocks,blumenthal2018}, bar-driven inflows \citep[][]{barnes1991,barnes1996,mihos1996,cox2008,blumenthal2018}, and shock-induced gas clumps that are deposited in the nuclei dynamical friction \citep[][]{blumenthal2018}. The exact inflow mechanism, efficiency, and timescale depends upon encounter geometry, galactic structure, and gas fraction, and could also be affected by stellar and SMBH feedback \citep[e.g.,][]{capelo2015,blumenthal2018}. Simulations predict that the large reservoirs of gas that collect within the nuclei can fuel rapid star formation \citep[][]{mihos1996,barnes1991,cox2008,capelo2015,capelo2017shocks,blumenthal2018,blecha2018} and -- at least for massive galaxies which host supermassive black holes (SMBHs) -- the growth of the central SMBHs \citep[e.g.,][]{hopkins2005,capelo2017duals,blecha2018}, which ignite as active galactic nuclei (AGNs). At high redshift, large-scale gas filaments may also play a key role in supplying and/or replenishing reservoirs of dense cold gas for triggering or sustaining nuclear activity \citep[e.g.,][]{steinborn2016}.

While the level of nuclear activity -- whether star formation, AGNs, or a combination of both -- fluctuates across the timescale of the merger, simulations consistently demonstrate that the activity can persist while the host galaxy orbits continue to decay, potentially peaking during the latest stages of the merger process when the nuclei of the galaxies begin to coalesce and merge into a single, morphologically disturbed spheroid \citep[e.g.,][]{vanwassenhove2012,capelo2015,blecha2018,chen2023}. It is this correlated evolution between the host galaxies and their central SMBHs that demands close attention, as it suggests that galaxy mergers could be one avenue for establishing the well-known scaling relations between the mass of the central SMBHs and the stellar velocity dispersions and luminosities of their host galaxies \citep[M-$\sigma$ and M-$L$ relations, e.g.,][]{gebhardt2000,ferrarese2000,gultekin2009,kormendy2013}. It is therefore of great importance to understand the possible connection between galaxy mergers and SMBH growth, and the relative importance of merger-triggered SMBH growth \citep[see an extensive discussion of the observational literature in Section 1 in][]{ellison2019}. 

One unique phase of galaxy mergers that we can study to better understand the AGN-merger connection are mergers hosting multiple AGNs (multi-AGNs). Multi-AGNs are expected to ignite, grow, and evolve alongside the host galaxies, evolving generally from dual AGNs/SMBHs, to gravitationally bound binary AGNs/SMBHs, and potentially to recoiling AGNs, each of which represent different evolutionary stages of the SMBHs over the merger sequence. %

\subsubsection{Dual AGNs}

In the case of two merging galaxies on kpc- and sub-kpc scales, simultaneously accreting SMBHs are referred to as ``dual AGNs''. Dual AGNs are generally predicted to show significant enhancements in mass accretion rates \citep[e.g.,][]{vanwassenhove2012,blecha2018,capelo2015,capelo2017duals,volonteri2022,chen2023}, high column densities and dust obscuration \citep[e.g.,][]{blecha2018,capelo2017duals,chen2023}, and substantial SMBH mass ratio evolution \citep[generally converging toward unity across the full population, e.g.,][]{capelo2015,chen2023}. The predicted behavior of these dual AGNs could potentially impact bound SMBH binary hardening timescales and the properties of gravitational waves (GWs) emitted during the bound binary and inspiral phases \citep[e.g.,][]{kelley2017}. Hence, dual AGNs are expected to be a critically important growth and evolutionary phase for the SMBHs as well as for their hosts, as the hosts simultaneously undergo their most dramatic evolution and perhaps receive feedback from the dual AGNs \citep[e.g.,][Sivasankaran, in preparation.]{li2021}.
%The SMBH mass ratios in these systems  \citep{chen2023}, 

Dual AGNs are predicted and observed to reside across the full merger sequence: cosmological simulations show -- among dual AGNs with separations $<30$\,kpc -- that $\sim50$\% of duals have separations $>10$\,kpc \citep{chen2023}, and a substantial number of dual AGNs and candidates have been observed with separations $>40-60$\,kpc \citep[e.g.,][]{koss2012,liu2011b,derosa2018,derosa2023,hou2020,barrows2023}. The majority of targeted and follow-up dual AGN searches, at least in the local universe, have focused, however, on late-stage mergers with separations $<10$\,kpc \citep[e.g.,][]{liu2010a,shen2011,liu2013,comerford2015,fu2015a,satyapal2017,pfeifle2019a,hou2019}. At least in the case of major mergers, dual AGNs are predicted to reach their peak activity levels during a heavily obscured phase when separations are $<10$\,kpc, and particularly close to the bound binary phase/coalescence of the SMBHs \citep[when the SMBH separations plunge below $1$\,kpc, e.g.,][]{capelo2015,blecha2018}. Indeed, there is some observational evidence that dual AGN fraction, X-ray luminosity, [O\,III] luminosity, and column density increases with decreasing pair separation \citep[e.g.,][]{liu2011b,koss2012,guainazzi2021,stemo2021,derosa2023,barrows2023}, but these relationships demand further examination across larger dual AGN samples and across a larger/more uniform parameter space (e.g., in terms of not only projected separation but also galaxy and merger morphology, mass ratio, etc.). In particular, the relationship between (spectroscopically derived) column density and pair separation is dominated by small sample statistics \citep{derosa2023} and is not as obvious as relationships derived, for example, from samples of single AGNs in LIRGs and ULIRGS, \citep[][]{ricci2017mnras}. Nevertheless, a significant number of known dual AGNs do show high column densities \citep[see discussion in][]{derosa2019,pfeifle2023c}, so dual AGNs may generally represent good examples of obscured AGN activity; larger, statistical complete samples are required for confirmation.

Simulations also predict that dual AGNs arise in both major (1:1-1:>4, or $\rm{q}>0.25$) and minor (1:$\leq$4, or $\rm{q}\leq0.25$) mergers \citep[e.g.,][]{capelo2017duals,chen2023}, with dual AGNs predicted to be observable over longer timescales in major mergers relative to minor mergers \citep{capelo2017duals}, while the most dramatic SMBH mass ratio evolution is expected in minor mergers \citep[e.g.,][]{capelo2015}. We do not yet have a statistically rigorous (or unbiased) census to determine the relative importance of major and minor mergers in dual AGN triggering; the majority of known dual AGNs (reported in various samples, each with their own complex selection function) appear to reside in major mergers \citep[e.g.,][]{koss2012}, though $\sim40$\% of dual AGNs identified at $z=3$ in the ASTRID cosmological simulation reside in minor mergers \citet{chen2023}, suggesting that we may be missing this statistically important population.

Despite the presumed importance of dual AGNs and extensive campaigns and pilot programs conducted to identify bona fide cases, relatively few confirmed dual AGNs exist in the literature; this is likely due to wavelength-dependent biases and technical challenges. The vast majority of dual AGN candidates have been selected using optical spectroscopic diagnostics \citep[e.g.,][]{hennawi2006,hennawi2010,wang2009,liu2010b,smith2010,liu2011b,ge2012,shi2014}, typically based on the Baldwin-Phillips-Terlevich \citep{baldwin1981} spectroscopic emission line ratios from distinct fiber spectra indicating the presence of multiple active nuclei \citep{liu2011b}, or based on the presence of double-peaked spectroscopic emission line AGNs exhibiting Seyfert-like line ratios in both emission line systems \citep[e.g.][]{wang2009,smith2010,liu2010b,ge2012}. Extensive follow-up campaigns focusing on double-peaked AGNs in the optical, X-rays, near-IR, and radio found few clear cases of dual AGNs; $<15\%$ of double-peaked AGNs in galaxy mergers coincide with dual AGNs \citep{fu2012,mullersanchez2015}, and most double-peaked systems are driven instead by jet-cloud interactions, rotating accretion disks, mass outflows, and other phenomenon \citep[e.g.,][]{rosario2010,crenshaw2010,mullersanchez2015,comerford2018}. On the other hand, hundreds of true dual AGNs and promising  candidates have been identified via distinct fiber spectra \citep[e.g.,][]{liu2011b}. A significant number of kpc- or sub-kpc scale dual SMBH candidates (wherein one SMBH is active) have been selected based on emission lines offset in velocity from the host galaxy \citep[originally referred to as ``offset AGNs''/candidates][]{comerford2009a,comerford2013,comerford2014}; recent work has shown that many of these offset velocity lines are likely due to shocked gas caused by outflows or inflows \citep[e.g.,][]{mullersanchez2016,comerford2017}. Since dual AGNs are predicted to show high gas columns and be dust enshrouded \citep[at least in late stage mergers][]{blecha2018}, a significant fraction of dual AGNs may be missed in the optical but accessible at other wavelengths, such as the mid-IR \citep{blecha2018,satyapal2017,pfeifle2019a} and the X-rays \citep[e.g.,][]{komossa2003,koss2012}, necessitating multiwavelength selection strategies.

Hard X-ray preselection, \citep{koss2011,koss2012}, mid-IR preselection \citep{satyapal2017,ellison2017,pfeifle2019a,pfeifle2019b}, and radio (pre-)selection \citep[][]{fu2015a,fu2015b,zhang2021a,zhang2021b} strategies have shown great promise in selecting dual AGNs and candidates in the local universe. In particular, serendipitous \citep{owen1985,parma1991,klamer2004} and systematic searches in the radio \citep{fu2015a,fu2015b} have begun to identify small but significant samples of dual radio AGNs and candidates. As in the case of double-peaked AGNs, follow-up (targeted or archival) observations have been a necessity for all of these (pre-)selection strategies (among others not listed here) to search for additional evidence in favor of multiple active nuclei in a given system \citep[e.g.,][]{koss2011,koss2012,fu2015b,satyapal2017,pfeifle2019a,pfeifle2019b,gross2019,zhang2021a,zhang2021b}. More recently, Wide-field Infrared Survey Explorer (WISE) mid-IR colors \citep[from ][]{assef2018} were used by \citet{barrows2023} to search for spatially-resolved mid-IR dual AGNs in earlier stage mergers in the SDSS; though promising, follow-up imaging and spectroscopy is still needed to confirm most of these mid-IR AGN pairs as bona fide dual AGNs. A similar effort to uncover spatially-resolved mid-IR dual AGNs using the criteria from \citep[][$W1-W2>0.8$]{stern2012} is currently underway in the DeCaLS footprint, wherein morphological evidence of mergers is a required component for selection (Pfeifle et al., in preparation).

At higher redshift, dual AGN candidates are typically selected via optical spectroscopy \citep[e.g.,][]{hennawi2006,inada2008,hennawi2010,more2016}, preselected via optical photometry or colors as quasar pairs \citep[and confirmed via follow-up  spectroscopy and/or imaging, e.g.,][]{hennawi2006,inada2008,hennawi2010,myers2007,myers2008,green2010,green2011,more2016,eftekharzadeh2017}, via varstrometry or multi-component systems \citep{shen2019,chen2022,mannucci2022,ciurlo2023}, and most recently via spatially-resolved infrared SEDs and imaging of X-ray-selected sources in the JWST Cosmos-Web field \citep{li2024}. Many high redshift dual AGNs and candidates have been identified as a result of gravitational lens surveys \citep[e.g.,][]{inada2008,inada2010,inada2012,more2016,anguita2018,lemon2018,lemon2020}, and in general high redshift dual AGN candidates often require extensive follow-up observations to rule out gravitational lensing or other alternative scenarios \citep[e.g.,][]{hennawi2006,inada2012,more2016,eftekharzadeh2017,lemon2018,lemon2019}.

\subsubsection{Binary AGNs}
Once the host galaxies have merged into a single spheroid, the SMBHs (previously at the centers of these galaxies) are expected to continue undergoing periods of correlated growth \citep{blecha2018} as they sink toward the center of the post-merger remnant galaxy from kpc-scales down to tens of parsec-scales via dynamical friction \citep{blecha2018,capelo2015,vanwassenhove2012}. At parsec to a few tens of parsec-scales (depending on the SMBH masses), the SMBHs form a gravitationally-bound binary system \citep[e.g.][]{begelman1980,rodriguez2006}, and this binary continues to harden via dynamical friction, scattering of stars from a loss cone, viscous drag within a circumbinary disk, and gravitational wave emission \citep[e.g.,][]{begelman1980,volonteri2003,kelley2017}, with these processes expected to dominate the binary hardening on scales of $\gtrsim1$\,pc, $\sim0.01-1$\,pc, $10^{-5}-0.01$\,pc, and $<10^{-5}$\,pc, respectively, \citep[based on results from Illustris,][]{kelley2017}. The coalescence of inspiraling SMBH binaries -- in addition to resulting in a single, more massive SMBH -- is expected to give rise to massive gravitational waves (GWs); inspiraling SMBHs are the suspected origin of the recently observed low frequency GW background signature detected in the NANOGrav 15 year PTA data set \citep{agazie2023}. Such GW signals from inspiraling intermediate-mass black hole binaries will be detectable in the next few decades with the Laser Interferometer Space Array  \citep[LISA,][]{amaro-seoane2023}. Binary AGNs (and the GWs of inspiraling pairs) are the observational tracers of these gravitationally-bound SMBH systems and -- by extension -- this key phase of hierarchical SMBH growth expected within our current cosmological paradigm. Only a single binary AGN has so far been confirmed in the literature \citep[with a separation of 7\,pc][]{rodriguez2006}, but myriad studies have sought binaries by searching for periodic signals in light curves \citep[e.g.,][]{graham2015a,graham2015b,charisi2016,tliu2016,tliu2019}, searching for velocity offset spectroscopic emission lines and/or displaying radial velocity changes over time \citep[e.g.,][]{shen2013,eracleous2012,runnoe2015,xliu2014,ju2013,wang2017}, or identifying candidates based on multiple mas-scale radio sources \citep[e.g.][]{xiangliu2014,gitti2013,kharb2017} or X-shaped radio sources \citep[e.g.,][]{lal2007}, among other techniques. These searches have recovered large samples of candidate binaries but no further convincing cases as of yet. 

\subsubsection{Recoiling AGNs}
When a gravitationally-bound pair of SMBHs inspirals and merges into a singular SMBH, the remnant SMBH is expected to receive a recoil ``kick’’ due to the anisotropic emission of GWs caused by asymmetries in the properties of the binary \citep[in terms of mass, spins and spin alignments, e.g.,][]{campanelli2007,schnittman2007a,blecha2011}. Such SMBH binary remnants have been predicted to receive a wide range of kick velocities that depend upon the initial mass ratio, spins, and spin orientations: in cases where the spins of the SMBHs become preferentially aligned due to torques from the circumbinary disk, the SMBHs may recieve kicks on the order $\lesssim500$\,km\,s$^{-1}$ \citep{bogdanovic2007}. For maximally spinning, equal mass, anti-aligned SMBH binaries, kick velocities of up to 4000\,km\,s$^{-1}$ may be expected \citep{campanelli2007}. \citet{blecha2011} showed that, assuming random spin magnitudes, 41.6\% and 13.5\% of major mergers are expected to produce recoiling AGNs with kick velocities $>500$\,km\,s$^{-1}$ and $>1000$\,km\,s$^{-1}$, respectively, while minor mergers preferentially result in lower velocity kicks ($<500$\,km\,s$^{-1}$). 

These recoiling SMBHs or AGNs (if actively fueling) may be observationally identifiable via kinematic and/or spatial offsets from the central nucleus of the host \citep[e.g.,][]{blecha2011,bonning2007,kim2016}, where the AGN lifetime for a recoiling SMBH is expected to be tied to the recoil speed \citep{blecha2011}. The large suite of hydrodynamic simulations of recoiling SMBHs/AGNs in \citet{blecha2018} predicted that recoiling SMBHs have different accretion histories than stationary SMBHs in normal galaxies (due to interruption of BH fueling), and this leads to an increase in intrinsic scatter in the M-$\sigma$ relation due to a mass deficit \citep[see also][]{sijacki2011}. Feedback from the recoiling AGN is also displaced due to the motion of the SMBH -- distributing the energy from the AGN across the inner kpc of the host galaxy -- reducing the quenching efficiency of the AGN and allowing the build up of more massive stellar cores via unhindered star formation \citep[][]{blecha2011}. 

Recoiling AGNs have most often been sought via the presence of emission line systems offset in velocity from host galaxies \citep[e.g.,][]{komossa2008,steinhardt2012,kim2016}, while smaller samples of candidates have been identified based on spatial offsets between the AGN position and the host nucleus \citep[e.g.,][]{civano2010,lena2014}. No recoils have so far been confirmed in the literature.

\subsubsection{A New Catalog for Multi-AGN Science}

Despite the large volume research that has been dedicated to multi-AGN science so far, myriad open questions still remain with regard to the relative importance, frequency, and evolutionary nature of different multi-AGN populations, as well as the relative importance of mergers for triggering and fueling AGN activity. Given these open questions, a need has arisen for the cataloging and creation of the first multi-AGN catalog; over the last four and a half years, we have crafted and curated The Big Multi-AGN Catalog (Big MAC), an archive of multi-AGN systems that is approximately literature complete from $\sim1970$ to $\sim2020$ and for now represents the largest accumulation of data and literature targeting multi-AGN systems. This library stands not only as an archive for multi-AGN science, but also as a service to the community to help clarify and define common terminology and methodology.

\subsection{Catalog Scope}
\label{sec:catalog_scope}

The Big MAC is the first literature complete, living library of multi-AGN systems; data release 1 (DR1) of the Big MAC was culled from  $\sim600$ literature articles published between $\sim1970-2020$. The catalog is designed to include all confirmed and candidate multi-AGN systems -- independent of redshift, mass range, and selection methodology -- that fall into one or more of the following system classes:
\begin{itemize}
    \item Dual AGNs/SMBHs and candidates (separations of $\sim$\,0.03-100\,kpc)
    \item Gravitationally-Bound Binary AGNs/SMBHs and candidates (separations $\lesssim$\,30\,pc)
    \item Recoiling AGNs/SMBHs and candidates (post-coalescence of a binary SMBH)
    \item N-tuple AGN systems, consisting of three or more physically-associated AGNs (gravitationally-interacting and/or close in projected, physical separation and/or velocity space)
\end{itemize}
We provide physically-motivated definitions for these objects in Section~\ref{sec:definingmultiagns}. The catalog is not designed to encompass confirmed or candidate gravitationally lensed quasars, though by design our catalog does recover some lens candidates that simultaneously represent high redshift dual AGN candidates. Similarly, AGN pairs with discordant redshifts (referred to here as ``projected AGNs'' or ``projected pairs'') are not scientifically relevant for the present catalog -- which focuses on physically associated multi-AGN -- and we therefore strictly exclude projected pairs. Within this work and catalog, we use ``AGNs'' to refer explicitly to accreting SMBHs.

\subsection{Outline for the Present Work}
\label{sec:outline}

Here in Section~\ref{sec:intro} we provided an introduction to multi-AGN science and motivated the purpose and creation of the multi-AGN catalog described throughout this work. In Section~\ref{sec:definingmultiagns}, we provide definitions for different classes of multi-AGN systems. In Section~\ref{sec:development}, we describe the assembly of both the multi-AGN literature library (Section~\ref{sec:library_assembly}) and construction and formatting of the catalog (Section~\ref{sec:catalog_construct}). We discuss our analysis of selection and confirmation methods in Section~\ref{sec:analysis}, wherein we illustrate the most common selection strategies for multi-AGNs. The general results of our selection and confirmation method analyses are presented in Section~\ref{sec:results} along with our takeaway results examining the distribution of redshifts, angular separations, and pair separations of multi-AGN systems in this catalog. 
 Section~\ref{sec:discussion} discusses the need for more diverse selection strategies (Section~\ref{sec:new_approaches}), the overlap between this catalog and current international celestial reference frame (ICRF) targets (Section~\ref{sec:ICRF}), and shortcomings in the published literature along with best practices for multi-AGN science articles moving forward (Sections~\ref{sec:best_practices}). Section~\ref{sec:conclusion} provides concluding remarks and plans for future work. The Appendix provides an exhaustive chronology of observational multi-AGN science, broken down by system class (dual AGNs, binary AGNs, recoiling AGNs, and N-tuple AGN systems) and selection methodology. While this last section serves as a summary of the multi-AGN literature, a complete synthesis of the theoretical literature for the field of multi-AGNs is beyond the scope of this work; we refer the reader to the most recent review article on multi-AGN science \citet{derosa2019} for a discussion on both theory and observations. Throughout this work and within the catalog, we adopt the following cosmological values: $H_0=70$~km~s$^{-1}$~Mpc$^{-1}$, $\Omega_{\rm{M}}=0.3$, $\Omega_\Lambda=0.7$.

\section{Defining Multi-AGN Systems}
\label{sec:definingmultiagns}

A variety of definitions have been employed in the literature for establishing samples of both galaxy mergers (pre- and post- coalescence) and multi-AGNs. Samples of galaxy mergers studied by \citet{ellison2008,ellison2010,ellison2011} required projected nuclear pair separations of $r_p<80\,h_{70}^{-1}$ kpc and velocity differences $|\Delta v|<200$\,km\,s$^{-1}$; \citet{barton2000} studied galaxy pairs based on $r_p<77\,$kpc and $|\Delta v|<1000$\,km\,s$^{-1}$; \citet{lambas2003} focused on galaxy pairs with $r_p<100\,h_{70}^{-1}$ kpc and $|\Delta v|<350$\,km\,s$^{-1}$, as these galaxy pairs showed enhanced star formation in their analysis; \citet{ellison2013} and \citet{satyapal2014} required $r_p<80\,h_{70}^{-1}$ kpc and $|\Delta v|<300$\,km\,s$^{-1}$; \citet{weston2017} considered only galaxy pairs with $r_p<100\,h_{70}^{-1}$ and $|\Delta v|<1000$\,km\,s$^{-1}$ (whenever two redshifts were available). In the dual AGN literature, \citet{liu2011b} defined dual AGNs as having pair separations $r_p<100\,h_{70}^{-1}$\,kpc and $|\Delta v|\leq600$\,km\,s$^{-1}$; \citet{koss2012} required dual AGNs to have $r_p\leq100$\,kpc and $|\Delta v|\leq300$\,km\,s$^{-1}$; \citet{burke2018} defined dual AGN and binary AGNs as having separations of $\leq10$\,kpc and $\leq10$\,pc, respectively; \citet{stemo2021} focused on dual AGNs with separations $r_p\leq20$\,kpc; and the review by \citet{derosa2019} quote dual AGNs as having separations $r_p<100$\,kpc. Additionally, recent theoretical works using cosmological simulations have consistently limited dual AGNs to separations $<30$\,kpc \citep[e.g.,][]{chen2023,saeedzadeh2024,volonteri2022}. While not every multi-AGN study should be required to examine dual AGNs across the merger sequence \citep[indeed, many dual AGN studies focus specifically on late-stage mergers with separations $r_p<10$\,kpc, e.g.,][]{liu2010a,liu2013,comerford2015,satyapal2017,pfeifle2019a}, it is prudent to develop a standard definition or set of guidelines for use in the field that offer statistical completeness for dual AGNs across the merger sequence. Likewise, myriad nomenclature have been used over the last several decades to refer to dual AGNs, binary AGNs/SMBHs, and recoiling SMBHs; a large -- though probably incomplete -- list of nomenclatures used in the literature for these different system classes is given in Table~\ref{table:aliases} (these lists were compiled during the literature review described in Section~\ref{sec:catalog_construct}). The large variety (and evolution) in nomenclature has made it difficult to keep track of relevant articles and samples (as discussed in Section~\ref{sec:uniform_names}). The definitions derived and nomenclature adopted in the following subsections (Sections \ref{sec:dualagnupperbound}-\ref{sec:definingrecoils}) offer general guidelines that the community can use going forward when assigning system class(es) in future sample(s). 

\begin{table}
\caption{Multi-AGN Nomenclature from the Literature}
\begin{center}
\hspace{-2.2cm}
\begin{tabular}{ccc}
\hline
\hline
\noalign{\smallskip}
Dual AGNs \\ %Binary Quasars & 
\noalign{\smallskip}
\hline
\noalign{\smallskip}
%Quasar pair      & Binary AGN(s)    & Binary SMBH(s)       \\
%Pairs of Quasars & AGN Binary(ies)  & SMBH pair(s)         \\
%Binary Quasar(s)   & Dual Active Galactic Nuclei   & SMBH Binary          \\
%Binary quasar    & Dual Active Galactic Nucleus  & SMBH Binaries        \\
%Quasar binary    & Dual AGN(s)      & Binary Supermassive Black Hole(s) \\
%Quasar binaries  & Double AGN(s)    & Supermassive Binary Black Hole(s) \\
%Dual quasar      & Two Active Nuclei& Supermassive Black Hole Binary    \\
%Dual quasars     & Two AGN(s)       & Supermassive Black Hole Binaries  \\
%Double quasars   & Dual Supermassive Black Holes & SBHB(s) \\
%Double quasar    & Dual SMBH(s)     & Massive Black Hole Binary         \\
%Pairs of QSOs    &     & Massive Black Hole Binaries       \\
%QSO pairs        &     & Recoiling Supermassive Black Hole \\
%Binary QSO       &     & Recoiling SMBH       \\
%Binary QSOs      &     & Recoiling Black Holes\\
%    &     & Recoiling AGN(s)                  \\
Binary AGN(s)  \\               
AGN Binary(ies)      \\         
Dual Active Galactic Nucleus(ei) \\  
%Dual Active Galactic Nucleus  \\
Dual AGN(s) \\                  
Double AGN(s)   \\              
Two Active Nuclei   \\          
Two AGN(s)     \\               
Dual Supermassive Black Holes \\
Dual SMBH(s)    \\              
Quasar Pair   \\                
Pairs of Quasars   \\           
Binary Quasar(s)   \\           
Quasar Binary(ies) \\           
Dual Quasar(s)    \\            
Double Quasar(s)  \\            
Pairs of QSOs   \\              
QSO Pairs      \\  
Binary QSO(s)   \\     
\noalign{\bigskip}
\hline
\hline
\noalign{\smallskip}
Binary AGNs/SMBHs \\ %Binary Quasars & 
\noalign{\smallskip}
\hline
\noalign{\smallskip}
Binary SMBH(s)     \\     
Binary AGN(s)    \\       
AGN Binary(ies)   \\      
SMBH Pair(s)   \\         
SMBH Binary(ies)   \\     
SBHB(s)        \\         
Binary Supermassive Black Hole(s) & \\
Supermassive Binary Black Hole(s) & \\
Supermassive Black Hole Binary(ies)    & \\
Massive Black Hole Binary(ies)  & \\
\noalign{\bigskip}
\hline
\hline
\noalign{\smallskip}
Recoiling AGNs/SMBHs \\ %Binary Quasars & 
\noalign{\smallskip}
\hline
\noalign{\smallskip}
\noalign{\smallskip}
Recoiling Supermassive Black Hole(s)       \\
Recoiling SMBH(s) \\
Recoiling Black Hole(s) \\
Recoiling AGN(s)       \\
\noalign{\smallskip}
\hline
\end{tabular}
\end{center}
\tablecomments{Various nomenclature used for dual AGNs, binary SMBHs, and recoiling SMBHs in the literature. This stacked set of three tables is broken up by system class, where our adopted nomenclature for these object classes are ``dual AGN,'' ``binary AGN,'' and ``recoiling AGN.'' Under each of the adopted nomenclatures, we list all nomenclature/aliases identified for each system class during the literature review described in Section~\ref{sec:catalog_construct}; these lists are likely incomplete. } 
\label{table:aliases}
\end{table}

\subsection{The Upper Bound on Dual AGN Separations}
\label{sec:dualagnupperbound}
%\label{sec:definingduals}
Before we can properly differentiate between the classes of dual AGNs (distinct pairs of AGNs, where each AGN is gravitationally bound to its own host) and gravitationally-bound binary AGNs and SMBHs (residing within a single, coalesced optical nucleus), we first must determine the expected upper bound on the parameter space of physical separations for dual AGNs.

When studying dual AGNs, the primary interest is in merger-driven AGN growth, which places a limit on the separation of the hosts. This is because the torques (which drive the gas inflow mechanisms) are a strong function of pericenter distance \citep[among other parameters, e.g.,]{capelo2015,blumenthal2018}; this can be seen in closed-box simulations of gas rich mergers (major and minor), where a passage at even 10 kpc has (at least modest) effects on the angular momentum of the gas in the outskirts of the galaxies \citep[e.g.,][]{capelo2015,capelo2017shocks} and can begin to drive gas inflow mechanisms \citep{blumenthal2018}.  Given that \citet{capelo2015} observed gas inflows for pericenter passages $<10$\,kpc and \citet{blumenthal2018} observed clumpy gas inflows for a passage at $\sim15$\,kpc, we conservatively assume gas inflows sufficient for fueling SMBH activity arise for pericenter passages of $\leq10$\,kpc.\footnote{Further theoretical efforts are required to better constrain gas inflow triggering distances.}

\begin{figure*}
    \centering
    \includegraphics[width=\linewidth]{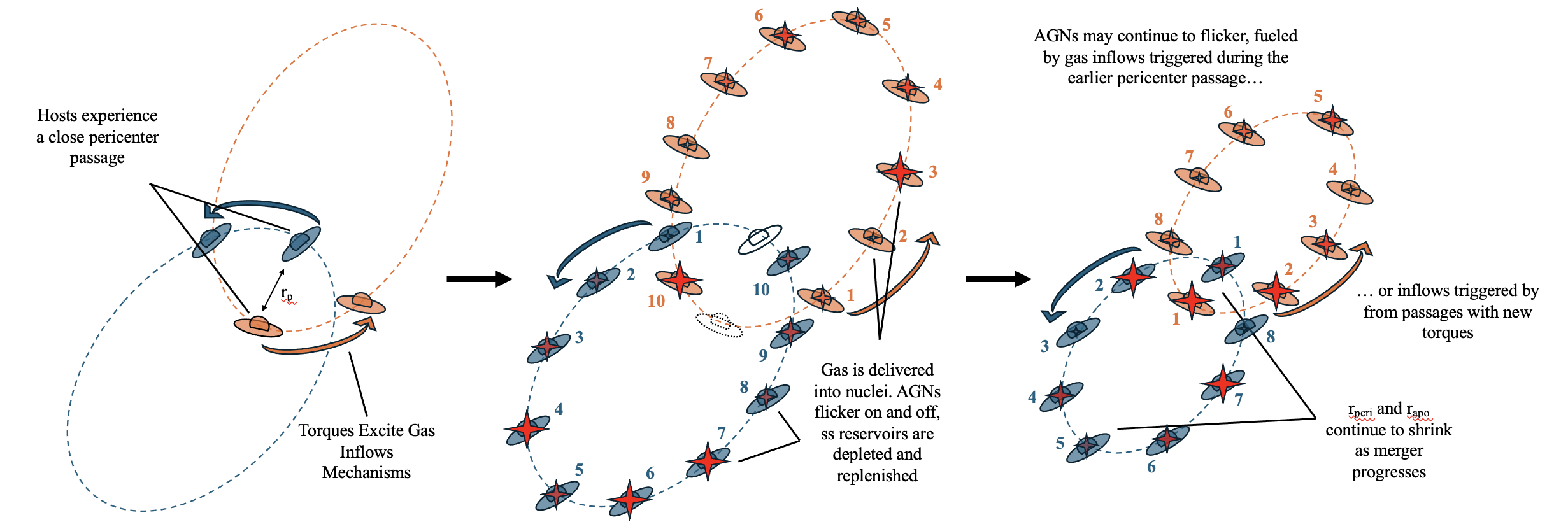}
    \caption{AGN fueling across the merger sequence and across individual host orbits. Here we provide a cartoon illustration of the first and second pericenter passages of a pair of interacting galaxies and the possible AGN activity levels of the SMBHs. Left: the two galaxies experience a close passage at $\sim10$\,kpc; torques induced during the interaction excite gas inflow mechanisms \citep[e.g., bars, shocks, clump inflows,][]{blumenthal2018}. Middle: Gas is delivered into the nuclei on varying timescales with varying efficiencies \citep[][]{capelo2017shocks,blumenthal2018}, SMBH fueling begins. AGNs will flicker on and off across the merger sequence due to changes in the nuclear gas reservoirs. AGNs are shown as red diamonds, and the size and opacity of the diamonds indicate the activity level of the AGNs (larger and more opaque diamonds are more luminous AGNs). The numbers indicate `snapshots' as in a cosmological simulation; AGN activity levels vary between snapshots. Right: As the merger progresses to the second (and later) pericenter passage, the orbits decay, bringing the galaxies and AGNs closer together. Gas inflows are expected to continue, either triggered from the prior passage or the new pericenter passage, and AGN activity is expected to continue, with different activity levels caught in different snapshots. The exact timescales for the orbital evolution of pericenter$_{\rm{current}}\rightarrow$apocenter$\rightarrow$pericenter$_{\rm{next}}$ once the galaxies have passed within $10$\,kpc is unique to a given merging pair of galaxies, but these timescales can range from as short as a few tens of Myr to hundreds of Myr. Note in these panels: dual AGNs are only visible during certain snapshots, to reflect realistic expectations for AGN duty cycles.}
    
    \label{fig:pericenterschematic}
\end{figure*}

The timescales and efficiencies associated with different gas inflow mechanisms \citep{blumenthal2018} could allow for time lags between torquing and AGN fueling and/or more variable AGN fueling \citep[``flickering'' with lifetimes on the order $\sim10^5$\,yr][]{schawinski2015,goulding2019}; hence, there remains the strong possibility that AGN(s) could ignite and be observable at any point between the close pericenter radius, the next apocenter radius, and the following pericenter radius (a timescale that can range from a few tens of Myr to hundreds of Myr; see simple schematic in Figure~\ref{fig:pericenterschematic}). This is consistent with the dual AGNs flickering on and off across the merger sequence in the ASTRID cosmological simulation \citep[e.g.,][]{chen2023}. A non-negligible number of confirmed dual AGNs have been identified in disrupted hosts with separations of $>30$\,kpc and up to even $\gtrsim60-90$ kpc \citep{koss2012,liu2011b,derosa2018,derosa2023,djorgovski1987a},  supporting this picture of AGN and dual AGN fueking across the full merger cycle. It follows that AGNs could be cyclically triggered either by clumpy gas inflows \citep[e.g.,][]{blumenthal2018} or via new inflows triggered by successive pericenter passages of the hosts; this picture of repeated/persistent AGN fueling in early merger-stages (and/or following successive pericenter passages) would be consistent with the AGN excesses recently found at large separations ($\gtrsim100$ kpc) in Illustris-TNG100 galaxy pairs \citep[][see comments on successive fueling in Section~3.2]{byrne-mamahit2024} and well as AGN excesses seen out to separations of $\sim40-50$\,kpc and $\sim60$\,kpc in mid-IR \citep{satyapal2014} and optical \citep{ellison2011} samples of AGNs in mergers, respectively. The key question, then, is: once a pair of galaxies have undergone a pericenter passage at a separation $\lesssim10$\,kpc, what is the typical distance the two galaxies reach before swinging back in for another pass? To answer this key question, we can use the three-dimensional reconstructed orbits of galaxy pairs from Illustris-TNG100 \citep{patton2024} to study the distribution of pericenters and apocenters of interacting galaxy pairs and derive a general upper bound -- independent of any AGN activity -- on the separation at which we should consider two AGNs in an interacting galaxy system to be merger-driven.

\begin{figure}
    \centering
    \includegraphics[width=\linewidth]{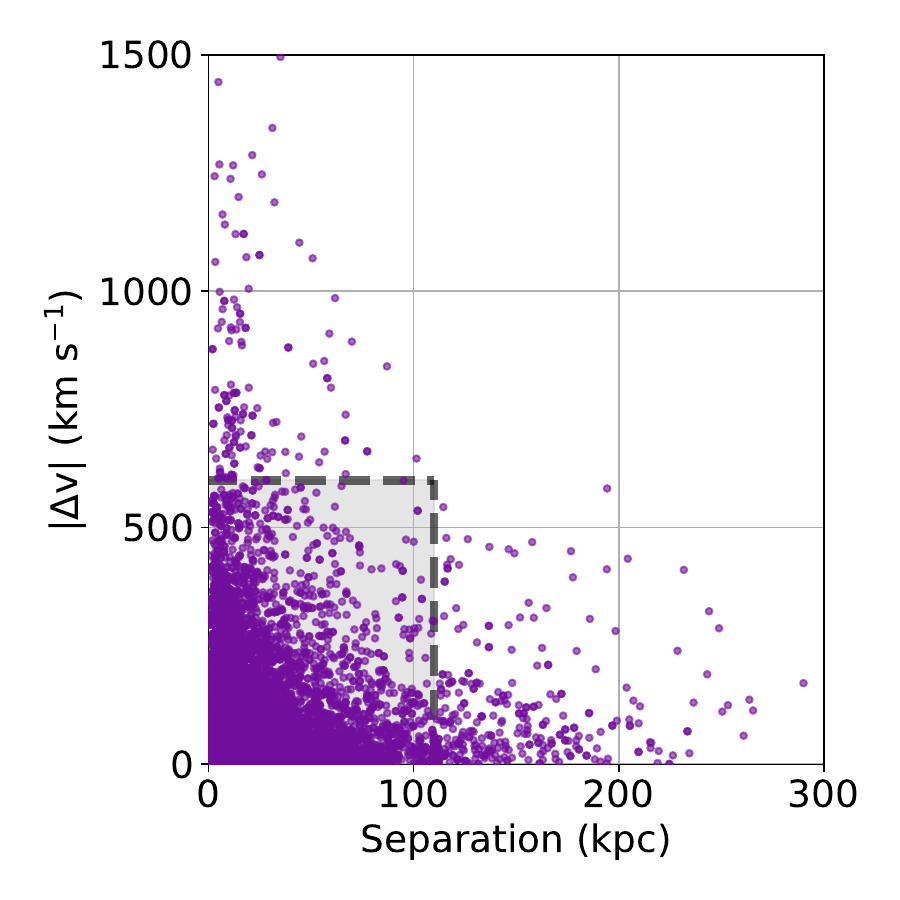}
    \caption{The distribution of three-dimensional distances of interacting galaxy pairs with reconstructed orbits in Illustris-TNG100 \citep[][]{patton2024} after a close passage $<10$\,kpc. The shaded region indicates our expectation for the parameter space of galaxy mergers likely to host dual AGNs (see Section~\ref{sec:definingduals}); the dashed black line ($\leq110$\,kpc and $\leq600$\,km\,s$^{-1}$) encloses 95\% of the sample of merging galaxy pairs.} 
    \label{fig:apocenterdist}
\end{figure}

From the sample of Illustris-TNG100 galaxies with close companions in \citet[][with mass ratios 1:1 to 1:10]{patton2024}, we first selected interacting pairs which recently (within the last 1\,Gyr) experienced a close encounter $r\leq10$\,kpc, where $r$ is the three-dimensional distance between the two galaxies. From this sample of 10027 galaxy pairs, we retrieved the distribution of three-dimensional distances $r$; these separations follow after the $r\leq10$\,kpc pericenter passages, and we show this distribution in Figure~\ref{fig:apocenterdist}. 95\% of systems have $r < 114.2$\,kpc (three dimensional), or projected (two dimensional) separations $r_p < 92.3$\,kpc. These separations  place a fiducial upper limit on the general expectation for separations of dual AGNs. In real samples of dual AGNs, a cut based on line-of-sight velocity difference is necessary to remove projected pairs of AGNs residing at different redshifts \citep[e.g.,][]{ellison2011,liu2011b,koss2012}; within this sample of simulated galaxy pairs, a cut in velocity difference of $\Delta v < 300$, $400$, $500$\,km s$^{-1}$ removes 9.2\%, 4.5\%, 2.7\% of the full sample, respectively. Coupling pair separation with a velocity difference like in the literature, 95\% of pairs have $\Delta v < 400$\,km s$^{-1}$ and $r_p < 205.2$\,kpc, 95\% have $\Delta v < 500$\,km s$^{-1}$ and $r_p < 126.9$\,kpc, 95\% have $\Delta v < 600$\,km s$^{-1}$ and $r_p < 109.1$\,kpc. In a realistic orbit, the galaxies should experience larger line-of-sight velocity differences at smaller pair separations and smaller velocity differences at larger pair separations; in a future release, we will introduce definitions for dual AGNs that incorporate velocity difference as a function of separation, rather than flat cuts in separation and velocity difference.

In this release of the Big MAC, we adopt an upper bound of 110\,kpc for dual AGN separations, informed by the projected separations observed in the sample of galaxy pairs in Illustris-TNG100 \citep{patton2024}. We further adopt a flat cut in (systemic) line-of-sight velocity difference of $\Delta v < 600$\,km s$^{-1}$, again informed by the sample of Illustris-TNG100 galaxy pairs. These operational definitions are open to revision in future data releases should future cosmological simulations demonstrate that a tighter upper bound on separation (e.g., 100\,kpc, 90\,kpc, etc.) or velocity difference (e.g., $\Delta v < 500$\,km s$^{-1}$, $< 300$\,km s$^{-1}$, etc.) are more representative of dual AGNs or mergers that trigger dual AGNs. With a statistically complete set of defining criteria for dual AGNs -- and upper limit on the parameter space of pair separations and velocity difference -- grounded in theoretical predictions from Illustris-TNG cosmological galaxy pairs \citep{patton2024}, we turn our attention to the transition point from dual AGNs to gravitationally-bound binary AGNs (Section~\ref{sec:transition_duals_to_binaries}).

\subsection{The Transition from Dual AGNs to Binary AGNs}
\label{sec:transition_duals_to_binaries}

The dynamical evolution of SMBH pairs, from the point at which two SMBHs become gravitationally-bound to the binary hardening, inspiral, and eventual coalescence, was first detailed in \citet{begelman1980}. \citet{volonteri2003} redescribed the mathematical framework of \citet{begelman1980} assuming a singular isothermal sphere (SIS) as a simple model to approximate the isotropic background of stars encountered by the SMBHs, with a stellar density of $\rho_{*}=\sigma^2/2\pi {\rm{G}}r^2$, where $\sigma$ is the stellar velocity dispersion.\footnote{In reality, such merging systems are neither isothermal nor virialized, so this only represents an idealized toy model.} One can determine the radius at which the two SMBHs become gravitationally-bound to one another by inserting the SIS into the separable differential equation:
\begin{equation}
    \frac{dM}{dr} = 4\pi r^2 \rho_{*}
\end{equation}
and integrating to obtain the following relation: 
\begin{equation}
    \label{eq:boundradius}
    r_{b} = \frac{G(m_{1}+m_{2})}{2\sigma^2} \approx 10\,\rm{pc}\frac{m_{1}+m_{2}}{10^8 M_{\odot}}\frac{\sigma^{-2}}{150\, \rm{km\,s}^{-1}}
\end{equation}
where the latter form of this expression -- contextualized in terms of physical units -- was given in \citet{xliu2018a}. In this form, the theoretical boundary dividing duals and binaries comes with the advantage of being expressed in observationally accessible terms (i.e., $\sigma$). Here, we aim to graphically illustrate the dividing line between duals and binaries, and describe how this toy model definition shifts depending on the properties of the host (e.g., $\sigma$ or stellar mass) and the properties of the SMBHs (e.g., binary mass).

The well-known M-$\sigma$ relationship relates the mass of the central SMBH and $\sigma$ of the host \citep{ferrarese2000,gebhardt2000,gultekin2009,kormendy2013}. Here we adopt the M-$\sigma$ relation from \citet[][Eq. 3 and 7 in Section 6.6]{kormendy2013}, given as:
\begin{equation}
    \label{eq:msigma}
    \frac{M_\bullet}{10^9\,M_{\odot}} = 0.309 \times \left(\frac{\sigma}{200\,\rm{km}\,\rm{s}^{-1}}\right)^{4.38}
\end{equation}

In our discussion here, we refer to the combined mass of the two SMBHs as the ``enclosed mass.'' We can make the simple assumption that given a central stellar velocity dispersion $\sigma$, we can determine the enclosed central mass of the binary and determine the radius at which a dual AGN/SMBH becomes a gravitationally bound binary AGN/SMBH. We make two key assumptions: (1) galaxies that are built up across cosmic time through subsequent mergers follow the M-$\sigma$ relation, and (2) the timescale for a dual AGN to sink via dynamical friction down to the gravitationally-bound binary regime is on the order of the timescale for the new host to settle sufficiently so as to allow an accurate measurement of the new host stellar velocity dispersion. %in a given gravitationally-bound binary

\begin{figure}
    \centering
    \includegraphics[width=\linewidth]{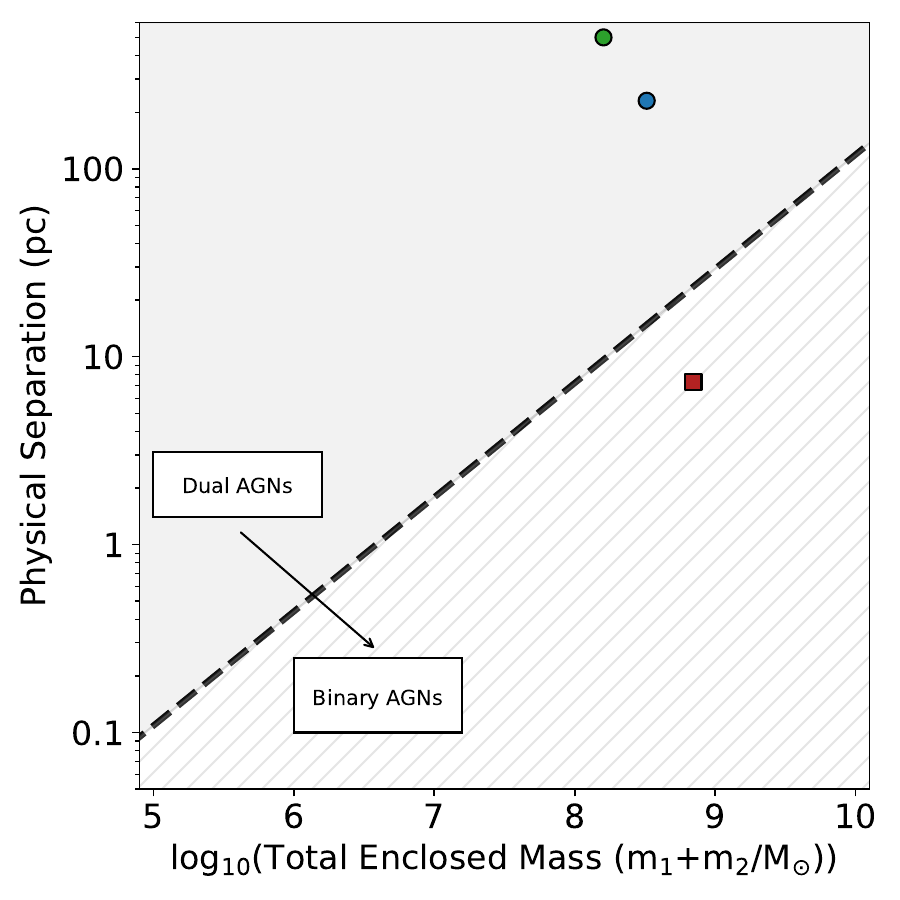}
    \caption{The transition from dual AGNs to binary AGNs. The logarithm of the total enclosed mass of a SMBH pair is shown on the x-axis, and the physical separation (in pc) is given on the y-axis. The dashed, black line corresponds to the radius, $r_b$, at which the two SMBHs in a pair become dominated by their mutual gravity, and hence a dual AGN/SMBH transitions to a binary AGN. Labels for ``Dual AGNs'' and ``Binary AGNs'' are used to indicate that dual AGNs are found above this transition line, and binaries are found below it. We plot several sub-kpc dual AGNs (blue: UGC 4211, \citealp{koss2023}, green: NGC 7727, \citealp{voggel2022}) and the single known binary AGN, CSO 0402+379 \citep{maness2004,rodriguez2006}, where the separations are the projected separations and the enclosed masses were either inferred from the stellar velocity dispersion (for CSO 0402+379) or direct dynamical or spectroscopic measurements \citep{voggel2022,koss2023}. }
    \label{fig:dual_v_binary}
\end{figure}

We use the M-$\sigma$ relation in Equation~\ref{eq:msigma} to derive a series of enclosed masses for a given list of $\sigma$ values, and we plug these enclosed masses into Equation~\ref{eq:boundradius} to derive the corresponding gravitationally-bound radii. We plot these radii in Figure~\ref{fig:dual_v_binary}, which illustrates the relationship between binary SMBH separation and enclosed SMBH mass (and hence, stellar velocity dispersion); the dashed black line shows the transition radius $r_b$, above which we would consider a SMBH pair with a given enclosed mass and separation a dual AGN, whereas below this line a pair of SMBHs would be considered a binary AGN for a given enclosed mass and pair separation. This toy model provides a simple method to differentiate between dual and binary AGNs when the separations approach this regime, and (when necessary) we use this toy model when differentiating between dual and binary AGN in the Big MAC. This toy model comes with the clear caveat that it is a simple approximation and intrinsically relies upon the assumption that the enclosed mass of a binary and the host $\sigma$ follow the M-$\sigma$ relation. For a given $\sigma$, this toy model would overestimate (underestimate) $r_b$ when the SMBHs are overmassive (undermassive) relative to their host galaxy \citep{steinborn2016,volonteri2022,chen2023}.

\subsection{Defining Dual AGNs}
\label{sec:definingduals}

Based upon our discussion in Sections~\ref{sec:dualagnupperbound} and \ref{sec:transition_duals_to_binaries}, the following criteria offers guidance for defining dual AGNs. In this data release, we define dual AGNs as any pair of AGNs in a physically interacting system with projected separations:
\begin{equation}
    \label{eq:dualsep}
    r_b \lesssim r_p < 110\,\h\,\rm{kpc}
\end{equation}
\noindent and with line-of-sight velocity differences:
\begin{equation}
    |\Delta v| \lesssim 600\,\rm{km}\,\rm{s}^{-1}
\end{equation}
\noindent where the velocity difference is calculated using the systemic redshifts of the host galaxies. $r_b$ in Equation~\ref{eq:dualsep} was defined in Equation~\ref{eq:boundradius} in Section~\ref{sec:transition_duals_to_binaries} and depends upon the masses of the SMBHs and the stellar velocity dispersion of the post-merger host galaxy: assuming the enclosed mass of the SMBH pair correlates with the host stellar velocity dispersion, $r_b\approx2.7$\,pc, $\approx9.4$\,pc, and $\approx32.9$\,pc for SMBH pairs with total masses of $10^7$ M$_{\sun}$, $10^8$ M$_{\sun}$, and $10^9$ M$_{\sun}$, respectively. These benchmark $r_b$ values may not be applicable to more complex systems where the SMBH pair is over-massive or under-massive relative to the host stellar velocity dispersion.

For cases in which broad emission lines are used to trace the redshifts of the AGNs, as in the case of high redshift dual AGNs \citep[e.g.,][]{hennawi2006,hennawi2010}, guidance on velocity differences comes in the form of:
\begin{equation}
    |\Delta v| < 2000\,\rm{km}\,\rm{s}^{-1}
\end{equation}
This velocity difference incorporates both the potential for high velocity blue shifts in the broad lines (such as those observed in CIV] that can be up to $
\sim1500$\,km\,s$^{-1}$, as well as the peculiar velocities of the host galaxies \citep[e.g.,][]{hennawi2006}. However, AGN pairs identified via broad emission lines (from individual fiber spectra or individual spectra obtained via follow-up spectroscopy) with $|\Delta v|<2000$\,km\,s$^{-1}$  should be considered candidate dual AGNs until follow-up spectroscopy can constrain the host redshifts and ensure the hosts are physically associated. Pairs of AGNs with  $|\Delta v|>2000$\,km\,s$^{-1}$ based on broad lines or $|\Delta v|>600$\,km\,s$^{-1}$ based on host systemic velocities should be considered ``projected'' AGNs rather than interaction-driven or interaction-associated dual AGNs. However, to avoid imposing possibly detrimental definitions, we do not provide a recommended terminology for these non-dual AGNs.

While the physically-motivated definitions for dual AGNs are designed to encompass the likely parameter space(s) of these systems -- based on the reconstructed orbits of merging galaxy pairs in Illustris-TNG100 \citep{patton2024} -- it is possible that some minor fraction of dual AGNs may not strictly abide by these guidelines. Dual AGN candidates that do not abide by these guidelines should be approached with appropriate caution, and unambiguous evidence is required to prove that such objects are indeed bona fide dual AGNs. The operational definition outlined here for dual AGNs is subject to change in future releases should future theoretical works provide more stringent constraints on the velocity difference and pair separation parameter spaces of dual AGNs. The term ``dual AGN'' should not be used to refer to AGN pairs with discordant redshifts, nor should it be used to refer to gravitationally-bound pairs of AGNs/SMBH (which are defined below in Section~\ref{sec:definingbinaries}).

For systems in which only a single SMBH is active and the other is inactive, we refer to these as dual SMBHs or single AGNs; we do not refer to these as offset AGNs \citep[e.g.,][]{comerford2013,chen2023}, as this latter terminology  can be confused with true kinematically/spatially offset AGNs  and does not sufficiently capture the activity/lack of activity within a given SMBH pair. 

\subsection{Defining Binary AGNs}
\label{sec:definingbinaries}

As discussed in Section~\ref{sec:transition_duals_to_binaries}, binary AGNs or binary SMBHs are pairs of SMBHs sufficiently close in separation to be dominated by their mutual spheres of influence and thus gravitationally-bound to one another. The radius at which two SMBHs become gravitationally-bound, $r_b$, is given by Equation~\ref{eq:boundradius} in Section~\ref{sec:transition_duals_to_binaries} and depends upon the masses of the two SMBHs and the stellar velocity dispersion of their host galaxy. Our toy model illustrating the point at which dual AGNs evolve into gravitationally-bound binary AGNs is shown in Figure~\ref{fig:dual_v_binary}. Assuming again that the enclosed binary mass correlates with the host $\sigma$, $r_b\approx2.7$\,pc, $\approx9.4$\,pc, and $\approx32.9$\,pc for binaries with total masses of $10^7$\,M$_{\sun}$, $10^8$\,M$_{\sun}$, and $10^9$\,M$_{\sun}$, respectively. In other words, more massive SMBH pairs naturally become gravitationally-bound at larger separations, whereas less-massive SMBH pairs must reach much closer separations to become bound; throughout this work, we refer to binaries as generally having separations $\lesssim30$\,pc. We emphasize again that these simple benchmark $r_b$ values are included for ease of reference and may not be applicable to more complex systems where the SMBH pair is over-massive or under-massive relative to the host \citep{steinborn2016,volonteri2022,chen2023}. The term ``binary AGN'' should not be used to refer to pairs of AGNs that are not gravitationally-bound to one another, i.e. should not be used to refer to dual AGNs.

\subsection{Defining Recoiling AGNs}
\label{sec:definingrecoils} 
Recoiling AGN/SMBH candidates are SMBHs that are the product of SMBH-SMBH mergers and therefore may manifest as spatially (in projected space) or kinematically (in velocity space) offset from their host nuclei due to the anisotropic emission of gravitational waves following the SMBH merger \citep[which result in a `kick' velocity][]{schnittman2007a,campanelli2007,blecha2011}. A large variety of possible kick speeds are possible for recoiling AGNs depending upon the properties of the SMBH binary prior to coalescence \citep{bogdanovic2007,blecha2011}; under specific conditions where the SMBHs are equal in mass, maximally spinning and spin-misaligned, recoil speeds can even reach 4000\,km\,s$^{-1}$ \citep[e.g.,][]{campanelli2007}. Recoiling SMBHs/AGNs could be capable of not only pc- and kpc-scale positional offsets and significant velocity offsets, but could be ejected from the host galaxy altogether in the highest velocity cases \citep[e.g.,][]{blecha2011}. 

Obviously, not all SMBHs that result from SMBH-SMBH mergers will receive significant enough kicks or have long enough AGN lifetimes to be selected as recoiling AGNs candidates \citep{blecha2011}; without a significant spatial or velocity offset, it may be difficult to differentiate between a SMBH that was the product of a SMBH-SMBH merger and a ``typical'' SMBH. However, those SMBHs with spatial and kinematic offsets can be selected, characterized, and used to place statistical constraints on the rates of SMBH-SMBH mergers and recoiling SMBHs. Recoiling AGNs are perhaps the most difficult objects to confirm; no recoiling SMBHs/AGNs have been confirmed in the literature so far, though many candidates exist \citep[e.g.,][]{komossa2008,civano2010,civano2012,steinhardt2012,kim2016,lena2014,cheung2007}.

It is theoretically possible for three-body interactions between an inner SMBH binary and a third intruding SMBH to result in a slingshot orbit of the third SMBH \citep[and possibly even ejection from the host, e.g.,][]{hoffman2007}, and such systems are also referred to as recoiling AGNs/SMBHs; it may be impossible to differentiate between recoiling SMBHs due to SMBH-SMBH mergers and those ejected as a result of a three-body interaction. Recoiling SMBHs or AGNs should generally be distinguishable from dual and binary AGNs because recoiling AGNs involve only a single SMBH rather than two distinct SMBHs. As such, it is not necessary to develop a rigorous method of differentiating between recoils and the other multi-AGN classes based on, for example, separations or velocity differences.

Due to the large range in kick velocities and spatial offsets possible for recoiling AGN candidates \citep[and other possible formation mechanisms, like three body interactions, e.g.,][]{hoffman2007}, we do not provide a rigorous definition for recoils based on nuclear offset or velocity offset; we instead adopt the very loose definition that recoiling AGNs are any AGN/SMBH exhibiting a kinematic and/or spatial offset from the host nucleus (caused by SMBH-SMBH mergers, triple body interactions, etc.).

\begin{table*}[t!]
\caption{Summary Table of Recommended Dual and Binary AGN Definitions from Section~\ref{sec:definingmultiagns}}
\begin{center}
\hspace{-2.2cm}
\begin{tabular}{cc}
\hline
\hline
\noalign{\smallskip}
System Class & Definition \\%& 
\noalign{\smallskip}
\hline
\noalign{\smallskip}
Dual AGN & 
    $\left\{
        \begin{array}{ccc}
                z_1\,\textrm{or}\,z_2\,\,\textrm{known},  &  0.03\,\textrm{kpc}\lesssim r_p\lesssim110\,\textrm{kpc},\,\,\,\, & \textrm{complete $z$ info still required}\\
                z_1\,\textrm{and}\,z_2\,\,\textrm{known, derived via narrow lines},  &  0.03\,\textrm{kpc}\lesssim r_p\lesssim110\,\textrm{kpc},\,\,\,\, & |\Delta v| < 600\,\textrm{km\,s}^{-1} \\
                z_1\,\textrm{and}\,z_2\,\,\textrm{known, derived via broad lines}, &  0.03\,\textrm{kpc}\lesssim r_p\lesssim110\,\textrm{kpc},\,\,\,\, & |\Delta v| < 2000\,\textrm{km\,s}^{-1}
    	\end{array}
    \right.$\\
%& Crab Nebula & Supernova remnant & 2 & Taurus & 8.4 \\

\noalign{\smallskip}
\noalign{\smallskip}

Binary AGN & 
    \textrm{Gravitationally-bound},\,\,\, $r_{b} \leq \frac{G(m_{1}+m_{2})}{2\sigma^2} \approx 10\,\rm{pc}\frac{m_{1}+m_{2}}{10^8 M_{\odot}}\frac{\sigma^{-2}}{150\, \rm{km\,s}^{-1}}$
 \\ 
\noalign{\smallskip}
\hline
\end{tabular}
\end{center}
\tablecomments{A summary of dual and gravitationally bound binary AGN definitions outlined in Section~\ref{sec:definingmultiagns}. We provide three separate definitions for dual AGNs: (1) One based only on separation of the nuclei/AGNs, but for which we emphasize that complete spectroscopic redshift information is needed; without unambiguous morphological evidence, spectroscopic redshifts are required. (2) A definition based on both separation and a velocity difference using redshifts derived via narrow emission lines (giving the systematic redshifts of the hosts). (3) A definition based on separation and velocity difference when the redshifts are derived from broad emission lines (and may not directly trace the host redshifts). In this last case, redshifts for the AGN hosts are still needed to ensure the AGNs are actually physically associated. We adopt $\sim0.03$\,kpc as the approximate dividing line between dual AGNs and binary AGNs, but in reality this value depends on the masses of the SMBHs and the velocity dispersion of the host (see Section~\ref{sec:definingbinaries}). Given the broad range in predicted properties of recoiling AGNs, we do not provide any specific, quantitative definitions for recoiling AGNs. }
\label{tab:multiAGN_definitions}
\end{table*}

\section{Catalog Development}
\label{sec:development}
\begin{figure*}[t!]
    \centering
    \includegraphics[width=0.98\linewidth]{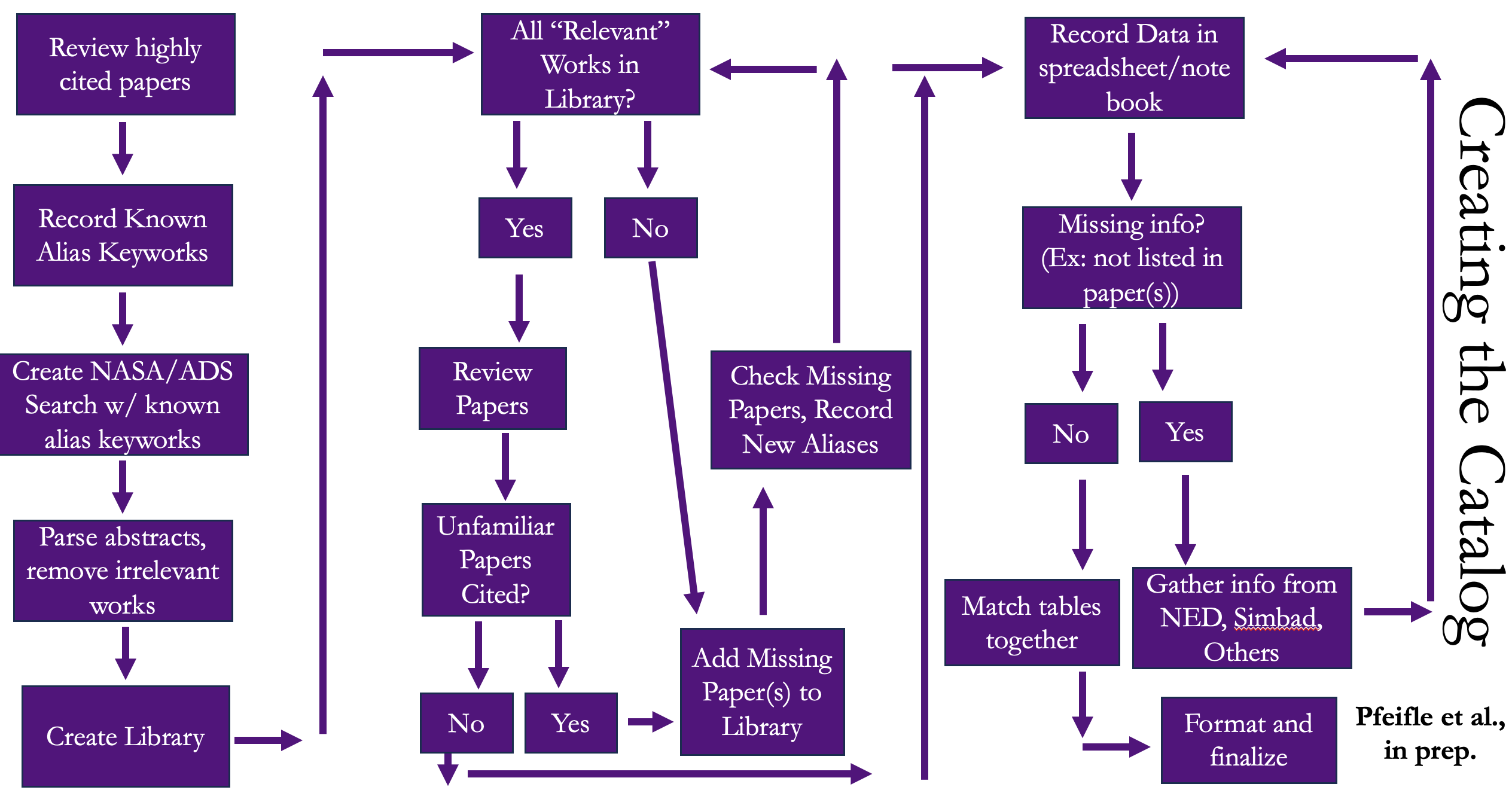}
    \caption{A flow-chart illustrating our systematic method of identifying papers in ADS related to observational searches and analyses of multi-AGN systems.}
    \label{fig:libraryassembly}
\end{figure*}

\subsection{Literature Library Assembly}
\label{sec:library_assembly}
%dear lord siri butcherd my voice to text here
In this section we discuss the particular manner in which the Big MAC ADS literature library was constructed. Due to the breadth of the multi-AGN literature, we converged on a systematic and iterative search strategy for refereed papers in the Astronomy ADS database that relied upon multi-AGN keyword searches (see Table~\ref{table:aliases}). This  systematic process is illustrated in the flow chart in Figure~\ref{fig:libraryassembly}, and the steps are outlined here: 

\begin{itemize}
    \item Step 1: A few dozen commonly-cited\footnote{Citation counts on the order of $\gtrsim80$.} papers (specifically known to the authors at the time) provided a small pool of naming schemes for different classes of multi-AGN systems (e.g., ``dual AGN,'' ``binary AGN,'' ``SMBH binary,'' etc.). We refer to these names as ``alias'' keywords, and with these alias keywords we initiated our ADS search(es). The complete lists of aliases identified during this entire process are shown in Table~\ref{table:aliases}. 
    \item Step 2: We searched for all refereed papers in the ADS Astronomy database that included instances of the known alias keywords associated with multi-AGN classes. During this step, we ensured that all papers known to the authors (regardless of citation count) were recovered; if known papers were not recovered, we checked these missing papers for additional aliases, added the aliases to the ADS search, and then repeated this step. This iterative process allowed us to cast a wider net that would result in an exhaustive and inclusive literature search.
    \item Step 3: We added all articles resulting from the ADS search(es) to an ADS library. These (iterative) searches returned more than 1300 articles. We then parsed the abstracts in the ADS library to remove miscellaneous works that did not report/contain multi-AGN candidates relevant for the construction of the catalog (i.e., theoretical works with no observational results, works entirely unrelated to multi-AGNs, and/or works that mention the aliases in passing but are not directly relevant). In some cases (on the order of 100-200 papers), if we could not ascertain from the abstract alone whether a paper focused on or discussed multi-AGNs in a meaningful manner, we instead skimmed through individual papers to determine their relevancy. 
    \item Step 4: Once the ADS library was sufficiently built up -- i.e. it seemed likely to contain the majority of the multi-AGN literature -- we began a full review of all papers in the library, starting from the earliest publication dates and moving toward the latest published articles. In a number of instances, we found that certain -- more general -- aliases led to the inclusion of irrelevant works: for example, ``double quasar'' was in some cases used when discussing gravitational lenses and physically associated quasar pairs \citep{porcas1979} as well as to refer to high redshift quasars in close (projected) proximity but at discordant redshifts \citep[e.g.,][]{stockton1972} . These papers had to be manually removed after review because it was not clear that these were irrelevant during our initial checks in Step \#3. During this review process, we often discovered papers in our ADS library that cited additional papers not recovered in our original searches. We were then required to iterate back through the previous steps, adding the additional aliases found in the newly identified papers to our list of aliases, searching ADS with the revised alias list, parse the abstracts of the newly recovered articles, and remove irrelevant articles. The construction of the catalog began alongside our full literature review.
\end{itemize}
The full literature library is housed in a public ADS library\footnote{ADS Library: https://ui.adsabs.harvard.edu/public-libraries/KbtXN9q4SzG5X-KJCJ3hxQ}, which can be accessed at all times from the public GitHub website\footnote{Landing page: https://thatastroguy.github.io/thebigmac/} and repository\footnote{Repository: https://github.com/thatastroguy/thebigmac} for the catalog.

\subsection{Catalog Construction}
\label{sec:catalog_construct}
In this section we describe the process of assembling the multi-AGN catalog. Once Step \#4 in Section~\ref{sec:library_assembly} (above) was underway (articles from the ADS library were being actively reviewed), the catalog containing relevant information for these multi-AGN systems could begin to take shape, and we describe this process here:
\begin{itemize}
    \item As articles were thoroughly reviewed, we parsed each paper for relevant information, including: system name and constituent AGN/SMBH names, coordinates, coordinate wavebands and sources, redshifts and redshift types, brightness measurements, selection methodologies, analysis methodologies, confirmation methodologies, angular and/or physical separations, velocity differences, and author information (author name and year, BibCodes, and DOI link).
    \item In a single spreadsheet designated for ``individual targets'' (as opposed to large samples), we recorded the information for single objects or small samples of objects that were the focus of one or a few related articles.
    \item Articles focusing on large samples of objects fall into three categories: (1) machine-readable tables were provided by the authors that could be read with Python without any formatting issues; (2) ``machine-readable'' tables were provided, but the information was not formatted in a completely machine-readable fashion (or, in some cases, the formats were not conducive to our objectives and/or column-wise operations); (3) machine-readable tables were not provided by the authors. In the case of (2), the data table(s) required manual adjustments and reformatting to make them machine-readable. In the case of (3), information needed to be manually copied from the in-text tables into a new data file and formatted into a delimited, machine-readable format. Once these machine-readable tables were assembled and formatted in a compatible fashion, they were loaded into distinct Jupyter Notebooks (typically based on selection method/object type) for further formatting and later matching. Table formatting and compatibility posed a considerable challenge during the matching process and required careful evaluation (and in some cases, reformatting) of all source catalogs and direct supervision of the source matching process. 
    \item In order to account for source overlap between different samples and independent works, we took a two-step approach to matching samples together: in the first step, large samples of objects selected using the same or similar methodologies and targeting the same kinds of objects were matched together first (based on names or coordinates). This initial matching process took place in distinct Jupyter Notebooks, where each Notebook typically focused on a single selection method/object type (for example: a Notebook was dedicated to binary AGN candidates selected based on velocity offset broad spectroscopic emission lines).  Objects selected in the same or similar fashion are likely to overlap between samples, so these Jupyter Notebooks were used to generate overarching, object-matched tables containing unique entries typically based on selection method. Once these selection method-based tables were created, we proceeded to the second step of the matching process: these large-sample tables along with the ``individual targets'' table were all coordinate matched against one another. This process was extensive and directly supervised, during which we iteratively adjusted the matching tolerance (from sub-arcsecond to tens of arcsecond tolerances in some cases) and compared angular separation and coordinates between matching objects to ensure all systems/nuclei were properly matched. During this process, duplicate targets were discarded, and for each matched object we appropriately assigned the associated (1) author name(s), BibCode(s), and DOI(s), and(2) selection and confirmation methodologies. 

    \item Once all tables were matched together into a single single table product, further formatting and final quality checks were required. This included final checks to: (1) track down or derive any information still missing (coordinates, redshifts and redshift types, names or naming conventions, angular or physical separations, etc.); (2) reclassify systems (e.g., relabeling ``binary AGNs'' as ``dual AGNs'' when the SMBHs are not gravitationally-bound); (3) introduce (subjective) confidence flags; (4) remove contaminants that do not meet our multi-AGN definitions outlined in Section~\ref{sec:definingmultiagns};  (5) organize and format the table to be machine-readable and as user-friendly for the community as possible. Additionally, (6) selection, analysis, and confirmation methods fo each system were organized into nested Python dictionaries using waveband:technique pairs for key:value access (see discussion in Section~\ref{sec:assigningselectionmethods}).
    
    \item As we will emphasize in Section~\ref{sec:discussion}, there was a prevalence of missing information in a substantial number of refereed multi-AGN literature articles, including: distinct nuclear coordinates or even general system coordinates, distinct nuclear and/or system redshifts, brightness measurements, selection methods, etc. It was necessary to track down this information using online archives such as NED\footnote{https://ned.ipac.caltech.edu}, Simbad\footnote{https://simbad.cds.unistra.fr/simbad/}, and/or parsing additional papers to retrieve these measurements. In some instances, such as where coordinates were missing for a few dual AGN systems, we directly retrieved the nuclear coordinates from archival optical imaging (e.g., SDSS, DeCaLs, etc.). %in a substantial number of articles
\end{itemize}

Table~\ref{tab:maintable} shows the column headings, data types, and descriptions for each column for the Big MAC DR1 table containing dual AGN, binary AGN, and recoiling AGN candidates. The N-tuple Big MAC tables were designed similarly but with additional columns to account for multi-AGN systems comprising three or more AGNs. 

Within the Big MAC, spectroscopic redshifts were prioritized  over photometric redshifts. Photometric redshifts were only used if no spectroscopic redshifts were available; redshift types are noted in the \texttt{z1\_type} and \texttt{z2\_type} columns (Table~\ref{tab:maintable}). In some cases, spectroscopic redshifts published in the original discovery papers were superseded by redshift measurements obtained with more recent observations \citep[e.g.,  RXJ162902.3+372434,][]{mason2000,inada2008}. Spectroscopic redshifts represent a crucial component in determining the projected, physical separations and velocity differences for a given multi-AGN system and therefore play an important role in assigning the appropriate system class (Section~\ref{sec:assignsystemclass}) and confidence flag (Section~\ref{sec:flagsystem}). As an example, the redshifts derived by \citet{inada2008} for RXJ162902.3+372434 revealed that the AGNs were not physically associated ($|\Delta v|>3000$\,km\,s$^{-1}$), contrary to \citet{mason2000}.

\begin{figure}[t!]
    \centering
    \includegraphics[width=0.99\linewidth]{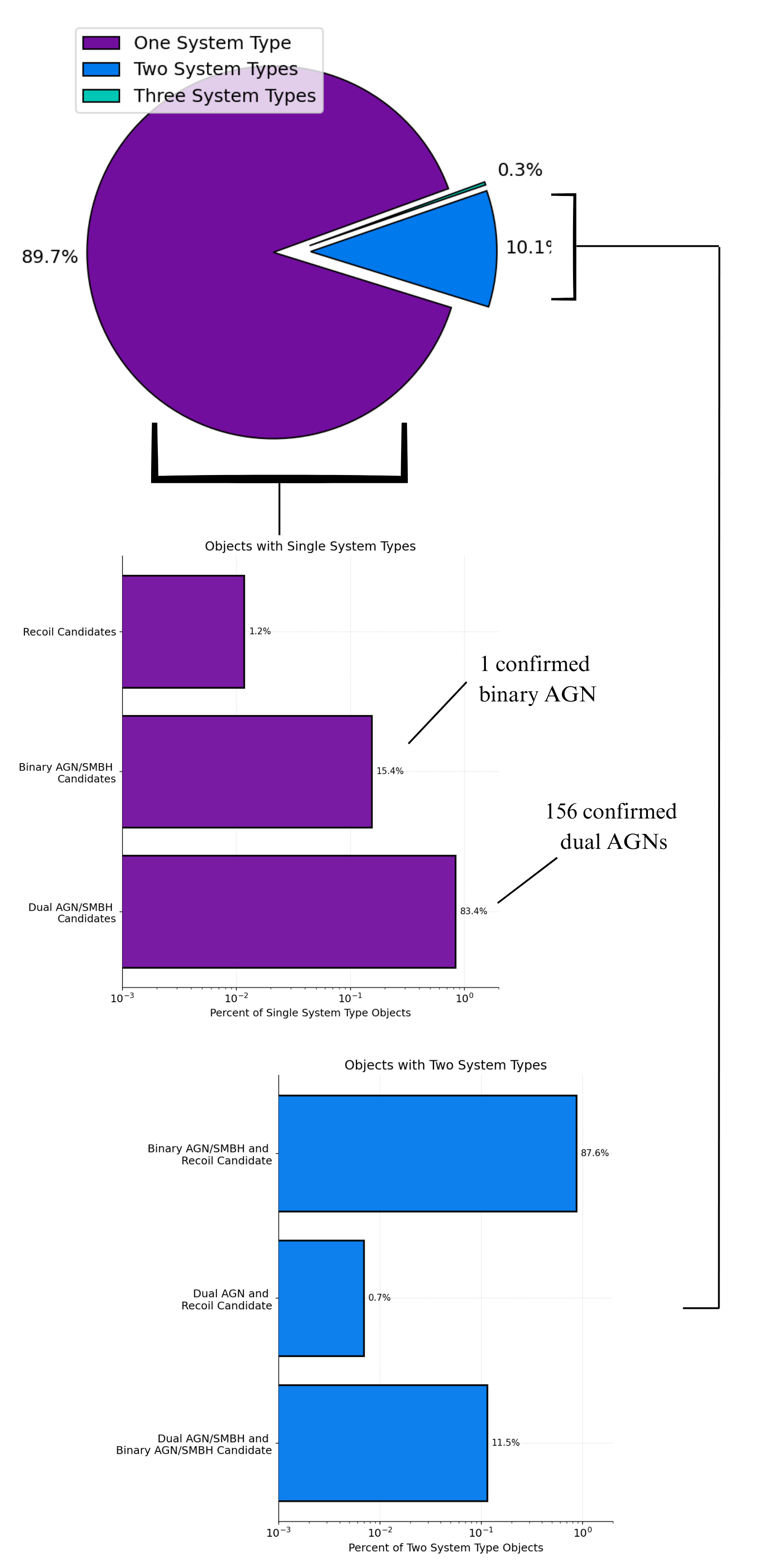}
    \caption{The breakdown of single or cross-listed system types in Big MAC DR1. The majority of systems within DR1 have been assigned a single system class ($89.7$\%); 10.15\% systems fall into two system classes, and 0.3\% systems fall into three system classes. } 
    \label{fig:class_breakdown}
\end{figure}

\begin{figure*}
    \centering
    \includegraphics[width=0.9\linewidth]{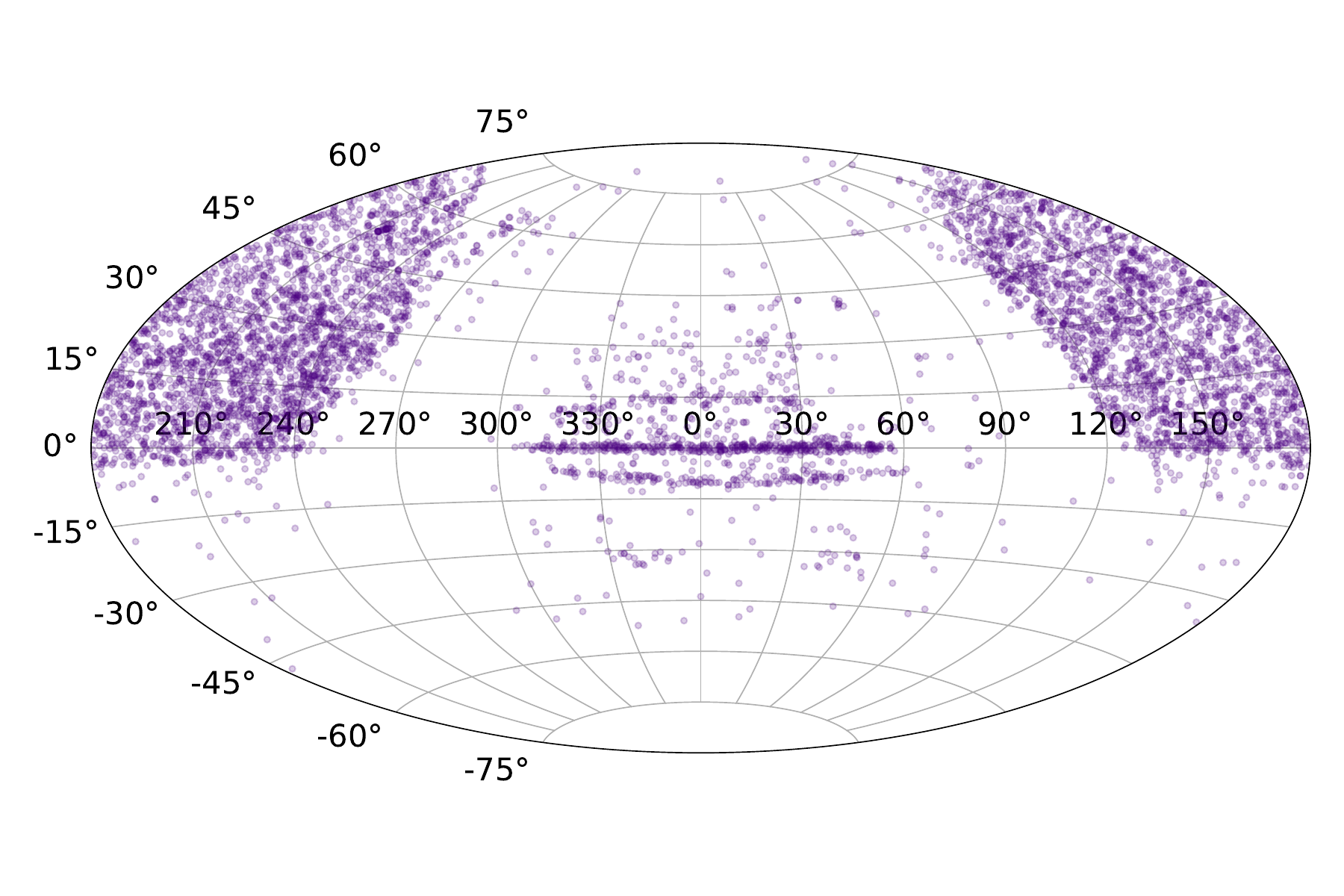}
    \caption{Sky distribution of Big MAC MAC DR1 systems. The current populations of multi-AGN systems are largely biased toward the SDSS field.}
    \label{fig:sky_dist}
\end{figure*}

\subsection{Assigning System Classes}
\label{sec:assignsystemclass}

For every multi-AGN system included within the catalog, we assign relevant ``system class'' flags based upon the physically-motivated definitions provided in Section~\ref{sec:definingmultiagns} as well as upon the choice of selection strategies used to identify the system (e.g., X-shaped radio sources could be formed via a recoiling AGN or an unresolved radio binary AGN). These multi-AGN system classes include: dual AGNs/SMBHs, binary AGNs/SMBHs, recoiling AGNs, and N-tuple AGNs. In $\sim10$\% of all cases, multi-AGN systems were identified as two independent classes (for example: a dual AGN and a binary AGN, based on different selection techniques), and$\sim0.3$\% of all systems were identified as three independent classes. To account for class cross-listings, each entry in the catalog can be assigned up to three distinct classes (a ``primary,'' ``secondary,'' and ``tertiary'' class). Entries with only a single system class are assigned a ``primary'' class, and the secondary and tertiary classes as flagged as `-99;' for entries with two system classes, the primary and secondary classes are assigned and the tertiary class is flagged as `-99;' entries with three system classes have all three classes assigned. For any given multi-AGN, the ``primary,'' ``secondary,'' and ``tertiary'' classes do not follow a rank order, instead these classes are assigned uniformly across the catalog: `dual AGNs' take precedence for the ``primary'' class when applicable, followed by `binary AGNs,' and finally by `recoil candidates.' The same ordering applies for the secondary class, where (if the primary class has been filled by a `dual AGN' class) the `binary AGN' class (when applicable) takes precedence over recoil candidates. Following this logic, in cases where an object is selected as a binary AGN and a recoil candidate, but not a dual AGN candidate, such a system would have `binary AGN' as the primary class and `recoil candidate' as the secondary class. The majority of objects, however, fall under only a single, ``primary'' system class. Each system class has an associated confidence flag that denotes the likelihood of the system class (see Section~\ref{sec:flagsystem}).

\subsection{A (Subjective) Confidence Rank System}
\label{sec:flagsystem}

Every multi-AGN entry has been assigned a confidence rank for each associated system class. As described in Section~\ref{sec:assignsystemclass}, most multi-AGN entries are associated with only a single system class (a `primary' class) and therefore have only a single confidence rank relevant to that system class. For multi-AGN entries that are associated with multiple classes (i.e. assigned a primary, secondary, and/or tertiary class), a confidence rank is provided for each of these individual system classes. These confidence ranks take into account a substantial amount of information: selection strategy, quality and quantity of analyses, specific analysis techniques, false-positive rate, and (when possible) firm evidence of confirmation and/or rejection.

Each rank indicates the likelihood that the null hypothesis -- that a given system does not constitute a multi-AGN -- can be effectively ruled out. More specifically, the null hypothesis represents the scenario in which the observed phenomenon used to select the system as a multi-AGN candidate can reasonably arise from physical mechanisms that do not involve the purported system class(es), i.e., not driven by dual AGNs, binary AGNs, or recoiling AGNs. In other words, the (subjective) confidence rank specifies the likelihood that the assigned system class is true. Here we outline the confidence ranks used within the catalog:

\begin{itemize}
    \item ``Confirmed'' (+1): Beyond a reasonable doubt, the system class should be considered firm based upon the published evidence in the literature. The null hypothesis can be safely rejected.
    \item ``Possible'' (+0.5): Some volume of evidence favors the system class, but the evidence does not prove beyond a reasonable doubt the validity of the system class. The ‘null’ hypothesis cannot be fully ruled out, but it is less likely than the proposed system class. In other words, a stronger case can be made for the proposed class than the null hypothesis.
    %suggests the presence of
    \item ``Candidate'' (0): 1) The entry was selected as the system class but no additional analysis (beyond the selection method) has been performed, where the selection method itself does not provide unambiguous evidence in favor of the proposed class. 2) Some evidence - either via selection methodology (without follow-up) or via follow-up observations/analysis - favors the system class, but no firm conclusions can be drawn. For rank 0 entries, the ‘null’ hypothesis and the proposed system class have approximately equal likelihood. Additional observations may be helpful.
    %exists to support the presence of
    \item ``Insufficient Evidence'' (-0.5): Some volume of data has been analyzed (selection methodology or follow-up observations), but the evidence presented is insufficient for the reported system class. The ‘null’ hypothesis is more likely – and in some cases, strongly – than the proposed system class. However, the system class cannot be ruled out (irrespective of the selection method), nor can it be ruled out that the phenomenon used to select the object is driven by processes unrelated to multi-AGNs.
    \item ``Unlikely'' (-1): Some volume of data has been analyzed (selection methodology or follow-up observations), but the ‘null’ hypothesis is overwhelmingly favored over the proposed system class based on the available evidence (including the phenomenon used to select the system). However, while we can prove the absence of evidence for the system class, the system class cannot -- in general -- be entirely ruled out\footnote{There are instances where the system class can be definitively ruled out: for example, an object claimed to be a dual AGN with a 5 kpc separation, but which shows zero evidence for a companion at 5 kpc, could be definitively ruled out.}. Nonetheless, the system class should, for all intents and purposes, be considered extremely unlikely.
\end{itemize}

To further elucidate how these subjective confidence ranks are employed in practice, we include below several examples of multi-AGN candidates from the catalog and what criteria were used to apply their confidence ranks. While these brief breakdowns are illustrative of our approach, thorough breakdowns such as these for each object in the catalog are well beyond the scope of this work and this version of the catalog; confidence rank breakdowns could be included in a future release.

\subsubsection{Example 1: NGC 6240, Dual AGN (Confirmed, Rank = 1)}

NGC 6240 is a well studied ultraluminous luminous infrared galaxy (ULIRG), and had been previously shown to host an X-ray AGN based on low-resolution ROSAT imaging \citep{komossa1998,schulz1998} and optical spectroscopic observations showed that one of the two observed nuclei showed LINER-like emission line ratios \citep[e.g.,][]{veilleux1995}, showing no unambiguous AGN activity in the optical. \textit{Chandra} X-ray imaging was used in \citet{komossa2003} to show for the first time that NGC 6240 hosted a bona fide dual AGN, with a nuclear separation $\sim900\,pc$. The spatially-resolved X-ray sources, each exhibiting Fe K$\alpha$ emission, provided unambiguous proof that NGC 6240 hosted a sub-kpc dual AGN. The observed X-ray properties could not be explained by the presence of only a single AGN. NGC 6240 has since become the poster child for dual AGNs, and it is a clear case of a rank $=1$ entry within this catalog.

\subsubsection{Example 2: NGC 7674, Dual AGN Candidate / Binary AGN Candidate (Possible, Rank = +0.5)}

NGC 7674 was noted as a binary AGN candidate by \cite{kharb2017}, who serendipitously identified two compact radio sources in VLBA 15 GHz radio imaging. The two radio sources are separated by 35\,pc ($\sim60$\,mas) and were not detected at 2 GHz, 5 GHz, or 8 GHz; a detection only at 15 GHz could suggest the source have inverted radio spectra. While  firm detections at lower frequencies are needed to rule out alternative scenarios and confirm this system as a dual or binary AGN (and thus reject the null hypothesis), we consider the detection of these radio sources at 15 GHz promising evidence for a dual or binary AGN system. We therefore apply a rank$=+0.5$ for this candidate system.

\subsubsection{Example 3a: Double-peaked Dual AGN Candidate with No Follow-up (Baseline Candidate, Rank = 0)}

J023831.82+001152.23 (J0238+0011) is a dual AGN candidate selected based on the presence of double-peaked optical spectroscopic emission lines observed in fiber spectra \citep{yuan2016}. Dual AGN candidates selected based on the presence of double-peaked optical emission lines require extensive observational campaigns to search for multiple optical and/or near-IR nuclei and identify definitive evidence of multi-AGN activity \citep[in, for example, the X-rays, near-IR, radio, and optical, e.g.,][]{shen2011,fu2011a,fu2011b,fu2012,mcgurk2011,comerford2011,comerford2015,mullersanchez2015}. J0238+0011, however, is representative of hundreds of dual AGN candidates selected via this technique but have not yet been studied in follow-up observations (at least prior to 2021). Some fraction of double-peaked sources do contain dual AGNs \citep[$\lesssim2$\% of all double-peaked AGNs, and $\lesssim15\%$ of double-peaked AGNs in galaxy mergers, e.g.,][]{fu2012}, but without further information for objects like J0238+0011, we simply assign a rank $=0$ (indicating a baseline candidate). We cannot reject the null hypothesis, but there is no additional evidence beyond the selection method to argue in favor of the proposed dual AGN scenario.

\subsubsection{Example 3b: 4C 48.29, Binary AGN / AGN Recoil Candidate (Baseline Candidate, Rank = 0)}
%, An X-shaped Radio Source

4C 48.29 is a radio source selected as a binary AGN/SMBH and/or recoiling SMBH candidate based upon the presence of an X-shaped radio source  \citep{miley1979,merritt2002,cheung2007,lal2007}, and is representative of several hundred X-shaped radio sources that have been identified and stand as binary or recoil candidates \citep[e.g.,][]{cheung2007,proctor2011,roberts2018,saripalli2018,yang2019,joshi2019}. A variety of competing hypotheses have been put forth to explain the formation of X-shaped radio sources \citep[see, e.g.,][]{merritt2002,lal2007,cheung2007,yang2019,gopalkrishna2012}, but a consensus has yet to reached \citep{gopalkrishna2012,yang2019}. Unresolved, binary radio AGNs \citep[e.g.,][]{lal2005,lal2007} or recoiling AGNs \cite[e.g.,][]{gopal2003} have been suggested to be possible formation mechanisms for these radio structures. A handful of double radio jet dual AGNs (each AGN powers a twin jet) have been resolved on tens to thousands of parsec-scales \cite[e.g., ][]{owen1985,parma1991,guidetti2008}, which display radio jets with curves, warps, and orientation changes on different scales; extrapolating to unresolved separations in the binary regime, such systems could be expected to manifest as X-shaped radio sources. Since the formation mechanisms for these radio sources are not yet understand, and that binaries and recoils stand as possible mechanisms, we flag 4C 48.29 (and those like it) as rank $=0$ baseline candidates. The available evidence does not allow us to reject the null hypothesis, nor does it provide substantial evidence in favor of the binary or recoil hypothesis.

\subsubsection{Example 4: UM 211, Binary AGN Candidate (Insufficient Evidence, Rank = -0.5)}

UM 211 is a binary AGN/SMBH candidate selected based upon an observed periodic signal in CRTS optical imaging \citep[based on $\sim 1.7$ cycles][]{graham2015b}, and is representative of a substantial fraction of periodicity-selected binary candidates. Extensive work has shown that the natural, stochastic variability in AGNs, so-called red noise, can mimic periodicities when observed over short temporal baselines, and hence can mimic a periodic signal that may be expected in the case of binary AGNs  \citep{vaughan2016}. In order to rule out red noise as the source of observed periodicities, a temporal baseline covering $>5$ cycles is required \citep[e.g.,][]{vaughan2016,tliu2019}. UM 211, like many binary AGN candidates identified in \citet{graham2015b} and \citet{charisi2016}, was selected based on periodic signals across only $\lesssim3$ cycles. Given that the minimum required temporal baseline has not been used in the selection of this candidate, we flagged UM 211 (and those like it) as having insufficient evidence (rank $=-0.5$) for the proposed system class. We cannot rule out the presence of a binary AGN candidate, but the null hypothesis -- that this object is a normal AGN exhibiting stochastic variability -- is a more likely explanation because a persistent periodic signal has not yet been proven to exist. With new observations covering a sufficient temporal baseline with many more cycles, UM 211 could be moved to either a rank $=0$ (baseline candidate) or rank $=+0.5$ (possible candidate) should the reported periodicity continue as expected.

\subsubsection{Example 5: LGGS J004527.30+413254.3, Binary AGN Candidate (Unlikely, Flag = -1)}
\citet{dorn2017} identified 
LGGS J004527.30+413254.3 as a background AGN at $z\approx0.215$ observed through the disk of M31, and reported that the object showed multiple periodicities in g-band Palomar Transient Factory imaging. While other phenomenon could also give rise to the reported periodicities, \citet{dorn2017} regarded LGGS J004+4132 as a promising SMBH binary candidate. However, \citet{barth2018} re-evaluated the optical light curve of this object and instead found no evidence for any periodicity in the light curve of LGGS J004+4132; in fact, \citet{barth2018} showed that the false alarm probabilities calculated by \citet{dorn2017} were underestimated by a factor of 100 for all reported periodicities. Given the complete lack of optical periodicity (which was used to select this candidate to begin with), we flag this candidate binary as ``unlikely'' (rank $=-1$) and we heavily favor the null hypothesis (that there is only a single AGN). However, we cannot definitively prove the absence of an unresolved SMBH binary.

\subsection{Miscellaneous Notes on Confidence Ranks}
\label{sec:confidence{misc}}

\subsubsection{Reliable Redshifts}
As mentioned in Section~\ref{sec:flagsystem}, redshifts - and specifically spectroscopic redshifts - played an integral role in the assignment of confidence ranks. For example, 2D projected separations and velocity differences are both a function of redshift and play a key role in applying our dual AGN criteria (Section~\ref{sec:definingduals}) to candidate systems. Similarly, when redshifts are not available for one or multiple AGNs in a given candidate system, firm conclusions cannot necessarily be drawn (barring unambiguous tidal features) as to whether the observed AGNs are physically associated or caught in projection. Redshift omission(s) necessarily lower the confidence in the system class. Even when photometric redshifts are available for the component AGNs in a candidate system, uncertainty can remain as to whether the photometric redshifts reflect the true redshifts; in the absence of clear morphological evidence (for the case of, e.g., dual AGNs), spectroscopic redshifts are generally required to ensure the component AGNs are physically associated and to confirm the multi-AGN classification.

\subsubsection{Fiber Spillover}
Another consideration that bears stating  explicitly is the potential for fiber or slit spillover in ground-based optical spectroscopic observations of multi-AGN candidates \citep{husemann2020,pfeifle2023b}. Poor atmospheric conditions during a set of observations can smear the emission across a larger angular size on the sky, allowing the emission from one object to be observed at the sky position of another object, leading to erroneous results that a single system hosts multiple AGNs. This fiber spillover has been shown to be a significant problem for closely-separated ($<\lesssim10$'') Type 2 Seyfert pairs \citep{husemann2020} and Type 1 AGN pairs \citep{pfeifle2023b} selected using the 3'' diameter SDSS fibers. Follow-up spectroscopy in both \citet{husemann2020} and \cite{pfeifle2023b} not only revealed that the distinct spectroscopic emission line fluxes and flux ratios change when the candidate AGNs are observed under better seeing conditions and with smaller apertures, but that spillover can remain an issue even in higher angular resolution slit spectroscopic data and with smaller extraction apertures \citep{pfeifle2023b}. Therefore, closely separated ($<10$'') dual or multi-AGN candidates identified via ground-based spectroscopy may be uncertain if the source spectra come from fibers with large angular sizes (e.g. SDSS), the seeing is poor or variable, or if at least one target shows evidence of extended emission (e.g. outflows, rotation, etc).  Additional observations at other wavelengths (e.g. high angular resolution radio or X-ray observations, narrow-band imaging) or spatially-resolved spectroscopic observations can provide a sanity check as to whether spillover or spectroscopic confusion may be occurring.

\subsubsection{Extended Narrow Line Regions}
Like fiber spillover, AGN extended narrow line regions (ENLRs) and cross-ionization also impact the selection, analysis, and confirmation of closely-separated multi-AGN candidates. Evidence for ENLRs was first uncovered by \citet{wampler1975} and have since been identified in a substantial number of AGNs \citep[see review by][]{stockton2006}. These ENLRs have a wide range of physical sizes, from a few kpc \citep[e.g.][]{fu2012,fu2018} and -- in some cases -- extending up to tens of kpc from the central AGN \citep[e.g.][]{greene2011,greene2012,stockton2006,hainline2016}, and they have been identified in candidate \citep[][$\sim$few kpc-scale]{greene2011,greene2012,fu2012,fu2018} and confirmed \citep[][$\sim30$ kpc-scale]{fu2012,hainline2016} dual AGN systems. Cross-ionization \citep{keel2019}, where one AGN in a galaxy pair ionizes the gas in the companion galaxy, is a specific kind of ENLR and can mimic the presence of a dual or multi-AGN \citep{fu2012,fu2018} and is often invoked as an alternative hypothesis for the observation of multiple AGN-like emission line regions \citep[e.g.,][]{fu2012,xliu2018b,fu2018,liu2010a,liu2011a}. To avoid erroneous bona fide dual AGN and multi-AGN classifications due to ENLRs and/or cross-ionization, we typically do not classify optically-selected multi-AGN candidates as confirmed systems (rank $=1$) unless the multiwavelength evidence is overwhelming or the separations of the AGNs are sufficiently large enough that ENLRs are a lesser concern. In the latter case, we typically adopt a physical, projected separation of $10$\,kpc, above which we allow for confirmed multi-AGNs based on optical spectroscopy alone so long as the optical emission line ratios indicate a Seyfert-Seyfert pair. Optical fiber spectra are the most susceptible to misclassification due to ENLRs, but long-slit spectra and IFU imaging can also be affected \citep{fu2018}; high resolution IFU imaging and spectroscopy is perhaps the best way to circumvent ENLRs in the optical. Confidence ranks for optically-selected close pairs can be adjusted in future data releases should future long-slit or IFU observations show that ENLRs are not a concern or if multiwavelength observations (for example, X-ray imaging) reveal the unambiguous presence of multiple AGNs.

\subsubsection{Ambiguity in Optical Emission Line Ratios}
One final issue related to the use of optical spectroscopy for multi-AGN identification is the general use of spectroscopic emission line ratios. Countless works have relied upon spectroscopic emission lines to select and study dual AGN candidates based on the BPT line ratios \citep{baldwin1981} of the individual nuclei \citep[e.g.,][]{wang2009,liu2011b,liu2011a,koss2012,comerford2012,liu2010a,comerford2015}, and while these searches have turned up candidate Seyfert-Seyfert pairs, they have also considered combinations such as Seyfert-Composite, Seyfert-LINER, LINER-LINER, LINER-Composite, and Composite-Composite as dual AGN candidates. Though AGNs are found in some Composites and LINERs, these two optical classes can also be driven purely by shocks \citep[e.g.,][]{allen2008,rich2014,rich2015} and in some cases by older stellar populations \citep[i.e., post-AGB stars,][]{singh2013}; we do not necessarily consider LINERs or Composites as bona fide AGNs without additional evidence, but we do flag such systems as ``possible'' dual AGN candidates (rank $=+0.5$).\footnote{Double-peaked composites and LINERs from \citet{ge2012} are not included in this catalog; only a small fraction \citep[$<15$\%][]{mullersanchez2015,fu2012} of double-peaked \textit{Seyferts} are expected to host dual AGNs, and as outlined in this section, Composites and LINERs do not all necessarily host AGNs.} Additionally, as mentioned above, these line ratios can be affected by fiber spillover due to poor atmospheric conditions, and these ratios can also be driven by a single AGN via  cross-ionization and/or ENLRs. It is important to consider both the angular and projected, physical separations of the systems -- in addition to the specific nuclear BPT optical classes -- when assigning confidence ranks based on emission line ratios.

\subsubsection{Periodicity}

Claimed periodicity and quasi-periodicity in the light curves of AGNs have been used in numerous works to select binary AGN candidates \citep[e.g.,][]{sillanpaa1988,graham2015b,charisi2016,tliu2016,ackermann2015}. However, nearly all binary AGN candidates selected based on periodicities have relied upon non-uniformly sampled data or rely upon small temporal baselines, meaning that only a few $<3$ cycles have been used to claim periodicity (or tentative periodicity). In this catalog, we assign rank $=-0.5$ (insufficient evidence) to objects selected via periodicity in short temporal baseline light curves ($<5$ cycles), because insufficient cycles have been observed to rule out red noise and prove the presence of a periodic signal \citep[e.g.,]{vaughan2016}. Sources with claimed quasi-periodicities, by extension, are ranked as $-0.5$ as well, because there is insufficient evidence to claim a binary, and the null hypothesis is far more likely. The confidence ranks of periodicity-selected  objects could be adjusted in future releases if longer baselines are probed and demonstrate the presence of real periodicities.

\subsubsection{X-ray Sub-binning}
Another, albeit rare, property we take into consideration is the binning of X-ray data from \textit{Chandra}. The \textit{Chandra} native pixel size of 0.5'' on a side generally sets the spatial resolution limit for \textit{Chandra}, but the on-axis PSF core is slightly smaller than the pixel size, enabling the use of sub-pixel alorigthms and binning during processing to improve the resolution of arcsecond-scale and sub-arcsecond-scale structure.\footnote{https://cxc.cfa.harvard.edu/ciao/why/acissubpix.html} However, it is important to point out that there is a known PSF asymmetry at sub-arcsecond scales (0.6''-0.8'') in \textit{Chandra} HRC imaging and a similar asymmetry in ACIS imaging; this asymmetry can mimic multiple sources at sub-arcsecond-scales\footnote{https://cxc.cfa.harvard.edu/ciao/caveats/psf\_artifact.html} \citep{fabbiano2011,koss2015}. It must be pointed out then that binning severely below the instrumental resolution ($0.5''$) runs the risk of erroneously identifying artificial structures (driven by an asymmetric PSF) as multiple, real X-ray sources when in fact there is only one source \citep{fabbiano2011,koss2015}. This is particularly true when sub-pixel binning is used on energy bands where the \textit{Chandra} effective area and PSF are poorer, such as Fe K imaging at energies $\sim6$\,keV, or when multiple observations are combined imprecisely (multiple sources could result from an incorrect accounting of offsets between images). Given the known dangers of sub-pixel binning when attempting to study $<1''$ and $<0.5$'' structures in \textit{Chandra} imaging, evidence of sub-arcsecond AGN multiplicity in sub-pixel binned \textit{Chandra} observations should be scrutinized judiciously using rigorous source detection algorithms such as BAYMAX \citep{foord2019,foord2020,foord2021} and should not be considered conclusive without substantial supporting evidence.

\begin{table*}[]
\caption{Description of Big MAC DR1 Catalog Columns}
\begin{center}
\hspace{-2.2cm}
\begin{tabular}{cccc}
\hline
\hline
\noalign{\smallskip}
Col. Number & Key & Type & Description \\
(1) & (2) & (3) & (4) \\
\noalign{\smallskip}
\hline
\noalign{\smallskip}
1  & \texttt{Primary System Type} & \texttt{str} & Primary assigned system class. \\
2  & \texttt{Secondary System Type} & \texttt{str} & Secondary assigned system class. \\
3  & \texttt{Tertiary System Type} & \texttt{str} & Tertiary assigned system class. \\
4  & \texttt{ST1 Confidence Flag} & \texttt{float} & Confidence rank for primary system type.\\
5  & \texttt{ST2 Confidence Flag} & \texttt{float} & Confidence rank for secondary system type.\\
6  & \texttt{ST3 Confidence Flag} & \texttt{float} & Confidence rank for tertiary system type.\\
7  & \texttt{Selection Method} & \texttt{dict, str} & Dictionary containing selection methods, organized by wavelength.\\
8  & \texttt{Analysis Method} & \texttt{dict, str} & Dictionary containing analysis methods, organized by wavelength.\\
9  & \texttt{Confirmation Method} & \texttt{dict, str} & Dictionary containing confirmation methods, organized by wavelength.\\
10 & \texttt{Literature Name} & \texttt{str} & Adopted literature name, e.g., from Simbad\\
11 & \texttt{Name1} & \texttt{str} & Designation of AGN 1.\\
12 & \texttt{z1} & \texttt{float} & Redshift of AGN 1.\\
13 & \texttt{z1\_type} & \texttt{str} & Redshift 1 type.\\
14 & \texttt{RA1} & \texttt{} & Right ascension (sexagesimal) of AGN 1.\\
15 & \texttt{Dec1} & \texttt{} & Declination (sexagesimal) of AGN 1.\\
16 & \texttt{RA1\_deg} & \texttt{float} & Right ascension (degrees) of AGN 1. \\
17 & \texttt{Dec1\_deg} & \texttt{float} & Declination (degrees) of AGN 1.\\
18 & \texttt{Coordinate\_waveband1} & \texttt{str} & Waveband of given coordinates for AGN 1.\\
19 & \texttt{Coordinate\_Source1} & \texttt{str} & Source of coordinates for AGN 1.\\
20 & \texttt{Equinox1} & \texttt{int} & Equinox for coordinate set 1.\\
21 & \texttt{Brightness1} &  \texttt{float} & Brightness measurement of AGN 1. \\
22 & \texttt{Brightness\_band1} & \texttt{str} & Wavelength (band) of brightness 1. \\
23 & \texttt{Brightness\_type1} & \texttt{str} & Type, or unit, for brightness 1. \\
24 & \texttt{Name2} & \texttt{str} & Designation of AGN 2. \\
25 & \texttt{z2} & \texttt{float} & Redshift of AGN 2. \\
26 & \texttt{z2\_type} & \texttt{str} & Redshift 2 type. \\
27 & \texttt{RA2} &  & Right ascension (sexagesimal) of AGN 2. \\
28 & \texttt{Dec2} &  & Declination (sexagesimal) of AGN 2. \\
29 & \texttt{RA2\_deg} & \texttt{float} & Right ascension (degrees) of AGN 2. \\
30 & \texttt{Dec2\_deg} & \texttt{float} & Declination (degrees) of AGN 2. \\
31 & \texttt{Equinox2} & \texttt{int} & Waveband of given coordinates for AGN 2. \\
32 & \texttt{Coordinate\_waveband2} & \texttt{str} &  Source of coordinates for AGN 2. \\
33 & \texttt{Coordinate\_Source2} & \texttt{str} & Equinox for coordinate set 2. \\
34 & \texttt{Brightness2} & \texttt{float} & Brightness measurement of AGN 2. \\
35 & \texttt{Brightness\_band2} & \texttt{str} & Wavelength (band) of brightness 2. \\
36 & \texttt{Brightness\_type2} & \texttt{str} & Type, or unit, for brightness 2. \\
37 & \texttt{dV} & \texttt{float} & Velocity difference between the AGN components. \\
38 & \texttt{Sep} & \texttt{float} & Angular ($\theta$) separation between the two AGN components. \\
39 & \texttt{Sep(kpc)} & \texttt{float} & Projected, physical separation between the two AGN components. \\
40 & \texttt{Paper(s)} & \texttt{str} & Discovery and key analysis articles. Format: Author+Year. \\
41 & \texttt{BibCode(s)} & \texttt{str} & NASA ADS BibCodes for discovery and analysis articles. \\
42 & \texttt{DOI(s)} & \texttt{str} & DOI links for discovery and analysis articles. \\
43 & \texttt{Legacy System Type} & \texttt{str} & Original system class used in the literature. \\
\noalign{\smallskip}
\hline
\end{tabular}
\end{center}
\tablecomments{Here we tabulate the names, data types, and descriptions for column information included in the Big MAC DR1. Note, here we illustrate the table for dual, binary, and recoiling AGNs and candidate. Multi-AGN tables follow an analogous setup.}
\label{tab:maintable}
\end{table*}

\subsection{Assigning Selection Methods}
\label{sec:assigningselectionmethods}

For each multi-AGN entry in the Big MAC, we include the specific wavebands and observational techniques used for selection. This required an in-depth review of every article to retrieve the selection methods used to identify each multi-AGN system or sample of systems.  We aimed for specific, but fairly coarse `wavelength' and `selection method' bins applicable to the various multi-AGN candidate populations in this catalog. For each entry in the Big MAC, we began by identifying the general, overarching selection waveband(s) and technique(s), for example: ``Optical Imaging'' or ``Optical Spectroscopy,'' where ``Optical'' is the waveband and ``imaging'' and ``spectroscopy'' are techniques. These general waveband-technique combinations were recorded for each entry, and we then -- wherever possible -- provided more specific techniques per waveband: for example, if an object selected using ``Optical Spectroscopy'' was selected specifically using fiber spectra, we also included the waveband-technique combination of ``Optical Fiber Spectroscopy.'' Our inclusion of selection techniques for Big MAC systems therefore comes with two caveats:
\begin{enumerate}
    \item Not all waveband-technique selection method combinations identify mutually exclusive samples of multi-AGN candidates; many techniques may be used in concert with, or are derived from, one another. For example,  ``Optical Fiber Spectroscopy'' necessarily implies the use of  `Optical Spectroscopy'' for selection (as alluded to above). Another example: for objects selected using optical spectroscopy and X-ray imaging, the overarching waveband-techniques ``Optical Spectroscopy'' and ``X-ray Imaging'' are listed simultaneously. As a more complex example: Seyfert 2s selected based on the presence of double-peaked optical emission lines in optical fiber spectra would have the following (non-mutually-exclusive) waveband-technique combinations listed under the selection method: ``Optical Double-Peaked Spectroscopic Emission Lines,'' ``Optical Fiber Spectroscopy,'' ``Optical Spectroscopic Emission Line Ratios,'' and ``Optical Spectroscopy.'' 
    \item Our ability to record specific waveband-technique combinations is limited to the information provided in the literature article for a given multi-AGN entry. As an example, consider an article that discusses a dual AGN candidate selected using optical spectroscopy: if the authors discuss a slit width, we can include the waveband-technique combinations ``Optical Spectroscopy'' and ``Optical Slit Spectroscopy'' as selection techniques. However, if the authors omitted information such as slit length we cannot ascertain whether the authors used long-slit spectroscopy or some other specific form of slit spectroscopy (and we therefore cannot include any more specific selection method techniques for that entry). When authors do not provide sufficient details for instrumental configurations, we default to the more general version of a technique. For example, if optical spectroscopy was used to identify a multi-AGN candidate, but no specific information was reported regarding what instrumental configuration was used, we simply defer to ``Optical Spectroscopy'' as the waveband-technique selection combination. 
\end{enumerate}
This latter caveat is important to bear in mind while examining the relative use of each selection technique (Sections~\ref{sec:dualselectmethods}-\ref{sec:recoilselectmethods}): because instrument configurations are not always sufficiently explicit, we may in some cases have accurate but not precise constraints on the selection techniques employed in a given article (i.e., ``Optical Spectroscopy'' may be the only combination listed, when in fact ``Optical Slit Spectroscopy'' was also applicable). For this reason, the percentages quoted in Sections~\ref{sec:dualselectmethods}-\ref{sec:recoilselectmethods} for more specific wave-band technique combinations should be considered accurate but not necessarily precise reflections of the selection methods employed in the literature. Selection methods for each multi-AGN entry in the catalog are organized in nested python dictionaries, with the waveband:technique format providing the baseline for the key:value pair format.%for example. slit vs. long-slit spectroscopy), and f

To provide additional clarity, we provide in Figure~\ref{tab:flowchartselect} illustrative examples of how we include selection techniques. Confirmation methods and analysis methods are assigned analogously to selection methods, but for confirmation methods we require the techniques listed to play a critical role in the confirmation of the individual AGNs. Like the selection methods, analysis and confirmation methods for each multi-AGN entry in the catalog are organized as nested python dictionaries within the catalog.

\begin{table*}[]
\caption{Illustrative Examples of Selection Method Assignment}
\begin{center}
\hspace{-2.2cm}
\begin{tabular}{ccc}
\hline
\hline
\noalign{\smallskip}
Case 1:  & Case 2:  & Case 3:  \\
IRAS 03219+4031  & J1502+1115  & J0122+0100 \\
\citep{koss2012}  & \citep{fu2011b}  & \citep{satyapal2017}  \\
\noalign{\smallskip}
\hline
\noalign{\smallskip}
%Initial Selection Method:  & 
Hard X-ray Preselection & Double-Peaked [O\,III] Lines & Mid-IR Preselection \\
$\downarrow$ & $\downarrow$ & $\downarrow$ \\

\begin{tabular}[c]{@{}c@{}}Optical Imaging \\ to Select Mergers\end{tabular}                & \begin{tabular}[c]{@{}c@{}}Near-IR Imaging \\ to Identify Two Nuclei\end{tabular}                                            & \begin{tabular}[c]{@{}c@{}}Optical Imaging \\ to Select Mergers\end{tabular} \\
$\downarrow$ & $\downarrow$ & $\downarrow$ \\
\begin{tabular}[c]{@{}c@{}}Optical Fiber Spectroscopy\\ to Determine BPT Ratios\end{tabular} &  \begin{tabular}[c]{@{}c@{}}Optical IFU Spectroscopy\\ and Imaging Used to \\ Identify Candidate Optical\\ Dual AGN\end{tabular}                                          & \begin{tabular}[c]{@{}l@{}}X-ray Imaging Used \\ to Identify Two Sources\end{tabular}                    \\
$\downarrow$ & $\downarrow$ & $\downarrow$ \\
\begin{tabular}[c]{@{}c@{}} X-ray Imaging to find \\ X-ray Source(s)\end{tabular}  &             \begin{tabular}[c]{@{}c@{}}Radio Imaging to \\ find radio Source(s)\end{tabular} & \begin{tabular}[c]{@{}c@{}} \dots \end{tabular}                                                                                    \\
$\downarrow$ & $\downarrow$ & $\downarrow$ \\
%``Full'' Selection Method: &
\begin{tabular}[c]{@{}c@{}} Hard X-ray: \{Imaging\}\\ Optical: \{Imaging, 
Spectroscopy,\\Fiber Spectroscopy,\\Spectroscopic Emission\\Line Ratios\}\\ X-ray:\{Imaging,\\ Spectroscopy\}\end{tabular}  & \multicolumn{1}{c}{\begin{tabular}[c]{@{}c@{}} Optical: \{Spectroscopy,\\Fiber Spectroscopy,\\Spectroscopic Emission\\Line Ratios, Double-peaked\\Spectroscopic Emission Lines,\\ IFU Imaging, IFU\\Spectroscopy\}\\ Near-IR:\{Imaging\}\\ Radio: \{Imaging, Double Sources\}\end{tabular}} & \begin{tabular}[c]{@{}c@{}} Mid-IR: \{Imaging, Colors\}\\ Optical: \{Imaging\}\\ X-ray:\{Imaging\}\end{tabular} \\
\noalign{\smallskip}
\hline
\end{tabular}
\end{center}
\tablecomments{Selection method assignment in the Big MAC. These three examples illustrate the general way we add/assign selection methods to the ``Section Method'' column for a given entry within the Big MAC. Each case begins with the ``initial'' selection strategy, and the arrows indicate the progression through additional, complementary methods used for the selection process. The final line shows the ``full'' selection strategy for these examples included in the Big MAC. }
\label{tab:flowchartselect}
\end{table*}

\section{Analysis}
\label{sec:analysis}

One of the most important aspects of the Big MAC DR1 is our ability to (1) quantify the variety of selection methods employed in the literature, (2) quantify the frequency with which different wavebands and techniques are employed in concert, and (3) quantify the overlap between independent selection methodologies that identify the same systems. Prior to studying and understanding the selection techniques that have recovered known multi-AGN systems, it is important to quantify the frequency with which different selection strategies have been attempted for all multi-AGN candidates up to 2020. Taking into account the basics of the selection functions present within the current literature (i.e., waveband choice and specific observational technique per waveband) elucidates the implicit wavelength- and technology-dependent biases in current samples of multi-AGNs and candidates. For example, optical spectroscopic selection may be biased against heavily dust-obscured AGNs \citep[][]{koss2012,satyapal2017,pfeifle2019a}, introducing a clear wavelength-dependent bias in the literature. With regard to technological biases, some selection methodologies are vastly under-subscribed due to observational feasibility and facility access; for example, X-ray imaging is in general underutilized, but this is to be expected due to the sample sizes of many dual AGN surveys and the observational demands of X-ray imaging for these objects (often requiring prohibitive per-source exposure times or program sizes). As another example, heavily obscured AGNs are biased against 2-10\,keV soft X-ray detection, but current hard X-ray observatories \citep[i.e., \textit{NuSTAR,}][]{harrison2013} can only spatially-resolve dual AGNs with large angular separations and are sensitive only to relatively bright hard X-ray AGNs (e.g., \citealp{koss2016}, \citealp{kosec2017}, \citealp{brightman2018}, \citealp{ptak2015}, \citealp{iwasawa2018}, \citealp{iwasawa2020}, \citealp{nardini2017}; see also discussion in \citealp{pfeifle2023c} and references therein), which introduces a clear technological (and obscuration) bias into our samples in the literature. In examining the selection functions of known multi-AGN candidates, we can simultaneously identify over- and under-subscribed selection methods (relative to other techniques) and identify candidate methodologies that may assist in circumventing our current observational and technological selection biases.

In Sections~\ref{sec:dualselectmethods}, \ref{sec:binaryselectmethods}, and \ref{sec:recoilselectmethods}, we discuss the selection methods used to identify dual AGNs, binary AGNs, and recoiling AGNs, respectively. Each of these three sections begins with a subsection describing the aggregate statistics on the frequency with which all  waveband+technique combinations were used in the literature for the given system class. These sections are followed by subsections that break these statistics down further by: (1) first limiting (or binning) the system class to only objects identified using a specific selection technique (we refer to this as the `primary' technique); then, (2) we identify other techniques that are commonly used as a part of/in concert with the primary technique (these could be derivatives of one another in the same waveband, i.e., optical spectroscopy and optical fiber spectroscopy, or could be methods using different wavebands, i.e., optical spectroscopy in concert with mid-IR imaging); then, (3) we show the techniques that were used to independently identify the same objects as the primary technique. This selection method breakdown allows us to study selection method frequency in finer detail, gaining insight into what overlap these techniques have within a particular waveband and across wavebands (for example: when optical spectroscopy was used, we can assess how often fiber spectroscopy is used relative to slit spectroscopy) and what methods have been used independently of the primary technique (for example: when optical spectroscopy was used to identify a set of objects, we can assess the fraction of objects that were independently identified via radio imaging).

\subsection{Dual AGN Selection Methods}
\label{sec:dualselectmethods}

\subsubsection{Dual AGN Aggregate Selection Method Statistics}
\label{sec:aggselmethodduals}
\begin{figure*}[t]
    \centering
    \includegraphics[width=1.0\linewidth]{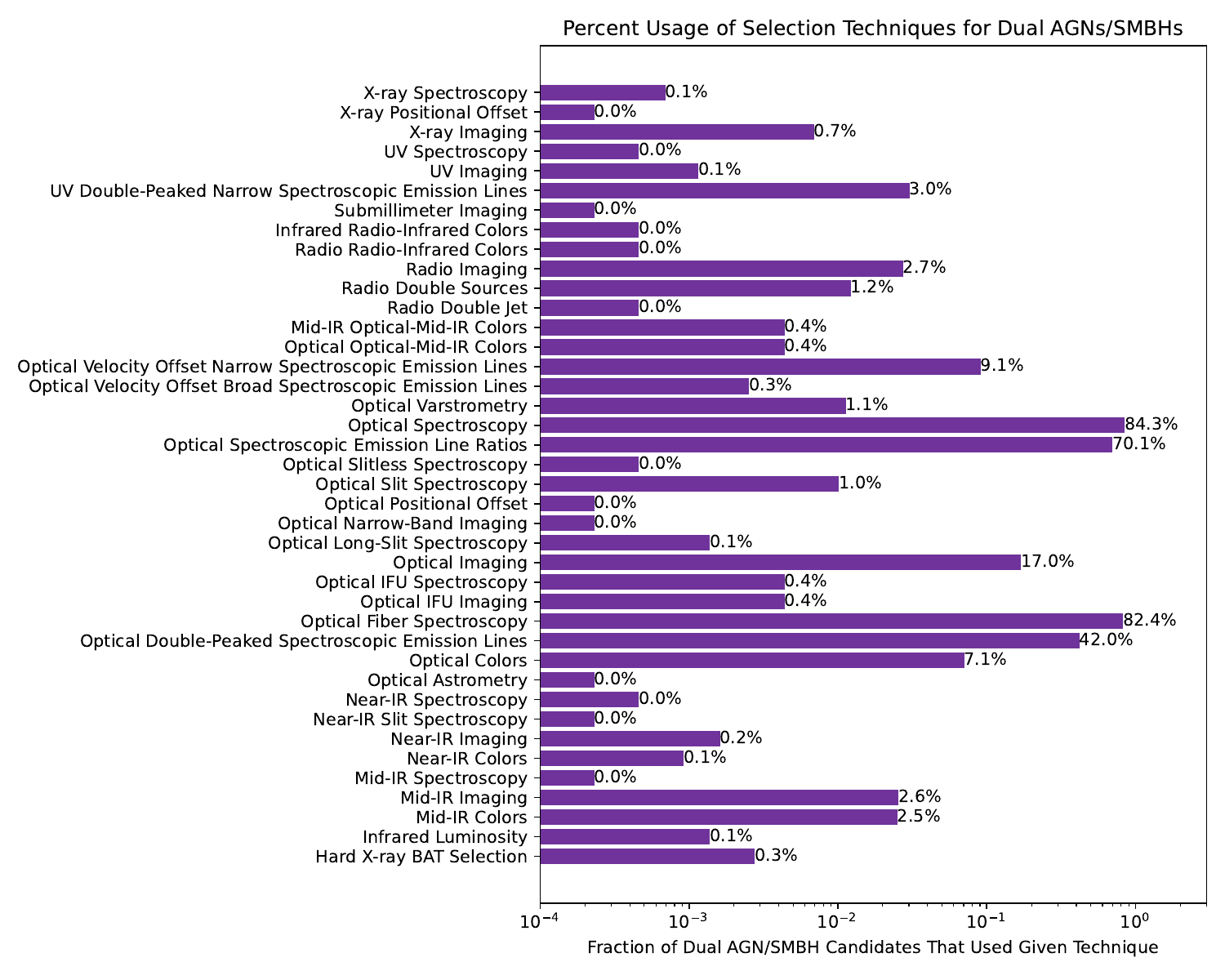}
    \caption{A breakdown of the selection methods used to identify dual AGNs and candidates in the literature. This chart shows the aggregate percent usage of all dual AGN selection techniques (normalized by the number of dual AGN candidates in the catalog); note, these selection methods are not all necessarily mutually exclusive (for example: selection based on double-peak optical spectroscopic emission lines necessarily requires the use of optical spectroscopy, and often specifically involves the use of optical fiber spectroscopy).}
    \label{fig:aggdualAGNselmethod}
\end{figure*}

The aggregate percent frequency with which specific selection methodologies are used for dual AGN systems is shown in Figure~\ref{fig:aggdualAGNselmethod}, where the percentages are normalized to the total number of dual AGN candidates in the DR1. The figure is organized by waveband and broken into various individual -- but not necessarily mutually-exclusive -- selection techniques; as discussed above, (a) certain techniques displayed in this figure are derived from/intrinsically rely upon other overarching techniques (e.g., double-peaked optical spectroscopic emission line selection requires the use of optical spectroscopy, optical spectroscopic emission lines ratios, and typically fiber spectroscopy as well), and (b) dual AGN candidates are often identified using multiple waveband selection techniques (e.g., dual AGN candidates selected via radio imaging or mid-IR imaging often require optical imaging as well). Figure~\ref{fig:aggdualAGNselmethod} reflects the variety of selection methods in the dual AGN literature, but it comes with two key caveats: 
\begin{itemize}
    \item It bears repeating that the selection techniques listed are not necessarily independent or unique, so Figure~\ref{fig:aggdualAGNselmethod} should not be construed as illustrating that each listed technique always identifies independent samples.
    \item The completeness of this figure is limited by the availability of information in refereed articles; the percentages laid out here should be considered accurate approximations but not necessarily precise (for example, some objects listed as using `optical slit spectroscopy' may have actually used `long-slit spectroscopy' but this information was not obvious or available for the given target and observation).
\end{itemize}
These same caveats apply also to Figure~\ref{fig:aggbinaryAGNselmethod} and Figure~\ref{fig:aggrecoilAGNselmethod} in Sections~ \ref{sec:binaryselectmethods} and \ref{sec:recoilselectmethods}.

Optical spectroscopy is by far the most common selection technique invoked in the literature to select dual AGN candidates (84.3\% of candidates are selected with it), which is driven primarily by the use of SDSS optical fiber spectra (82.4\%) for the selection of (1) kpc-scale dual AGNs based on double-peaked optical spectroscopic emission lines -- with optical spectroscopic emission line ratios indicative of optical AGNs -- observed in a single fiber \citep[][though this last work used LAMOST fiber spectra instead of SDSS]{wang2009,liu2010b,smith2010,ge2012,pilyugin2012,lyu2016,yuan2016,shi2014}, and (2) optical AGN pairs or groups identified from distinct fiber spectra showing spectroscopic emission line ratios indicative of optical AGNs within multiple, distinct galaxy nuclei \citep[e.g.,][]{liu2011b}. Comparing the frequency of optical spectroscopy with optical fiber spectroscopy, other optical spectroscopic techniques (long-slit, slit, slitless, IFU, etc.) make up $<2$\% of the objects selected using optical spectroscopy (though, this is due to the saturated use of SDSS fiber spectra for large samples of objects, which we will discuss below). 70.1\% of all dual candidates are selected using optical spectroscopic emission line ratios, either from fiber spectra or some other type of spectroscopy (e.g., slit, slit-less, etc.), and 42\% of all candidates are selected based upon double-peaked optical lines \citep[all of which depend upon spectroscopic emission line ratios, e.g.,][]{wang2009,liu2010b,smith2010,ge2012,pilyugin2012,lyu2016,yuan2016,shi2014}. Velocity offset narrow spectroscopic emission lines \citep{comerford2011,comerford2013,comerford2014}, used to select so-called ``offset AGNs,'' or dual SMBHs comprised of active-inactive SMBH pairs, are used to select 9.1\% of all candidates.

Optical imaging was used during the selection process for a significant fraction of candidate dual AGNs (17\%), where it is used for selection based on optical color, astrometry, PSF shape, etc. \citep[7.1\%, e.g.,][]{weedman1982,hennawi2006,hennawi2010,eftekharzadeh2017,myers2007,green2011,findlay2018,inada2008,inada2010,inada2012,hwang2020,orosz2013,agnello2018,spiniello2018,rusu2019,lemon2018,lemon2020,lemon2019,more2016} and/or for morphological characterization of the host galaxies \citep[e.g.,][]{satyapal2017,pfeifle2019a,fu2015a,fu2015b,fu2018,teng2012,green2011,koss2012,koss2011,junkkarinen2001,tadhunter2012}. Mid-IR imaging and/or colors (2.6\%, Table~\ref{tab:citsec411}, Row 3), radio imaging (2.7\%, Table~\ref{tab:citsec411}, Row 4), near-IR imaging (0.2\%, Table~\ref{tab:citsec411}, Row 4), hard X-ray imaging (0.3\%, Table~\ref{tab:citsec411}, Row 7), and soft X-ray imaging (0.7\%, Table~\ref{tab:citsec411}, Row 8) each were used to select small but notable samples of dual AGN candidates. About half of all radio imaging was used to identify dual radio sources in dual AGN candidates (Table~\ref{tab:citsec411}, Row 5).

\begin{table*}[t!]
\caption{Citation Table for Section~\ref{sec:aggselmethodduals}}
\begin{center}
\hspace{-2.2cm}
\begin{tabular}{cccc}
\hline
\hline
\noalign{\smallskip}
Row \# & Method & \% & Citations\\
\noalign{\smallskip}
\hline
\noalign{\smallskip}
1 & Optical Imaging & 2.6\% & \hspace{-10mm}\makecell[l]{e.g., \citet[][]{satyapal2017,pfeifle2019a,fu2015a,fu2015b,fu2018}\\ \citet{teng2012,green2011,koss2012,koss2011}\\ \citet{junkkarinen2001,tadhunter2012} \\ \citet[e.g.,][]{weedman1982,hennawi2006,hennawi2010}\\ \citet{eftekharzadeh2017,myers2007,green2011}\\ \citet{findlay2018,inada2008,inada2010,inada2012,hwang2020}\\ \citet{orosz2013,agnello2018,spiniello2018}\\ \citet{rusu2019,lemon2018,lemon2020,lemon2019,more2016}} \\
2 & Optical Colors & 7.1\% & \hspace{-10mm}\makecell[l]{e.g., \citet[][]{weedman1982,hennawi2006,hennawi2010}\\ \citet{eftekharzadeh2017,myers2007,green2011}\\ \citet{findlay2018, inada2008,inada2010,inada2012,hwang2020}\\ \citet{orosz2013,agnello2018,spiniello2018}\\ \citet{rusu2019,lemon2018,lemon2020,lemon2019,more2016} \\ } \\
3 & Mid-IR imaging and/or colors & 2.6\% & \hspace{-10mm}\makecell[l]{\citet{findlay2018,satyapal2017,pfeifle2019a}\\ \citet{ellison2017,agnello2018,spiniello2018}\\ \citet{lemon2018,lemon2019,lemon2020,schechter2017,torres2018}} \\
4 & Radio imaging & 2.7\% & \hspace{-10mm}\makecell[l]{\citet{fu2015a,fu2015b,orosz2013}\\ \citet{owen1985,klamer2004,dutta2018}\\ \citet{tadhunter2012,wong2008,koopmans2000}\\ \citet{heckman1986,gitti2006,gitti2013,xiangliu2014}\\ \citet{herzog2015,stockton2004,mu1998}\\ \citet{hewitt1987,mullersanchez2015}}\\
5 & Dual Radio Sources & X\% & \hspace{-10mm}\makecell[l]{\citet{fu2015a,fu2015b,koopmans2000,owen1985}\\ \citet{klamer2004,gitti2006,gitti2013,xiangliu2014}\\ \citet{mullersanchez2015}} \\
6 & Near-IR imaging & 0.2\% & \hspace{-10mm}\makecell[l]{\citet{imanishi2014,imanishi2020,graham1990}\\ \citet{stockton2004,heidt2003,u2013}}\\
7 & Hard X-ray imaging & 0.3\% & \citet{koss2011,koss2012} \\
8 & Soft X-ray imaging & 0.7\% & \hspace{-10mm}\makecell[l]{\citet{komossa2003,bianchi2008,piconcelli2010}\\ \citet{jim2007,koss2012,satyapal2017}\\ \citet{pfeifle2019a,ellison2017,liu2013} \\\citet{hou2019,wang2010,stockton2004}\\ \citet{ballo2004,guainazzi2005,torres2018}\\ \citet{kim2015,barrows2016,wong2008}\\ \citet{eckert2017,iwasawa2011a,gitti2006}\\ \citet{mu2001,lena2018}}\\
\noalign{\smallskip}
\hline
\end{tabular}
\end{center}
%\tablecomments{}
\label{tab:citsec411}
\end{table*}

Having examined the aggregate frequencies with which different selection methods are used within the dual AGN literature, we now turn our attention in the following subsections (Sections~\ref{sec:dualbreakoptspec}-\ref{sec:dualbreakradioimaging}) to finer quantitative breakdowns of these selection methodologies. In each subsection, we focus on one particular, overarching selection technique (for example: ``Optical Spectroscopy''), and we then assess the breakdown of derivative and/or complementary selection methods used in concert with the overarching technique. Additionally, we also assess the fraction of these objects that are also independently identified using methods that do not rely upon the overarching technique. These overarching selection techniques are either used significantly in the literature or warrant specific mention. Each bar in every chart is normalized to the total number of systems required to have the `overarching' selection technique in their selection function. All percentages quoted in the text are relative to the total number of objects selected via the overarching technique and -- as emphasized above -- do not necessarily represent mutually-exclusive fractions of the subsample in question, so caution should be exercised when comparing percentages against one another. Note, Sections~\ref{sec:binaryselectmethods} and \ref{sec:recoilselectmethods} (and all subsections therein) are organized in identical fashions to this section (Section~\ref{sec:dualselectmethods} and subsections \ref{sec:aggselmethodduals}-\ref{sec:dualbreakradioimaging}). 

\subsubsection{Dual AGN Selection Method Breakdown: Optical Spectroscopy}
\label{sec:dualbreakoptspec}
%\label{sec:selmethodcomparisons}

\begin{figure*}
    \centering
    \includegraphics[width=1.0\linewidth]{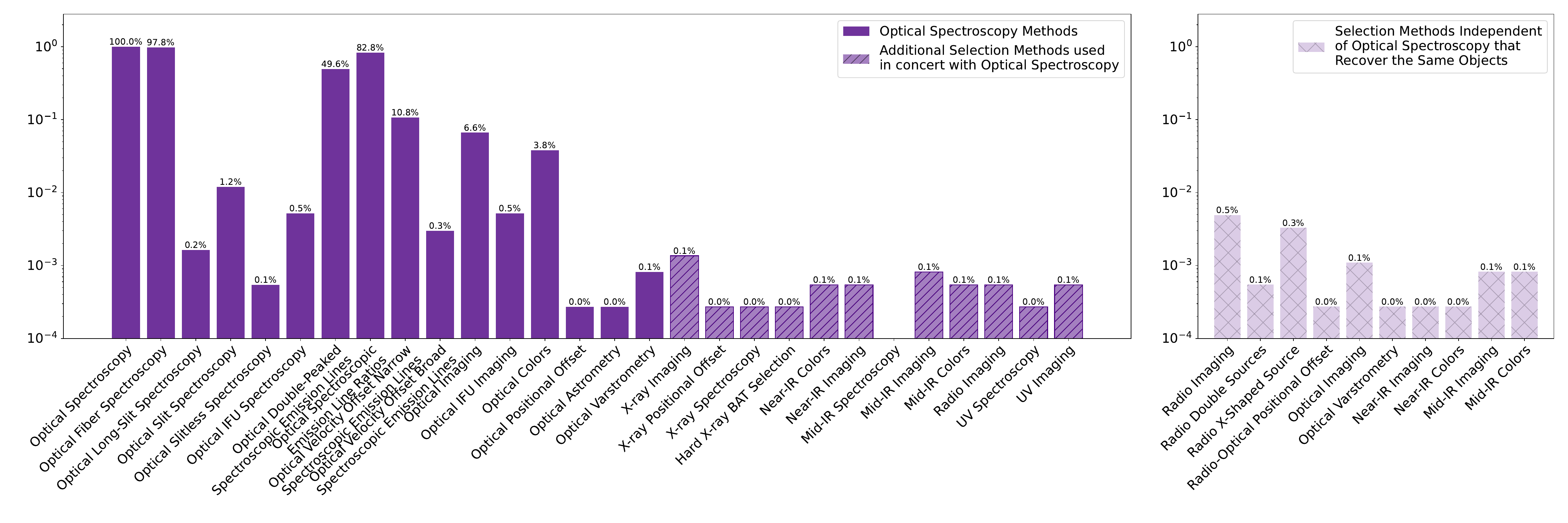}
    \caption{Dual AGN candidates that required optical spectroscopy for selection. A breakdown of dual AGN selection strategies used in concert with (left) or independently (right) of optical spectroscopy.}
    \label{fig:selection_duals_optical_spec}
\end{figure*}

In Figure~\ref{fig:selection_duals_optical_spec} we show the distribution of selection methods for objects that required optical spectroscopy (3671 objects) as a component of the selection function. The vast majority of these optical spectroscopically-selected objects used fiber spectroscopy (97.8\%, Table~\ref{tab:citsec412}, Row 1) and optical spectroscopic emission line ratios (82.8\%, Table~\ref{tab:citsec412}, Row 2) as a part of the selection methodology, though a small fraction of those objects that required fiber spectroscopy focused on higher redshift AGNs (Table~\ref{tab:citsec412}, Row 1) and did not rely upon spectroscopic emission line ratios. Double-peaked spectroscopic emission lines (which were commonly quoted as possibly hosting kpc-scale dual AGNs) were used for the selection of essentially half (49.6\%, Table~\ref{tab:citsec412}, Row 3) of all optical spectroscopically-selected dual AGN candidates. Additionally, 10.8\% (0.3\%) of these optically-selected objects relied upon the presence of velocity offset narrow (broad) spectroscopic emission lines for selection (Table~\ref{tab:citsec412}, Row 4), and smaller fractions of objects included the use of IFU (0.5\%, Table~\ref{tab:citsec412}, Row 1), long-slit (0.2\%, Table~\ref{tab:citsec412}, Row 6), slit (1.2\%, Table~\ref{tab:citsec412}, Row 7; this may or may not include long-slit spectroscopy, but it is not always clear in the literature), and slitless (0.1\%, Table~\ref{tab:citsec412}, Row 8) spectroscopy. Optical imaging was used in concert with optical spectroscopy for 6.6\% of objects (Table~\ref{tab:citsec412}, Row 9), where 0.5\% of objects used IFU imaging (Table~\ref{tab:citsec412}, Row 10), and optical colors were used for 3.8\% of objects (Table~\ref{tab:citsec412}, Row 11). A large number of additional techniques have been used in concert with optical spectroscopy - such as optical astrometry/varstrometry \cite{shen2019,hwang2020}, optical positional offsets \citep[][]{sun2016}, X-ray imaging and X-ray-optical positional offsets \citep{lena2018,barrows2016}, hard X-ray BAT selection \cite{koss2012}, near-IR imaging and colors \citep{altamura2020}, mid-IR imaging and/or colors \citep{ellison2017,goulding2019,altamura2020}, and radio imaging 
\citep{mu1998,heckman1986,evans2008,tadhunter2012} - but these various techniques each are only used for $\leq0.1$\% of objects selected via optical spectroscopy.
%\textcolor{violet}{Check these two...}

Small fractions of these optical spectroscopically-selected objects have also been independently identified using a small variety of multiwavelength techniques. These independent techniques include selection based on radio imaging (0.5\%), which includes selection based on double radio sources \citep[0.1\%,][]{fu2015a,fu2015b} and radio-optical positional offsets \citep[$<0.1$\%,][]{orosz2013}, optical imaging \citep[0.1\%, e.g.,][]{fu2015a,fu2015b,pfeifle2019a,orosz2013,hwang2020}, mid-IR imaging and color selection \citep[each $0.1$\%,][]{satyapal2017,pfeifle2019a}, and near-IR imaging and color selection \citep[each $<0.1$\%,][]{imanishi2014,imanishi2020}.

\begin{table*}[t!]
\caption{Citation Table for Section~\ref{sec:dualbreakoptspec}}
\begin{center}
\hspace{-2.2cm}
\begin{tabular}{cccc}
\hline
\hline
\noalign{\smallskip}
Row \# & Method & \% & Citations\\
\noalign{\smallskip}
\hline
\noalign{\smallskip}
%Binary Quasar & SDSSJ0812+2520A & 105.8 & ICRF 
1 & Fiber Spectroscopy & 97.8\% & \hspace{-10mm}\makecell[l]{\citep{liu2010b,liu2011b,smith2010,wang2009}\\ \citet{ge2012,yuan2016,lyu2016}\\ \citet{kim2020,shi2014,comerford2014}\\ \citet{barrows2016,teng2012,hennawi2006,hennawi2010}\\ \citet{inada2008,inada2010,inada2012,pindor2006,more2016} } \\
2 & \hspace{-10mm}\makecell[l]{Optical Spectroscopic\\ Emission Line Ratios} & 82.8\% & \hspace{-10mm}\makecell[l]{\citep{liu2010b,liu2011b,smith2010,wang2009}\\ \citet{ge2012,yuan2016,lyu2016}\\ \citet{kim2020,shi2014,comerford2014}\\ \citet{barrows2016,teng2012} }\\
3 & \hspace{-10mm}\makecell[l]{Optical Double-peaked\\ spectroscopic emission lines} & 49.6\% & \hspace{-10mm}\makecell[l]{\citet{liu2010b,smith2010,wang2009}\\ \citet{ge2012,shi2014,kim2020}\\ \citet{lyu2016,yuan2016}} \\
4 & \hspace{-10mm}\makecell[l]{Velocity offset narrow (broad)\\ spectroscopic emission lines} & 10.8\% (0.3\%) & \hspace{-10mm}\makecell[l]{\cite{comerford2014,comerford2013}\\ \citet{barrows2016,sun2016,komossa2008}\\ \citet{boroson2009} }\\
1 & Optical IFU Spectroscopy & 0.5\% & \hspace{-10mm}\makecell[l]{\citet{fu2018,ellison2017,balmaverde2018}} \\
5 & Optical Long-Slit Spectroscopy & 0.2\% & \hspace{-10mm}\makecell[l]{\citet{dutta2018,keel1985,bothun1989}\\ \citet{keel2019,tadhunter2012}} \\
5 & Optical Slit Spectroscopy & 1.2\% & \hspace{-10mm}\makecell[l]{\citet{barrows2016,comerford2009a}\\ \citet{heckman1986,altamura2020}\\ \citet{sekiguchi1992,jim2007}\\ \citet{hagen1996}} \\
5 & Optical Slitless Spectroscopy & 0.1\% & \hspace{-10mm}\makecell[l]{\citet{crampton1988,djorgovski1984}} \\
2 & Optical Imaging & 6.6\% & \hspace{-10mm}\makecell[l]{e.g., \citet{more2016,inada2008,inada2010,inada2012}\\ \citet{tadhunter2012,menezes2014}\\ \citet{balmaverde2018,woo2014,hewett1998}\\ \citet{cai2018,arrigoni2019b}\\ \citet{altamura2020,bothun1989}\\ \citet{sekiguchi1992,ellison2017}\\ \citet{fu2018,huang2014,teng2012}\\ \citet{jim2007,mu1998}\\ \citet{gregg2002,green2010}\\ \citet{djorgovski1984,hagen1996}\\ \citet{hennawi2006,hennawi2010,weedman1982}\\ \citet{fu2015a,fu2015b,sun2016} }\\
3 & Optical IFU Imaging & 0.5\% & \hspace{-10mm}\makecell[l]{\citet{fu2018,ellison2017,balmaverde2018}\\ \citet{menezes2014} } \\ 
4 & Optical Colors & 3.8\% & \hspace{-10mm}\makecell[l]{ \citet{more2016,inada2008,inada2010,inada2012}\\ \citet{altamura2020,eftekharzadeh2017}\\ \citet{djorgovski1984,weedman1982}\\ \citet{hennawi2006,myers2007,green2011} }\\
\noalign{\smallskip}
\hline
\end{tabular}
\end{center}
%\tablecomments{}
\label{tab:citsec412}
\end{table*}

\subsubsection{Dual AGN Selection Technique Breakdown: Optical Imaging}
\label{sec:dualbreakoptimg}
%\label{sec:selmethodcomparisons}

We show the distribution of selection methods for objects that required optical imaging (738 objects) as a component of the selection function in Figure~\ref{fig:selection_binary_optical_imaging}. There is a large variety of selection techniques that either rely upon optical imaging or are used in concert with optical imaging; here we will discuss those techniques which were used during the selection of at least $\sim2\%$ of the objects that required optical imaging. Optical colors have been frequently used (41.5\% of objects, Table~\ref{tab:citsec413}, Row 1). Optical imaging also played a key role in the morphological selection of objects in ongoing interactions or mergers (Table~\ref{tab:citsec413}, Row 2), though we do not include ``optical morphological classification'' as a selection technique in this data release. Optical spectroscopy has been used for 31.7\% of these objects (Table~\ref{tab:citsec413}, Row 3), specifically fiber spectroscopy (28.2\%, Table~\ref{tab:citsec413}, Row 4). In 2.6\% of cases, optical IFU imaging and spectroscopy played a role in the selection function of objects (Table~\ref{tab:citsec413}, Row 5). Smaller fractions of objects, in concert with optical imaging, used radio imaging (13.8\%, Table~\ref{tab:citsec413}, Row 9) -- often to search for double radio sources (7.0\%, Table~\ref{tab:citsec413}, Row 10) or offsets between radio and optical positions \citep[6.5\%][]{orosz2013} -- or used mid-IR imaging and colors to select objects (14.5\%, Table~\ref{tab:citsec413}, Row 7) and mid-IR-to-optical colors (2.6\%, Table~\ref{tab:citsec413}, Row 8). Finally, optical varstrometry played a role in the selection of 6.6\% of these objects (Table~\ref{tab:citsec413}, Row 6). A substantial number of other techniques have been used in concert with optical imaging as well -- including optical spectroscopic techniques \citep[e.g., slit or long-slit spectroscopy, velocity offset broad lines, and/or spectroscopic emission line ratios,][]{hennawi2006,fu2018,teng2012,djorgovski1984,jim2007,sekiguchi1992,huang2014}, infrared imaging and luminosity selection \citep[e.g.][]{mudd2014,wang2010,sekiguchi1992}, near-IR imaging and/or colors \citep{heidt2003,altamura2020}, and optical positional offsets and astrometry -- but each of these selection techniques were used in a severe minority of cases ($<2\%$).

A very small fraction of objects (1.9\%) were selected using optical spectroscopy completely independent of efforts that require optical imaging; fiber spectroscopy was used to select 1.5\% of candidates independently \citep{liu2011b,ge2012}, spectroscopic emission line ratios were used for 1.2\% \citep{liu2011b}, while double-peaked emission lines \citep[0.1\%,][]{ge2012} and offset narrow emission lines \citep[0.1\%,][]{comerford2014} selected only a few objects independently of optical imaging efforts.

\begin{figure*}
    \centering
    \includegraphics[width=1.0\linewidth]{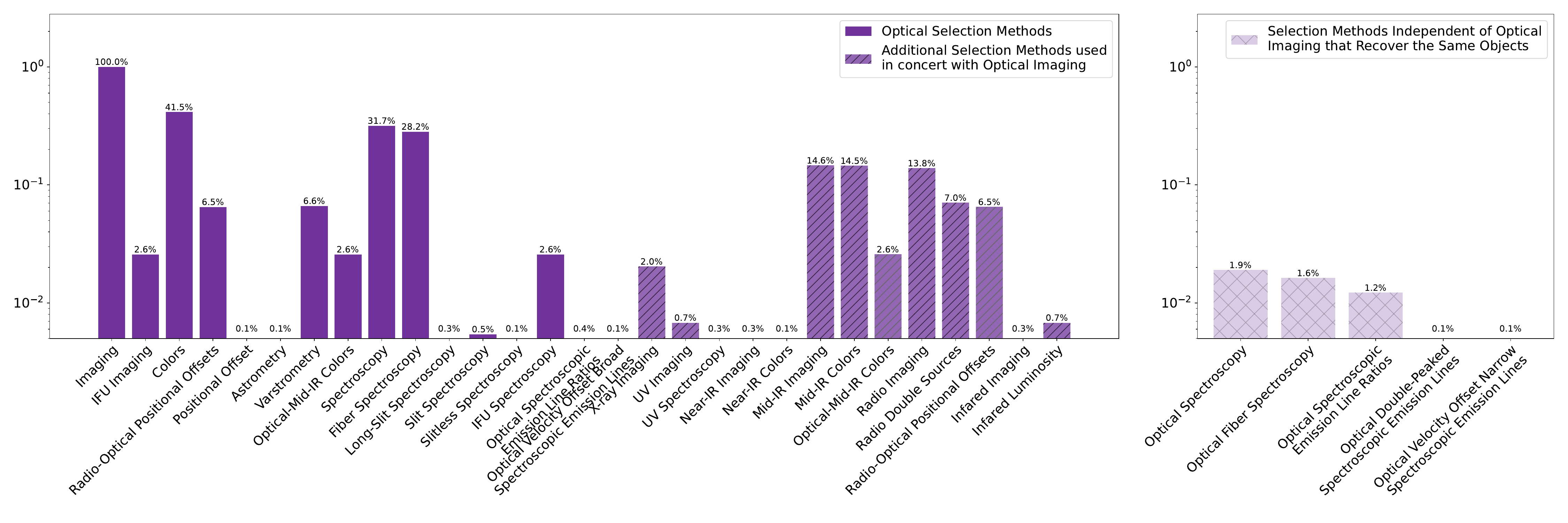}
    \caption{Dual AGN candidates that required optical imaging for selection. A breakdown of dual AGN selection strategies used in concert with (left) or independently (right) of optical imaging.}
    \label{fig:selection_duals_optical_imaging}
\end{figure*}

\begin{table*}[t!]
\caption{Citation Table for Section~\ref{sec:dualbreakoptimg}}
\begin{center}
\hspace{-2.2cm}
\begin{tabular}{cccc}
\hline
\hline
\noalign{\smallskip}
Row \# & Method & \% & Citations\\
\noalign{\smallskip}
\hline
\noalign{\smallskip}
%Binary Quasar & SDSSJ0812+2520A & 105.8 & ICRF 
1 & Optical colors & 41.5\% & \hspace{-10mm}\makecell[l]{e.g., \citet{hennawi2006,eftekharzadeh2017} \\ \citet{weedman1982,myers2007,hennawi2010} \\ \citet{findlay2018,inada2008,inada2010,inada2012,more2016} \\ \citet{djorgovski1984,green2010,brotherton1999} \\ \citet{meylan1989,agnello2018,spiniello2018} \\ \citet{lemon2018}} \\
2 & Morphological Class & \dots & \hspace{-10mm}\makecell[l]{e.g., \citet{fu2015a,fu2015b,satyapal2017,pfeifle2019a} \\ \citet{ellison2017,mudd2014,junkkarinen2001}} \\
3 & Optical Spectroscopy & 31.7\% & \hspace{-10mm}\makecell[l]{e.g., \citep{fu2015a,fu2018,teng2012,hennawi2006,hennawi2010} \\ \citet{hagen1996,weedman1982,more2016} \\ \citet{inada2008,inada2012,eftekharzadeh2017} \\ \citet{green2010,djorgovski1984} \\ \citet{jim2007,gregg2002,mu1998} \\ \citet{sandrinelli2014b,huang2014,ellison2017} \\ \citet{sekiguchi1992,bothun1989} \\ \citet{arrigoni2019b,cai2018,sun2016} \\ \citet{hwang2020}} \\
4 & Optical Fiber Spectroscopy & 28.2\% & \hspace{-10mm}\makecell[l]{e.g., \citet{hennawi2006,hennawi2010,more2016} \\ \citet{sun2016,lemon2019,inada2008,inada2010,inada2012} \\ \citet{hwang2020,teng2012,huang2014} \\ \citet{arrigoni2019b,cai2018} } \\
5 & Optical IFU Imaging/Spectroscopy & 2.6\% & \hspace{-10mm}\makecell[l]{\citet{ellison2017,fu2018,menezes2014} \\ \citet{balmaverde2018}} \\
6 & Optical Varstrometry & 6.6\% & \hspace{-10mm}\makecell[l]{\citep{shen2019,hwang2020,popovic2012}} \\
7 & Mid-IR Imaging and Colors & 14.5\% & \hspace{-10mm}\makecell[l]{\citet{findlay2018,pfeifle2019a,ellison2017} \\ \citet{altamura2020,satyapal2017,schechter2017} \\ \citet{assef2016,agnello2018,spiniello2018} \\ \citet{lemon2018,lemon2019,lemon2020}}\\
8 & Mid-IR-to-Optical Colors & 2.6\% & \hspace{-10mm}\makecell[l]{\citet{agnello2018,assef2016,lemon2018,lemon2019}}\\
9 & Radio imaging & 13.8\% & \hspace{-10mm}\makecell[l]{e.g., \citet{fu2015a,fu2015b,orosz2013,mu1998} \\ \citet{tadhunter2012}}\\
10 & Dual Radio Sources & 7.0\% & \hspace{-10mm}\makecell[l]{\citet{fu2015a,fu2015b}} \\
11 & Near-IR imaging & 0.2\% & \hspace{-10mm}\makecell[l]{\citet{heidt2003,altamura2020} }\\
12 & Infrared imaging and luminosity & 0.3\% & \hspace{-10mm}\makecell[l]{e.g., \citet{mudd2014,wang2010} \\ \citet{sekiguchi1992}} \\
\noalign{\smallskip}
\hline
\end{tabular}
\end{center}
%\tablecomments{}
\label{tab:citsec413}
\end{table*}

\subsubsection{Dual AGN Selection Technique Breakdown: Mid-IR Imaging}
\label{sec:dualbreakmidirimg}
%\label{sec:selmethodcomparisons}

\begin{figure*}
    \centering
    \includegraphics[width=1.0\linewidth]{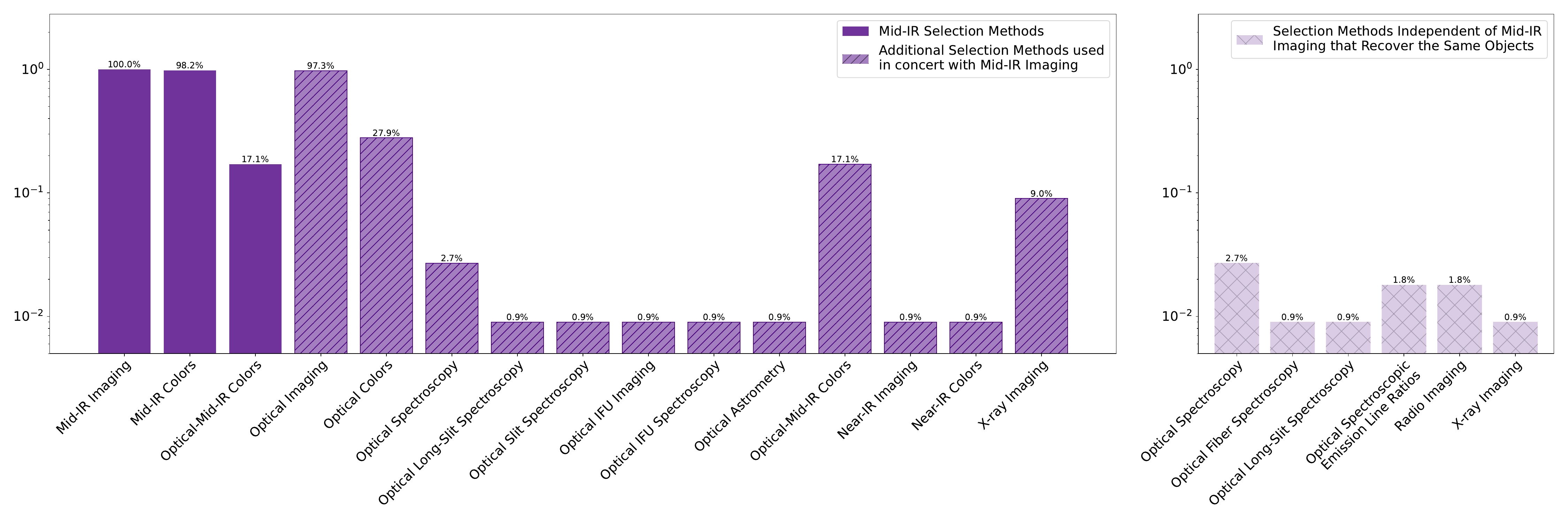}
    \caption{Dual AGN candidates that required mid-IR imaging for selection.  A breakdown of dual AGN selection strategies used in concert (left) or independently (right) of mid-IR imaging.}
    \label{fig:selection_duals_mir_imaging}
\end{figure*}

In Figure~\ref{fig:selection_duals_mir_imaging} we show the distribution of selection methods for objects that required mid-IR imaging (111 objects). Naturally, 98.2\% of these objects relied upon mid-IR color selection \citep[][]{satyapal2017,pfeifle2019a,findlay2018,ellison2017,altamura2020,schechter2017,assef2016,agnello2018,spiniello2018,lemon2018,lemon2019,lemon2020,goulding2019} and optical imaging was used in concert with the mid-IR imaging for 97.3\% of all mid-IR selected objects in these works. Of this 97.3\%, optical colors played a role in the selection of 27.9\% of objects \citep{findlay2018,altamura2020,schechter2017,agnello2018,spiniello2018,lemon2018,lemon2020}, while mid-IR-to-optical colors played a role in the selection of 17.1\% of objects \citep{agnello2018,lemon2018,lemon2019,assef2016}. In 9\% of systems, X-ray imaging also played a key role in the selection function \citep[where these objects were first selected using mid-IR colors and optical imaging for morphological selection,][]{satyapal2017,pfeifle2019a}. In a small fraction of objects, optical spectroscopy (2.7\%) - including IFU spectroscopy \citep{ellison2017} and slit spectroscopy \citep{altamura2020} - and near-IR imaging \citep[0.9\%][]{altamura2020} also played roles in the selection of these systems. 

A small handful of techniques were also used to independently identify small fractions of these objects: optical spectroscopy was used to independently identify 2.7\% of these mid-IR-selected candidates \citep{bianchi2008,liu2011b,fu2018,dutta2018}, while radio imaging \citep{dutta2018} and X-ray imaging \citep{bianchi2008,comerford2015,barrows2016} were used independently to identify small percentages of these objects. 

\subsubsection{Dual AGN Selection Technique Breakdown: X-ray Imaging}
\label{sec:dualbreakxrayimg}
%\label{sec:selmethodcomparisons}

\begin{figure*}
    \centering
    \includegraphics[width=1.0\linewidth]{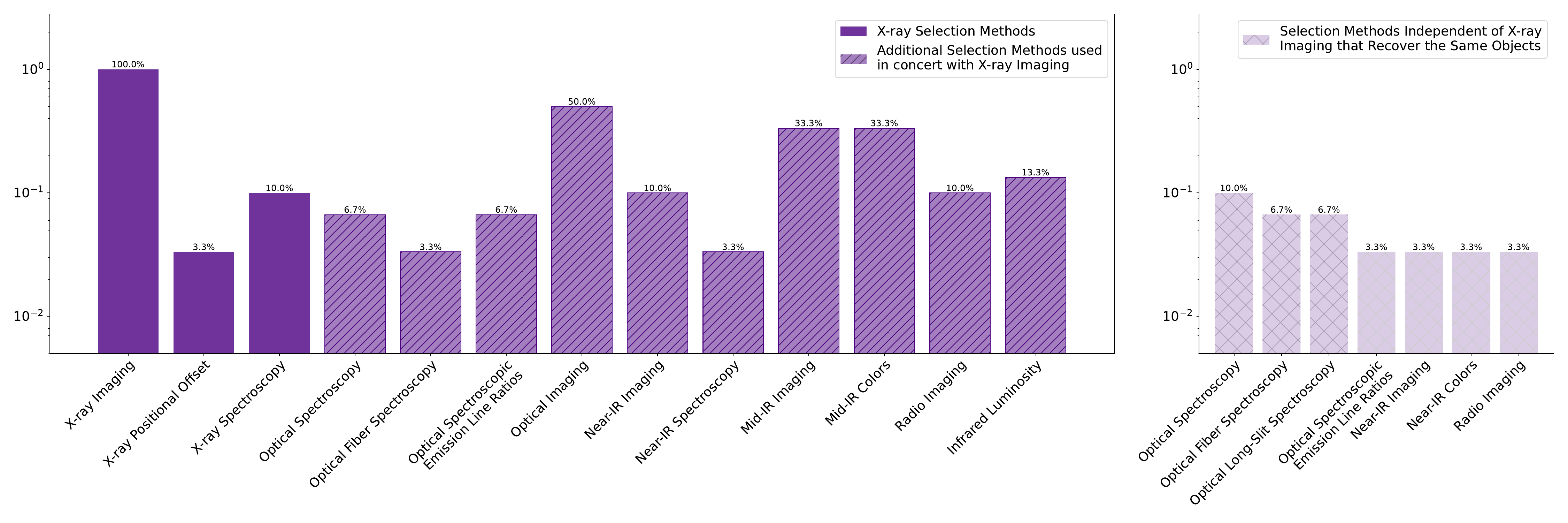}
    \caption{Dual AGN candidates that required X-ray imaging for selection. A breakdown of dual AGN selection strategies used in concert with (left) or independently of (right) X-ray imaging.}
    \label{fig:selection_duals_xray_imaging}
\end{figure*}

In Figure~\ref{fig:selection_duals_xray_imaging} we show the distribution of selection methods for objects that required X-ray imaging for selection (30 objects). These  include a variety of objects selected either because of a previously known, unresolved X-ray source in low resolution imaging \citep[e.g.,][]{mu2001}, or based on the identification of one or two X-ray sources directly or possibly indicative of AGNs \citep{komossa2003,iwasawa2011a,satyapal2017,pfeifle2019a,stockton2004,ballo2004,guainazzi2005,fabbiano2011,torres2018,kim2015,eckert2017,comerford2015,wang2010,wong2008,lena2014,liu2013,hou2019,hou2020,bianchi2008}.\footnote{Note: Not all of these sources were convincing candidates: NGC 3393 was selected by \citet{fabbiano2011} based on sub-pixel binned \textit{Chandra} imaging, and \citet{koss2015} later showed this system did not host a dual AGN.} 10\% of these objects include the use of X-ray spectroscopy for selection \citep[e.g.,][]{kim2015,iwasawa2011a,piconcelli2010,bianchi2008,komossa2003,guainazzi2005}, and a small fraction \citep[3.3\%,][]{barrows2016} were selected based on X-ray positional offsets from optical positions. 

Optical imaging was used in concert with the X-ray imaging for $\sim50\%$ of these objects to select and/or study the morphology of the host system \citep{satyapal2017,pfeifle2019a,comerford2015,barrows2016,wang2010,torres2018,mu2001}. 6.7\% of X-ray selected objects were paired with optical spectroscopy \citep{lena2018,barrows2016,piconcelli2010}; among objects where the specific spectroscopic technique was known, 3.3\% of objects were paired specifically with fiber spectroscopy \citep{lena2018,barrows2016}. In all cases where optical spectroscopy was used in concert with X-ray imaging, optical spectroscopic emission line ratios were used as a part of the selection technique. Roughly a third (33\%) of these X-ray-selected objects also used mid-IR imaging and colors in concert with the X-ray imaging to select candidates \citep{satyapal2017,pfeifle2019a}. Near-IR imaging (10\%) and spectroscopy (3.3\%) were used for a small number of objects \citep{mu2001,stockton2004,u2013}, and in a few cases infrared luminosity (used to select objects as LIRGs; 13.3\%) and X-ray imaging were paired to  select the objects as dual AGNs and candidates \citep{wang2010,torres2018}. 10\% of X-ray selected objects also used radio imaging as a part of the selection function \citep{stockton2004,wong2008,gitti2006,gitti2013}. 

A small number of waveband techniques were used to independently identify some of these same objects: optical spectroscopy was used independently to identify 10\% of these objects \citep[][]{liu2011b,dutta2018}, while near-IR imaging/colors \citep{imanishi2014} and radio imaging \citep{dutta2018} were each used to independently select 3.3\% of these X-ray selected objects. 

\subsubsection{Dual AGN Selection Technique Breakdown: Radio Imaging}
\label{sec:dualbreakradioimaging}
%\label{sec:selmethodcomparisons}

\begin{figure*}
    \centering
    \includegraphics[width=1.0\linewidth]{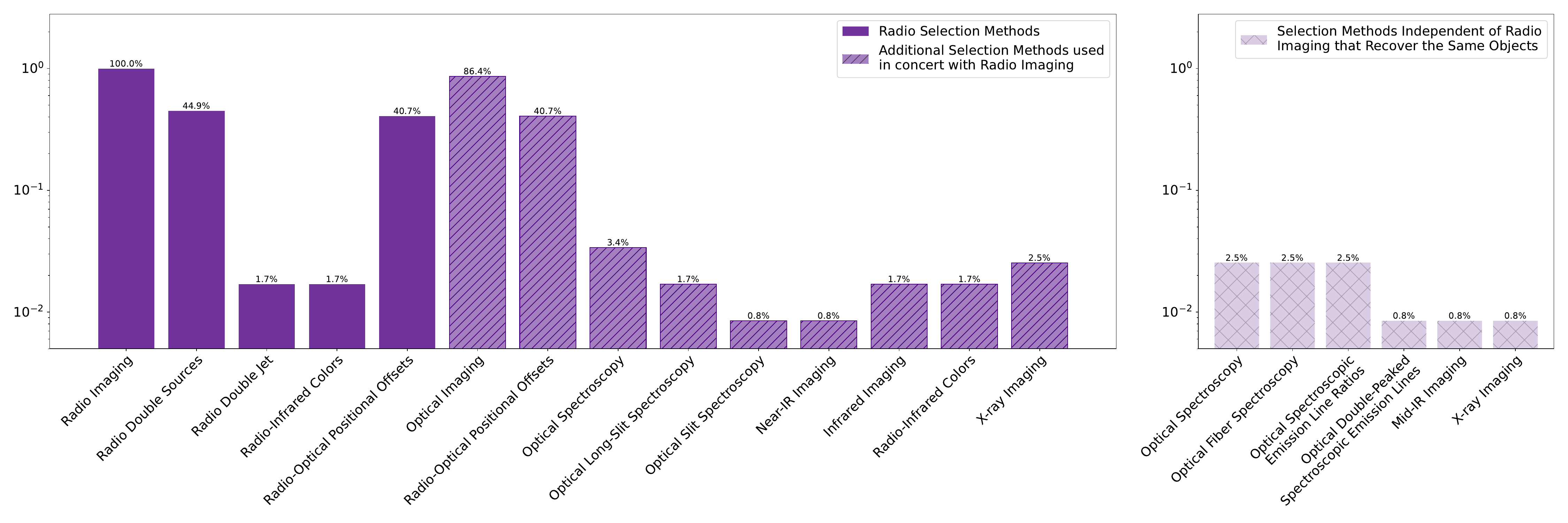}
    \caption{Dual AGN candidates that required radio imaging for selection. A breakdown of dual AGN selection strategies used in concert with (left) or independently of (right) radio imaging.}
    \label{fig:selection_dual_radio_imaging}
\end{figure*}

In Figure~\ref{fig:selection_dual_radio_imaging} we show the distribution of selection methods for objects that required radio imaging for selection (118 objects). As expected, optical imaging is often (86.4\%) used in conjunction with radio imaging to study host morphology and nuclear positions \citep{tadhunter2012,fu2015a,fu2015b,mu1998} -- such as when double radio sources are identified \citep[44.9\% of all objects,][]{fu2015a,fu2015b} -- or to study radio-optical positional offsets \citep[40.7\% of objects,][]{orosz2013} that could be due to a multiplicity of sources. Radio imaging has also been used to identify a small number of sources hosting (or possibly hosting) double radio jets \citep[1.7\%, e.g.,][]{owen1985,klamer2004,xiangliu2014}, one of which is a bona fide dual AGN \citep[3C 75,][]{owen1985}. A small number of objects were selected based on radio-infrared colors \citep[1.7\%,][]{herzog2015}, or used X-ray imaging \citep[2.5\%,][]{wong2008,gitti2006,gitti2013,stockton2004} as a complementary selection step. 3.4\% also used optical spectroscopy as a complementary strategy to derive redshifts and/or optical classes \citep{dutta2018,fu2015b,mu1998,heckman1986,tadhunter2012}.

A small fraction of these objects selected via radio imaging were also selected independently via optical spectroscopy \citep[2.5\%,][]{liu2011b}, and specifically fiber spectroscopy used to select on spectroscopic emission lines \citep[2.5\%,][]{liu2011b,ge2012} and/or double-peaked emission lines \citep[0.8\%,][]{ge2012}. In a similar vein, a few candidates were independently selected using X-ray imaging and mid-IR colors \citep[0.8\%,][]{satyapal2017}.

\subsection{Binary AGN Selection Methods}
\label{sec:binaryselectmethods}

\subsubsection{Binary AGN Aggregate Selection Method Statistics}
\label{sec:aggselectmethodbins}

\begin{figure*}[t]
    \centering
    \includegraphics[width=1.0\linewidth]{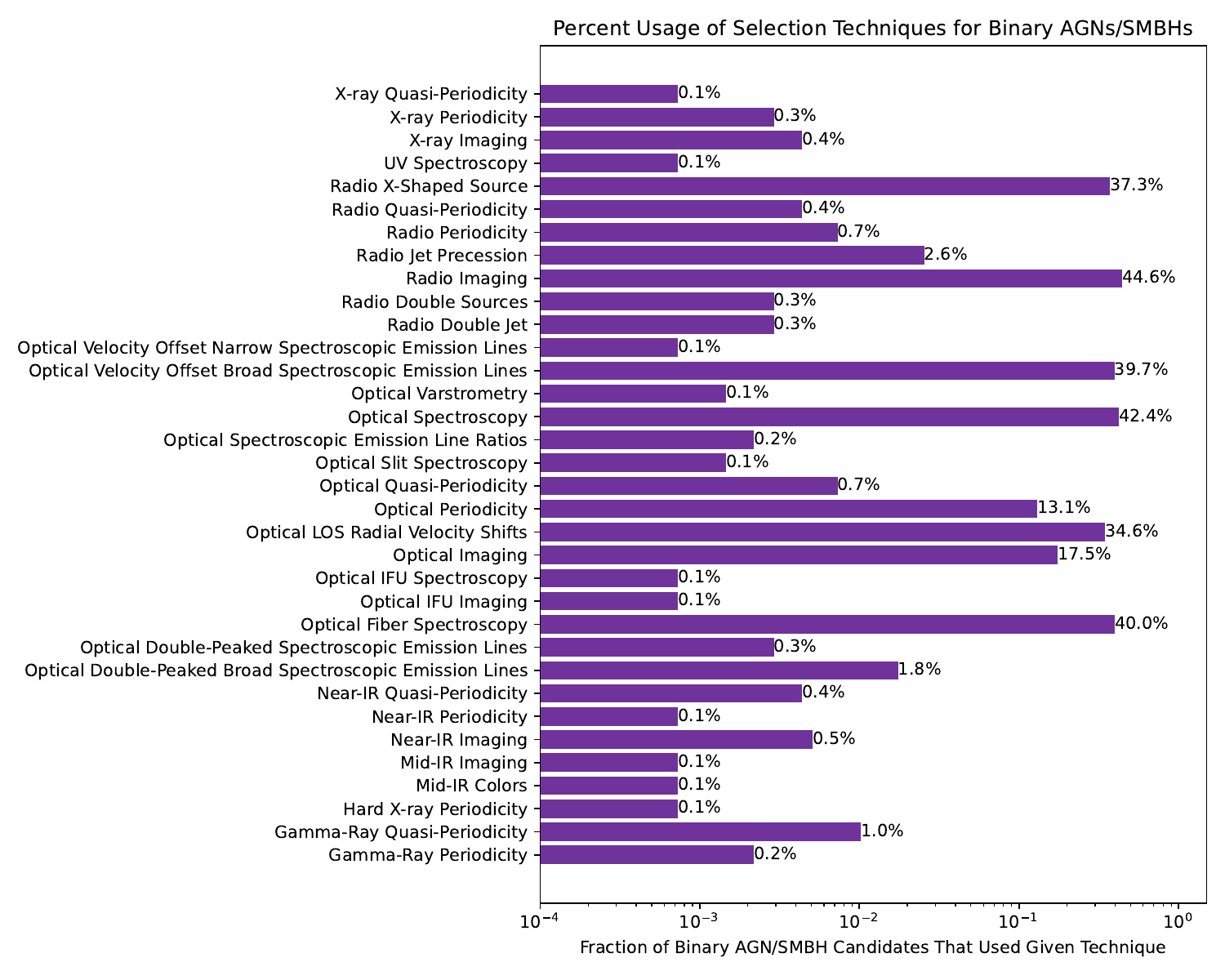}
    \caption{A breakdown of the selection methods used to identify binary AGNs/SMBHs and candidates in the literature. This chart shows the aggregate percent usage of all binary AGN selection techniques (normalized by the number of binary AGN candidates in the catalog); note, these selection methods are not all necessarily mutually exclusive (for example: selection based on optical periodicity necessarily requires optical imaging).}
    \label{fig:aggbinaryAGNselmethod}
\end{figure*}

The aggregate percent frequency with which specific selection methodologies are used for binary AGN and binary SMBH systems is shown in Figure~\ref{fig:aggbinaryAGNselmethod} in an identical fashion to Figure~\ref{fig:aggdualAGNselmethod} as described in Section~\ref{sec:aggselmethodduals}. 

Binary AGNs and SMBHs have the most variety in selection methods when compared to dual AGN candidates and recoiling AGN candidates (Figure~\ref{fig:aggbinaryAGNselmethod}). 42.4\% of binary candidates are selected via optical spectroscopy (Table~\ref{tab:citsec421}, Row 1); 40\% of all candidates are specifically selected via SDSS fiber spectra (Table~\ref{tab:citsec421}, Row 2). Optical spectra are predominantly used to identify broad spectroscopic emission lines that are offset in velocity from the rest frame of the host galaxy (39.7\%, Table~\ref{tab:citsec421}, Row 3) as well as to search for evidence of line-of-sight radial velocity shifts of these broad lines across different epochs (34.6\%, Table~\ref{tab:citsec421}, Row 4). Double-peaked and/or multiple broad spectroscopic emission line systems are also used to select a small fraction of binary AGN candidates (1.8\%, Table~\ref{tab:citsec421}, Row 5).

Radio imaging is used to select a similarly large fraction of all binary AGN candidates (44.6\% of all candidates). This is primarily driven by the selection of X-shaped radio sources (37.3\%, Table~\ref{tab:citsec421}, Row 7); the formation mechanism for X-shaped radio sources is still debated, but one formation mechanism could involve the presence of an unresolved binary AGN where each AGN emits distinct jets. Cases of jet precession have been reported in the literature and have been attributed to the possible presence of SMBH binaries (2.6\%, Table~\ref{tab:citsec421}, Row 8), and several radio imaging-dependent techniques have also been used to search for and/or argue for the presence of SMBH binaries, including mas-scale dual radio sources (0.3\%, Table~\ref{tab:citsec421}, Row 9), double radio jets (0.3\%, Table~\ref{tab:citsec421}, Row 10), and radio periodicities and quasi-periodicities (0.7\% and 0.4\%, Table~\ref{tab:citsec421}, Row 11 and 12).

Optical imaging was used for the selection of a significant fraction of all binary candidates (17.5\%). This imaging is predominantly used to select objects that exhibit signs of optical periodicities (13.1\%, Table~\ref{tab:citsec421}, Row 14) or 
 quasi-periodicities (to a much lesser extent, 0.7\%, Table~\ref{tab:citsec421}, Row 15) in their optical light curves. The search for periodicities is not limited to the optical band, however, with many works selecting small samples of objects based on Gamma-Ray periodicities or quasi-periodicities (0.2\% and 0.1\%, Table~\ref{tab:citsec421}, Row 16 and 17), near-IR periodicities and quasi-periodicities (0.1\% and 0.4\%, respectively, Table~\ref{tab:citsec421}, Row 18 and 19), X-ray periodicities and quasi-periodicities (0.3\% and 0.1\%, respectively, Table~\ref{tab:citsec421}, Row 20 and 21), as well as radio periodicities (as discussed previously). While binary SMBHs are often invoked to explain periodicities and quasi-periodicities, such reported periodicities are often derived from non-uniformly sampled data and/or do not cover sufficient temporal baselines/cycles. Given that red noise can mimic periodic signals on shorter temporal baselines and that $>5$ cycles are needed to rule out red noise, the vast majority of binary candidates selected via periodicities may turn out to be false positives, as demonstrated in small samples of re-examined binary candidates \citep[e.g.,][]{vaughan2016}. Only a single binary AGN candidate selected via optical periodicity \citep[OJ 287,][]{sillanpaa1988} has remained a convincing binary AGN candidate upon repeated inspection.

\begin{table*}[t!]
\caption{Citation Table for Section~\ref{sec:aggselectmethodbins}}
\begin{center}
\hspace{-2.2cm}
\begin{tabular}{cccc}
\hline
\hline
\noalign{\smallskip}
Row \# & Method & \% & Citations\\
\noalign{\smallskip}
\hline
\noalign{\smallskip} 
1 & Optical Spectroscopy & 42.4\% & \hspace{-10mm}\makecell[l]{e.g., \citet[][]{barrows2011,tsalmantza2011,shen2013} \\ \citet{ju2013,eracleous2012,runnoe2015,runnoe2017a} \\ \citet{xliu2014,wang2017,boroson2009} \\ \citet{komossa2008,peterson1987,puchnarewicz1996} \\ \citet{stockton1991,zheng1991,eracleous1994} \\ \citet{gezari2007,lewis2010,doan2020} \\ \citet{du2018,zheng2016,menezes2014} \\ \citet{bon2012,jliu2016}\\ } \\
2 & Optical Fiber Spectroscopy & 40\% & \hspace{-10mm}\makecell[l]{e.g., \citet[][]{barrows2011,boroson2009} \\ \citet{komossa2008,tsalmantza2011,shen2013} \\ \citet{ju2013,wang2017,xliu2014} \\ \citet{eracleous2012}\\ }\\
3 & \hspace{-10mm}\makecell[l]{Velocity Offset broad \\ spectroscopic emission lines} & 39.7\% & \hspace{-10mm}\makecell[l]{e.g., \citet[][]{ju2013,wang2017,xliu2014} \\ \citet{eracleous2012,komossa2008,barrows2011} \\ \citet{boroson2009}} \\
4 & LOS radial velocity shifts & 34.6\% & \hspace{-10mm}\makecell[l]{e.g., \citet[][]{eracleous2012,shen2013,decarli2013} \\ \citet{ju2013,xliu2014,runnoe2015,runnoe2017a} \\ \citet{wang2017,guo2020}\\ }\\
5 & \hspace{-10mm}\makecell[l]{Double-peaked and/or \\multiple broad spectroscopic\\ emission lines}  & 1.8\% & \hspace{-10mm}\makecell[l]{\citet[][]{boroson2009,gezari2007,lewis2010} \\ \citet{doan2020,jliu2016,eracleous1994} \\ \citet{du2018,zheng1991,peterson1987} \\ \citet{barrows2011,halpern1988,eracleous1997}\\ }\\
6 & Radio Imaging & 44.6\% & \hspace{-10mm}\makecell[l]{$\dots$\\ }\\
7 & X-shaped radio sources  & 37.3\% & \hspace{-10mm}\makecell[l]{e.g., \citet[][]{cheung2007,proctor2011,roberts2018} \\ \citet{yang2019}\\ }\\
8 & Jet Precession & 2.6\% & \hspace{-10mm}\makecell[l]{e.g., \citet[][]{krause2019,britzen2019,britzen2019,britzen2015,roland2013}\\ }\\
9 & mas-scale dual radio sources & 0.3\% & \hspace{-10mm}\makecell[l]{ \citet[][]{maness2004,rodriguez2006,xiangliu2014}\\ }\\
10 & Double Radio Jets & 0.3\% & \hspace{-10mm}\makecell[l]{\citet[][]{xiangliu2014,qian2019}\\ }\\
11 & Radio Periodicity & 0.7\% & \hspace{-10mm}\makecell[l]{e.g., \citet[][]{ciaramella2004,rieger2007,romero2003}\\ }\\
12 & Radio Quasi-Periodicity & 0.4\% & \hspace{-10mm}\makecell[l]{e.g., \citet[][]{ciaramella2004,rieger2007,romero2003}\\ }\\
13 & Optical Imaging & 17.5\% & \hspace{-10mm}\makecell[l]{$\dots$\\ }\\
14 & Optical Periodicities & 13.1\% & \hspace{-10mm}\makecell[l]{\citep[e.g.,][]{graham2015a,graham2015b,charisi2016,sillanpaa1988} \\ \citep{webb1990,rieger2007,bon2012,oknyanskij2016}\\ \citep{romero2003,liu2015}\\ }\\
15 & Optical Quasi-Periodicities & 0.7\% & \hspace{-10mm}\makecell[l]{\citep[e.g.,][]{sandrinelli2016a,sandrinelli2016b,zhang2014,yang2020}\\ }\\
16 & Gamma-Ray periodicities & 0.2\% & \hspace{-10mm}\makecell[l]{\citep[][]{ackermann2015,pzhang2017a,pzhang2017b,pzhang2017c,yang2020} \\ \citep{zhou2018}\\ }\\
17 & Gamma-Ray Quasi-Periodicities & 0.1\% & \hspace{-10mm}\makecell[l]{\citep[][]{ackermann2015,pzhang2017a,pzhang2017b,pzhang2017c,yang2020} \\ \citep{zhou2018}\\ }\\
18 & Near-IR Periodicities & 0.1\% & \hspace{-10mm}\makecell[l]{\citep[e.g.,][]{sandrinelli2016a}\\ }\\
19 & Near-IR Quasi-Periodicities & 0.4\% & \hspace{-10mm}\makecell[l]{\citep[e.g.,][]{sandrinelli2016a}\\ }\\
20 & X-ray Periodicities & 0.3\% & \hspace{-10mm}\makecell[l]{\citep[e.g.,][]{rani2009,serafinelli2020}\\ }\\
21 & X-ray Quasi-Periodicities & 0.1\% & \hspace{-10mm}\makecell[l]{\citep[e.g.,][]{rani2009,serafinelli2020}\\ }\\
\noalign{\smallskip}
\hline
\end{tabular}
\end{center}
%\tablecomments{}
\label{tab:citsec421}
\end{table*}

\subsubsection{Binary AGN Selection Technique Breakdown: Optical Spectroscopy}
\label{sec:binarybreakoptspec}

\begin{figure*}
    \centering
    \includegraphics[width=1.0\linewidth]{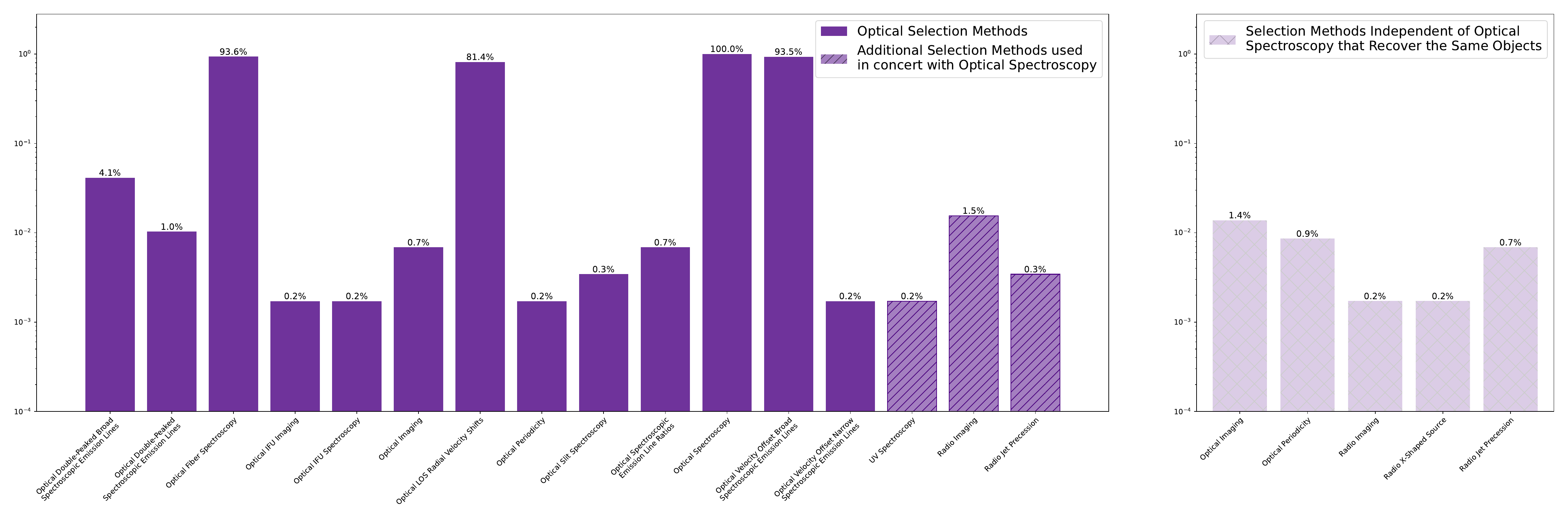}
    \caption{Binary AGN candidates that required optical spectroscopy for selection.  A breakdown of binary AGN selection strategies used in concert with (left) or independently of (right) optical spectroscopy}
    \label{fig:selection_binary_optical_spec}
\end{figure*}

In Figure~\ref{fig:selection_binary_optical_spec} we show the distribution of selection methods for objects that required optical spectroscopy as a component of the selection function (582 objects). These systems primary rely upon optical fiber spectroscopy (93.6\%, Table~\ref{tab:citsec422}, Row 1) and the observation of velocity offset broad spectroscopic emission lines ($\sim90$\%, Table~\ref{tab:citsec422}, Row 2) and radial velocity shifts of broad lines (83\%, Table~\ref{tab:citsec422}, Row 3) for their selection. Smaller percentages of objects also used techniques including slit spectroscopy (0.3\%, Table~\ref{tab:citsec422}, Row 4), IFU spectroscopy and imaging (0.2\%, Table~\ref{tab:citsec422}, Row 5), and selection based on the presence of double-peaked broad emission lines (4.1\%, Table~\ref{tab:citsec422}, Row 6), double-peaked narrow emission lines (1.0\%, Table~\ref{tab:citsec422}, Row 7), velocity offset narrow emission lines (0.2\%, Table~\ref{tab:citsec422}, Row 8), and optical spectroscopic emission line ratios (0.7\%, Table~\ref{tab:citsec422}, Row 9). Optical imaging has been used in concert with optical spectroscopy for 0.7\% of objects (Table~\ref{tab:citsec422}, Row 10), and 0.2\% of objects were selected based on optical periodicity in optical imaging (Table~\ref{tab:citsec422}, Row 11). Finally, radio imaging played a role in the selection of 1.5\% of these objects (Table~\ref{tab:citsec422}, Row 12).

A few percent of objects in this subsample were selected independently of optical spectroscopy. Optical imaging was used to independently select 1.4\% of objects \citep{bon2016,graham2015b,oknyanskij2016,li2019,li2016}, and 0.9\% of objects were selected based on optical periodicities \citep{oknyanskij2016,li2016,li2019,graham2015b}. Radio imaging \citep[0.2\%][]{blundell2001,vol2010,krause2019,proctor2011} was also used independently of optical spectroscopy to select objects based on X-shaped radio sources \citep[0.2\%,][]{proctor2011} or potential radio jet precession \citep[0.7\%,][]{krause2019,blundell2001}.

\begin{table*}[t!]
\caption{Citation Table for Section~\ref{sec:binarybreakoptspec}}
\begin{center}
\hspace{-2.2cm}
\begin{tabular}{cccc}
\hline
\hline
\noalign{\smallskip}
Row \# & Method & \% & Citations\\
\noalign{\smallskip}
\hline
\noalign{\smallskip}
%Binary Quasar & SDSSJ0812+2520A & 105.8 & ICRF 
1 & Optical Fiber Spectroscopy & 93.6\% & \hspace{-10mm}\makecell[l]{e.g., \citet[][]{xliu2014,shen2013,ju2013,wang2017}\\ \citet{tsalmantza2011,barrows2011,eracleous2012} \\ \citet{zheng2016,boroson2009,komossa2003}\\ } \\
2 & \hspace{-10mm}\makecell[l]{Velocity offset broad\\ spectroscopic emission lines} & $\sim90$\% & \hspace{-10mm}\makecell[l]{e.g., \citet[][]{eracleous2012,tsalmantza2011,ju2013}\\ \citet{xliu2014,wang2017,decarli2010b} \\ \citet{boroson2009,komossa2008,barrows2011}\\ }\\
3 & \hspace{-10mm}\makecell[l]{LOS Radial Velocity\\ Shifts of Broad Lines} & 83\% & \hspace{-10mm}\makecell[l]{\citet[][]{xliu2014,shen2013,ju2013,wang2017}\\ \citet{bon2012,zheng1991,eracleous2012} \\ \citet{runnoe2015,runnoe2017a,gaskell1996binaryqso}} \\
4 & Optical Slit Spectroscopy & 0.3\% & \hspace{-10mm}\makecell[l]{e.g., \citet[][]{puchnarewicz1996}}\\
5 & \hspace{-10mm}\makecell[l]{Optical IFU Spectroscopy\\ and Imaging} & 0.2\% & \hspace{-10mm}\makecell[l]{e.g., \citet[][]{menezes2014}}\\
6 & \hspace{-10mm}\makecell[l]{Optical Double-Peaked\\ Broad Spectroscopic\\ Emission Lines} & 4.1\% & \hspace{-10mm}\makecell[l]{e.g., \citet[][]{eracleous1994,lewis2010,gezari2007} \\ \citet{zheng1991,halpern1988,halpern1992} \\ \citet{peterson1987,barrows2011}\\ }\\
7 & \hspace{-10mm}\makecell[l]{Optical Double-Peaked\\ Narrow Emission Lines} & 1.0\% & \hspace{-10mm}\makecell[l]{\citet[][]{kharb2019,nandi2017}\\ }\\
8 & \hspace{-10mm}\makecell[l]{Optical Velocity Offset\\ Narrow Emission Lines}  & 0.2\% & \hspace{-10mm}\makecell[l]{\citet[][]{kharb2020}\\ }\\
9 & \hspace{-10mm}\makecell[l]{Optical Spectroscopic\\ Emission Line Ratios} & 0.7\% & \hspace{-10mm}\makecell[l]{\citet[][]{nandi2017,kharb2020}\\ }\\
10 & Optical Imaging & 0.7\% & \hspace{-10mm}\makecell[l]{\citet{zheng2016,menezes2014,bon2012,li2019}\\ }\\
11 & Optical Periodicity & 0.2\% & \hspace{-10mm}\makecell[l]{\citet[e.g.,][]{zheng2016,bon2012,li2019}\\ }\\
12 & Radio Imaging & 0.7\% & \hspace{-10mm}\makecell[l]{\citet{pesce2018,kharb2014,zheng1991} \\ \citet{eracleous1994,kharb2014,gezari2007}\\ }\\
\noalign{\smallskip}
\hline
\end{tabular}
\end{center}
%\tablecomments{}
\label{tab:citsec422}
\end{table*}

\subsubsection{Binary AGN Selection Technique Breakdown: Optical Imaging}
\label{sec:binarybreakoptimaging}

\begin{figure*}
    \centering
    \includegraphics[width=1.0\linewidth]{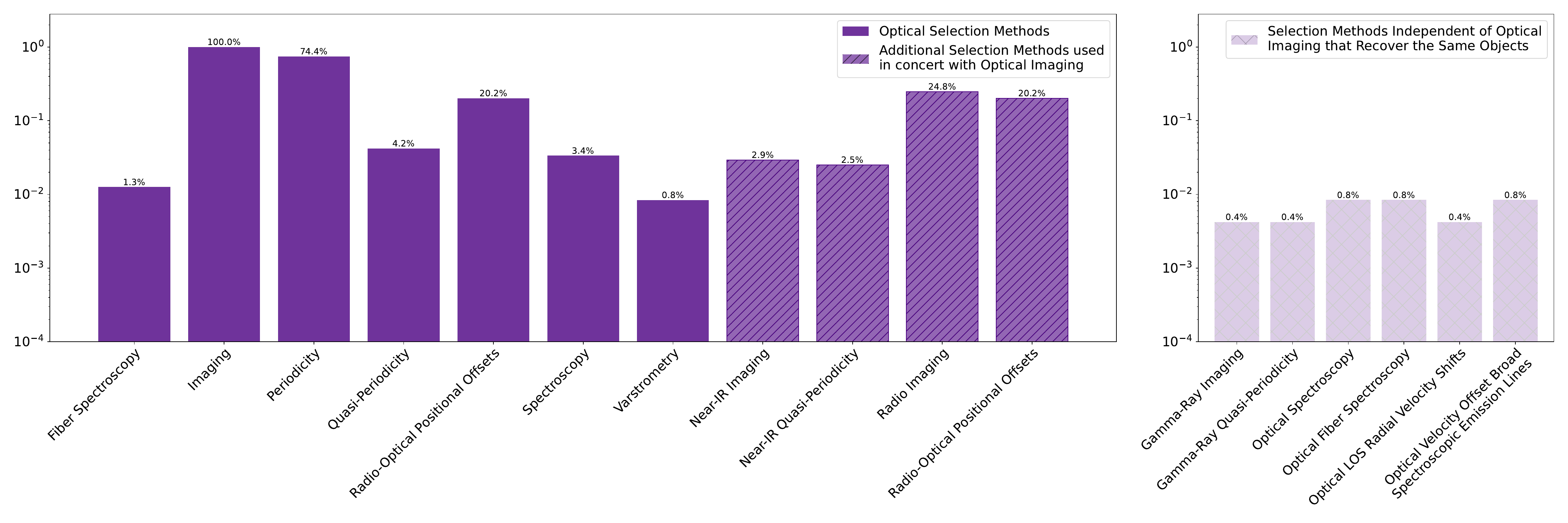}
    \caption{Binary AGN candidates that required optical imaging for selection. A breakdown of binary AGN selection strategies used in concert with (left) or independently of (right) optical imaging.}
    \label{fig:selection_binary_optical_imaging}
\end{figure*}

In Figure~\ref{fig:selection_binary_optical_imaging} we show the distribution of selection methods for systems that required optical imaging as a component of the selection function (238 objects). Most of these systems (74.4\%) were selected based on the presence of optical periodicities reported in optical imaging \citep[e.g.,][]{graham2015b,charisi2015,liu2015,tliu2019,zheng2016,sillanpaa1988,rieger2007,bon2012,webb1990}, a small fraction relied upon optical quasi-periodicities \citep[4.2\%, e.g.,][]{yang2020,sandrinelli2016a}, and a few objects were selected based on optical varstrometry \citep{popovic2012}. Radio imaging was used in conjunction with optical imaging for 24.8\% of binary candidates  \citep{orosz2013,rieger2007,romero2003,qian2013,qian2007,roland2013,ciaramella2004,caproni2004a}, and was predominantly used to identify radio-optical positional offsets \citep[20.2\%,][note, these offsets could be due to dual AGNs, binary AGNs, or other phenomenon such as lenses]{orosz2013}. A small number of articles used optical spectroscopy \citep[3.4\%, e.g.,][ and specifically fiber spectroscopy for 1.3\% of objects]{bon2012,peterson1987,menezes2014} to search for radial velocity shifts \citep{bon2012}, or double-peaked/multiple broad emission line systems \citep{peterson1987,boroson2009}. Finally, near-IR imaging and quasi-periodicities were coupled with optical imaging for 2.5\% of these objects \citep[e.g.,][]{sandrinelli2016a,volvach2016b,sandrinelli2014a}.

Small fractions of these systems were also independently identified without the use of optical imaging. Gamma-Ray imaging was used to independently identify 0.4\% of these objects based on reported Gamma-Ray quasi-periodicities \citep{ackermann2015,sandrinelli2016a,sandrinelli2016b,pzhang2017a,pzhang2017b,pzhang2017c,sandrinelli2018b}.  Optical spectroscopy -- and specifically fiber spectroscopy \citep[0.8\%,][]{xliu2014,boroson2009} -- was used to search for velocity offset broad spectroscopic emission lines and potential radial velocity shifts of these offset broad lines, which could arise due to SMBH binary orbital motion.

\subsubsection{Binary AGN Selection Technique Breakdown: Radio Imaging}
\label{sec:binbreakradioimg}

\begin{figure*}
    \centering
    \includegraphics[width=1.0\linewidth]{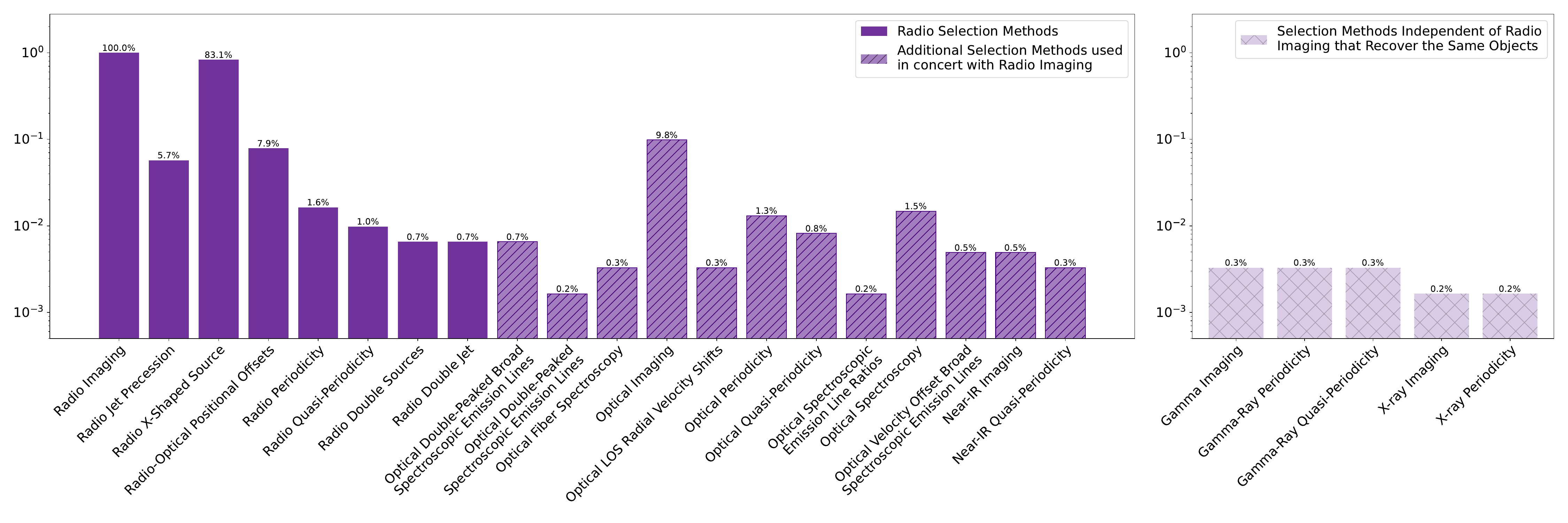}
    \caption{Binary AGN candidates that required radio imaging for selection. A breakdown of binary AGN selection strategies used in concert with (left) or independently of (right) radio imaging.}
    \label{fig:selection_binary_radio_imaging}
\end{figure*}

In Figure~\ref{fig:selection_binary_radio_imaging} we show the distribution of selection methods for binary AGN candidates that required radio imaging as a component of the selection function (610 objects). The majority of these candidates were selected based upon the presence of X-shaped radio sources \citep[e.g.,][]{cheung2007,proctor2011,yang2020}. The formation mechanism for X-shaped radio sources is still heavily debated \citep[see][and references therein]{yang2019}, but one proposed formation route is the presence of an unresolved binary AGN, where each AGN is emitting its own distinct radio jets \citep{lal2005,lal2007}. Given that we have observed bona fide dual AGNs at kpc separations emitting dual radio jets \citep{owen1985,klamer2004} and there is one confirmed binary radio AGN \citep{rodriguez2006} on parsec-scale separations (within which one radio AGN is emitting a jet structure), unresolved binary radio jet AGNs represent a reasonable formation mechanism for these structures. Note, another possible formation mechanism is the rapid reorientation of the radio jet axis following the merger of a binary SMBH (hence, coinciding with a recoiling AGN, see Section~\ref{sec:app_recoil_xshaped_select}). Small fractions of radio imaging-based searches selected objects based on potential jet precession \citep[5.7\%,][]{krause2019,britzen2015,britzen2017,britzen2019,caproni2017,qian2017,roland2013,qian2013,caproni2004a,blundell2001,conway1995,caproni2013}, dual radio sources and/or double jets \citep[0.7\% each,][]{maness2004,rodriguez2006,xiangliu2014,qian2019}, and radio periodicities and quasi-periodicities \citep[1.6\% and 1\%, respectively, e.g.,][]{ciaramella2004,rieger2007,bhatta2018,vol2010,romero2003,li2006,liu2006,vol2013,volvach2007,volvach2016a,volvach2016b,vol2014,vol2010}. Optical imaging has been paired with radio imaging for 9.8\% of objects in this subsample, and was most often used to identify radio-optical positional offsets \citep[7.9\% of objects,][]{orosz2013}. In a minority of objects, optical spectroscopy (1.5\% of objects) was used in concert with radio imaging; these spectroscopic techniques included selection using fiber spectroscopy \citep[0.3\%,][]{zhou2004,pesce2018}, double-peaked broad spectroscopic emission lines \citep[0.7\%,][]{gezari2007,jliu2016,doan2020,zheng1991,gaskell1996binaryqso,shapovalova2001,eracleous1994} and narrow emission lines \citep[0.2\%,][but this turned out not to be a binary AGN]{zhou2004,jaiswal2019}, velocity offset broad emission lines \citep[0.5\%,][]{du2018}, and spectroscopic emission line ratios \citep[0.2\%,][]{zhou2004}. Note, several works that selected binary candidates based on double-peaked broad emission lines ultimately disfavored the binary hypothesis \citep[e.g.][]{eracleous1994,eracleous1997,shapovalova2001,gezari2007,jliu2016,doan2020}.
Only small fractions of these objects were selected independently of radio imaging, using Gamma-Ray imaging and (quasi-)periodicities (0.3\% in each case) and X-ray imaging and periodicities (0.3\%).

\subsubsection{Binary AGN Selection Technique Breakdown: Miscellaneous Techniques}
\label{sec:binarybreakmisc}

A small number of other techniques have been used, including near-IR imaging \citep[e.g.][]{volvach2016b,sandrinelli2016a,sandrinelli2014a} X-ray imaging \citep[][]{fliu2014,rani2009,serafinelli2020,li2015} to to search for and select binary AGN candidates based on periodicities or quasi-periodicities in the X-ray and near-IR \citep[e.g.,][]{rani2009,serafinelli2020,li2015,volvach2016b,sandrinelli2016a,sandrinelli2014a}. These small samples of objects often overlap with the optical imaging and Gamma-Ray imaging selection methods.% discussed in Sections~\ref{sec:binarybreakoptimaging} and \ref{sec:binarybreakgammaray}. 

\subsection{Recoiling AGN Selection Methods}
\label{sec:recoilselectmethods}

\subsubsection{Recoiling AGN Aggregate Selection Method Statistics}
\label{sec:aggselectmethodrecoils}

\begin{figure*}[t]
    \centering
    \includegraphics[width=1.0\linewidth]{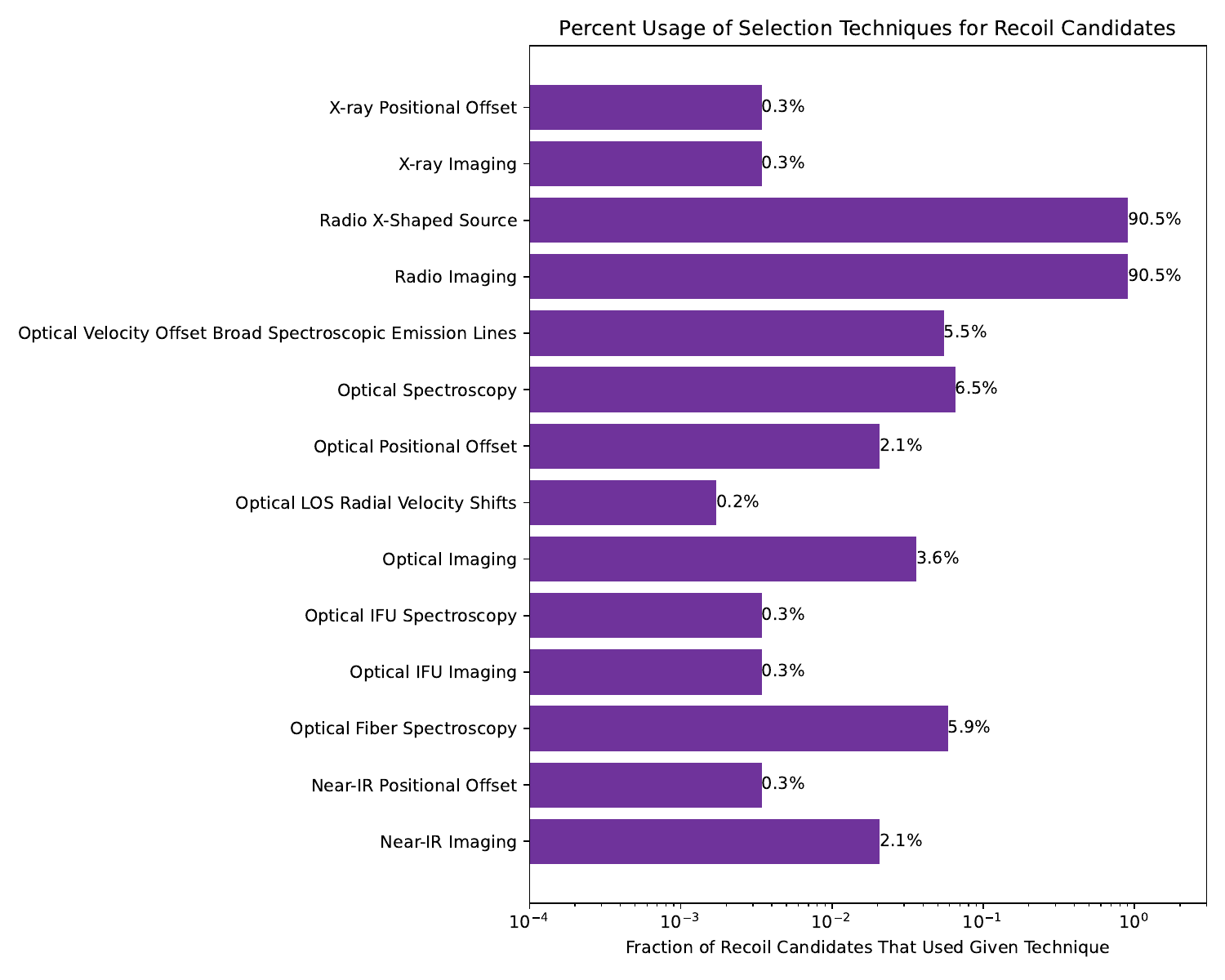}
    \caption{A breakdown of the selection methods used to identify recoiling AGNs and candidates in the literature. This chart shows the aggregate percent usage of all recoil AGN selection techniques (normalized by the number of recoil candidates in the catalog); note, these selection methods are not all necessarily mutually exclusive (for example: selection based on X-shaped radio sources necessarily requires radio imaging).}
    \label{fig:aggrecoilAGNselmethod}
\end{figure*}

The aggregate percent frequency with which specific selection methodologies were used for recoiling AGN and SMBH systems is shown in Figure~\ref{fig:aggrecoilAGNselmethod} in an identical fashion to how this breakdown was shown for dual AGNs and binary AGNs in Figures~\ref{fig:aggdualAGNselmethod} and \ref{fig:aggbinaryAGNselmethod} and as described in Section~\ref{sec:aggselmethodduals}. The overwhelming majority (90.5\%) of recoil candidates are selected based upon the presence of X-shaped radio sources (or candidate sources) \citep[e.g.,][]{cheung2007,proctor2011,yang2020} and hence require radio imaging for selection (90.5\%); the origin of X-shaped radio sources is still debated in the literature, but possible formation mechanisms related to multi-AGN systems include (1) rapid jet reorientation following the coalescence of a SMBH binary \citep[and hence, association with a potentially recoiling AGN, e.g.,][]{merritt2002,dennett2002} and (2) unresolved, binary radio AGNs emitting distinct double radio jets \citep[e.g.,][]{lal2005,lal2007}.

Optical spectroscopy was used to select 6.5\% of known recoiling AGN candidates, and most cases used optical SDSS fiber spectra \citep[5.9\% of all candidates,][]{komossa2008,jonker2010,steinhardt2012,kim2016,dkim2017,kim2018}, with a strong emphasis on the search for broad spectroscopic emission lines significantly offset in velocity from the host galaxy \citep[5.5\% of all candidates,][]{komossa2008,steinhardt2012,kim2016}. Optical imaging (3.6\%), X-ray imaging (0.3\%), and near-IR imaging (0.3\%) have all been used to search for and/or identify optical, X-ray, or near-IR positional displacements (2.1\%, 0.3\%, and 0.3\%, respectively) between the AGNs and their host galaxy nuclei that might be indicative of kicked AGNs. Optical imaging was also used to examine host galaxy morphologies \citep[e.g,][]{sun2016}. No recoiling AGNs have so far been confirmed.

\subsubsection{Recoiling AGN Selection Technique Breakdown: Radio Imaging}
\label{sec:recoilbreakradioimg}
%\label{sec:selmethodcomparisons}

\begin{figure}
    \centering
    \includegraphics[width=1.0\linewidth]{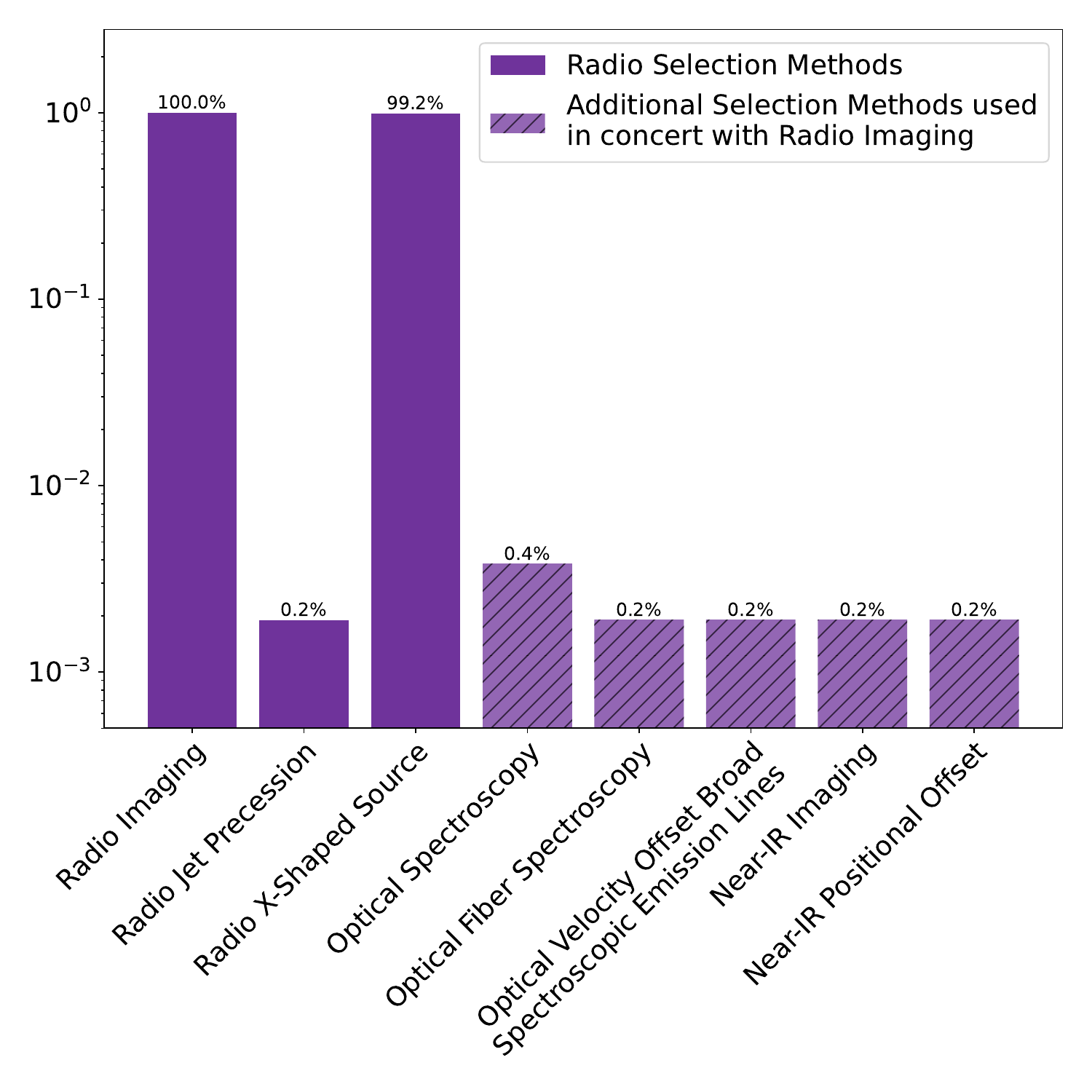}
    \caption{Recoiling AGN candidates that required radio imaging for selection. A breakdown of recoiling AGN selection strategies used in conjunction with radio imaging. None of these objects were selected independently of radio imaging using other techniques.}
    \label{fig:selection_recoil_radio_imaging}
\end{figure}

In Figure~\ref{fig:selection_recoil_radio_imaging} we show the distribution of selection methods for objects that required radio imaging for selection (527 objects). Among these objects, almost all candidates (99.2\%, or 523 total objects) were selected based upon the presence of X-shaped radio sources \citep[e.g.,][]{merritt2002,dennett2002,cheung2007,proctor2011,roberts2018,saripalli2018,yang2019,joshi2019}. Only one candidate (0.4\%) was selected/followed-up based on reported jet precession \citep[][]{robinson2010}. Among candidates selected using radio imaging, very small fractions of these objects also used other techniques in combination with the radio imaging, including optical imaging (0.2\%, or one object), optical spectroscopy \citep[0.4\%, or two objects,][]{robinson2010,pesce2018}, and near-IR imaging \citep[0.2\%, or one object,][]{markakis2015}. None of these objects were selected as recoil candidates via other wavebands and techniques independently of radio imaging.

\subsubsection{Recoiling AGN Selection Technique Breakdown: Optical Spectroscopy}
\label{sec:recoilbreakoptspec}

\begin{figure}
    \centering
    \includegraphics[width=1.0\linewidth]{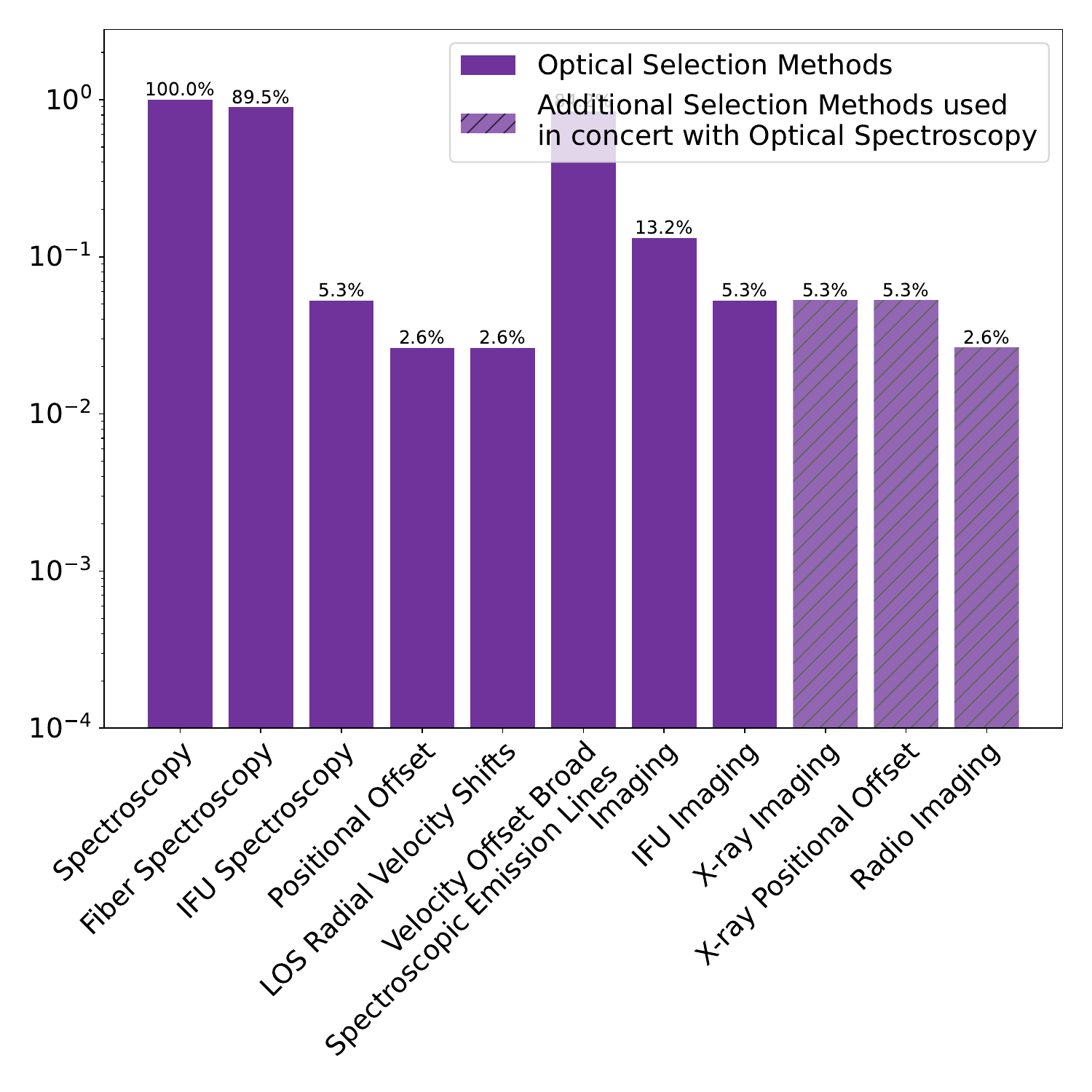}
    \caption{Recoiling AGN candidates that required optical spectroscopy for selection. A breakdown of recoiling AGN selection strategies used in concert with optical spectroscopy. None of these objects were selected independently of optical spectroscopy using other techniques.}
    \label{fig:selection_recoil_optical_spec}
\end{figure}

In Figure~\ref{fig:selection_recoil_optical_spec} we show the distribution of selection methods for objects that required optical spectroscopy as a component of the selection function (38 total). The vast majority of these objects were selected using optical fiber spectroscopy (89.5\%), with a strong emphasis on searches for velocity offset broad spectroscopic emission lines \citep[84.2\%, e.g.,][]{komossa2008,steinhardt2012,kim2016,dkim2017,kim2018}, a potential kinematic imprint of SMBHs. A small number of objects were selected via IFU spectroscopy and imaging \citep[5.3\%,][]{menezes2014,menezes2016}, optical positional offsets \citep[2.6\%, some of which were identified within the IFU imaging][]{menezes2014,menezes2016,sun2016}, or as a result of line-of-sight radial velocity shifts of broad emission lines over time \citep[2.6\%, or one object,][]{dkim2017}. Similarly, a small fraction of these objects also used complementary X-ray imaging \citep[5.3\%, to detect optical-X-ray positional offsets,][]{jonker2010} and radio imaging \citep[2.6\%,][]{pesce2018} in concert with the optical spectroscopy. None of these objects were selected as recoil candidates via other techniques completely independently of optical spectroscopy.

\subsubsection{Recoiling AGN Selection Technique Breakdown: Optical Imaging}
\label{sec:recoilbreakoptimg}

\begin{figure}
    \centering
    \includegraphics[width=1.0\linewidth]{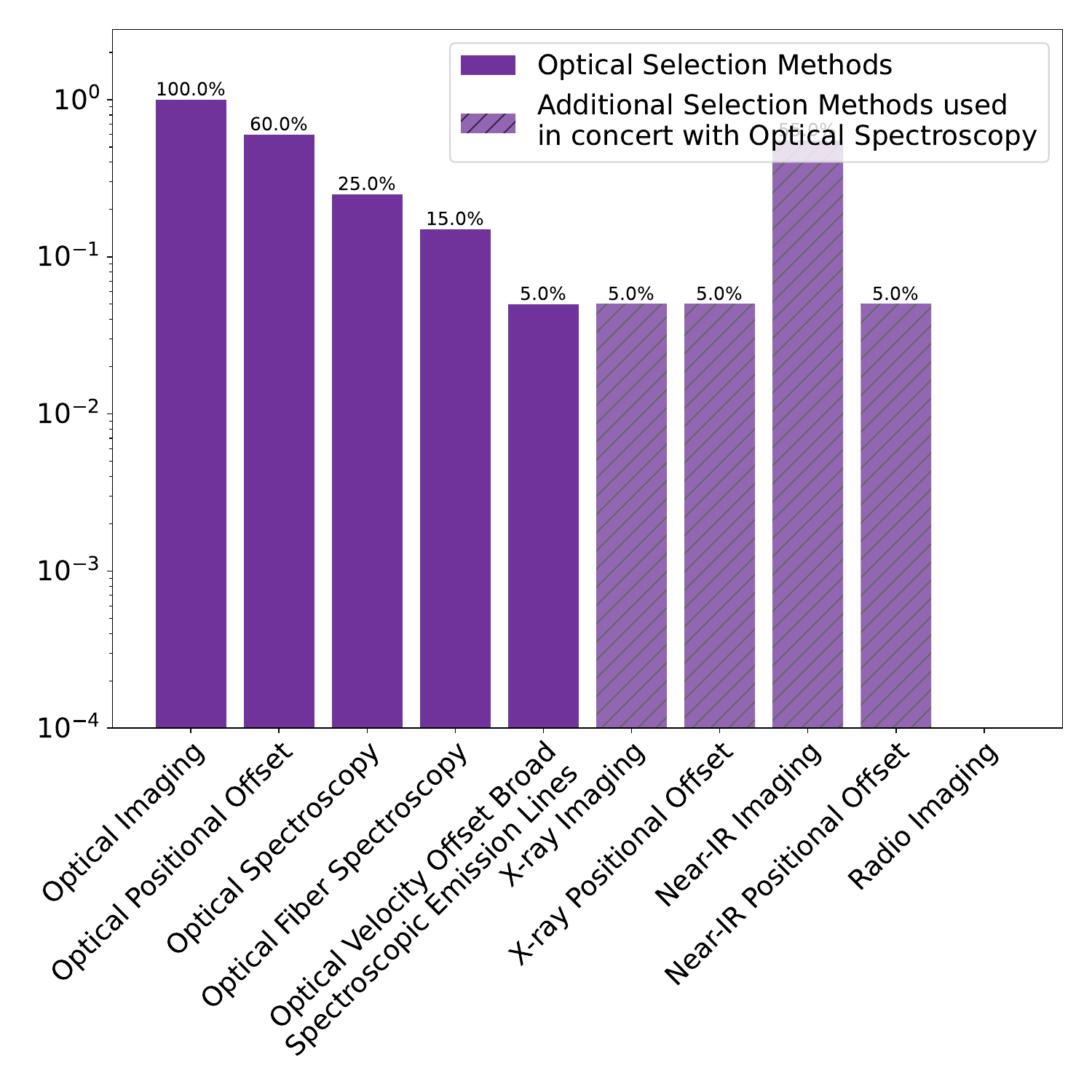}
    \caption{Recoiling AGN candidates that required optical imaging for selection. A breakdown of recoiling AGN selection strategies used in concert with optical imaging. None of these objects were selected independently of optical imaging using other techniques.}
    \label{fig:selection_recoil_optical_imaging}
\end{figure}

In Figure~\ref{fig:selection_recoil_optical_imaging} we show the distribution of selection methods for objects that required optical imaging for selection (22 objects). Roughly half of these objects were selected based on optical positional offsets \citep[54.5\%, e.g.,][]{batcheldor2010,lena2014}. About a third (31.8\%) of the objects also used complementary optical spectroscopy \citep[18.2\% used fiber spectroscopy, e.g.,][]{sun2016,koss2014,jonker2010} as a part of the selection strategy, and in a small number of cases to search for velocity offset broad emission lines \citep[9.1\%, e.g.,][]{sun2016}. X-ray imaging was the only non-optical technique used in concert with optical imaging, and was used in one case (4.5\%) to select a recoil candidate based on optical-X-ray positional offsets \citep[][but the recoil hypothesis was eventually ruled out for this object]{jonker2010,heida2015}. None of these objects were selected as recoil candidates via techniques completely independent of optical imaging.

\subsubsection{Recoiling AGN Selection Technique Breakdown: Near-IR and X-ray Imaging}
\label{sec:recoilbreaknearirxrayimg}
%\label{sec:selmethodcomparisons}

Only a small number of recoil candidates required either near-IR ($\sim 6-8$ objects) or X-ray imaging ($2$ objects) as a part of the selection method, so we have omitted figures that illustrate the selection method breakdowns for these techniques, but we do briefly discuss these techniques here. The two candidates that required the use of X-ray imaging were selected based on X-ray-to-optical positional offsets and therefore both required optical imaging as well \citep{jonker2010,dkim2017}. These objects also depended upon the use of optical fiber spectroscopy as a part of the selection strategy, and one of them was further selected based upon the presence of a velocity offset broad optical spectroscopic emission line \citep{dkim2017}. Neither of these objects were selected completely independently of X-ray imaging. %\textcolor{violet}

\subsection{Multi-AGN Confirmation Methods}
\label{sec:dualsconfirmmethods}
We begin this subsection by discussing the methods used to select specifically those dual AGNs which have been confirmed in the literature, and then we will discuss the various confirmation methods used to confirm said dual AGNs. Given the fact that only a single binary AGN has been confirmed in the literature and no recoils have been confirmed, we focus primarily on dual AGNs in this section. Note again: this section is an overview of the existing literature and carries with it all selection biases inherent to the literature. This analysis is more a reflection of the efforts of the community in the field of multi-AGNs than the properties or selectability of dual AGNs. % and N-tuple AGNs

\begin{figure*}[t]
    \centering
    \includegraphics[width=1.0\linewidth]{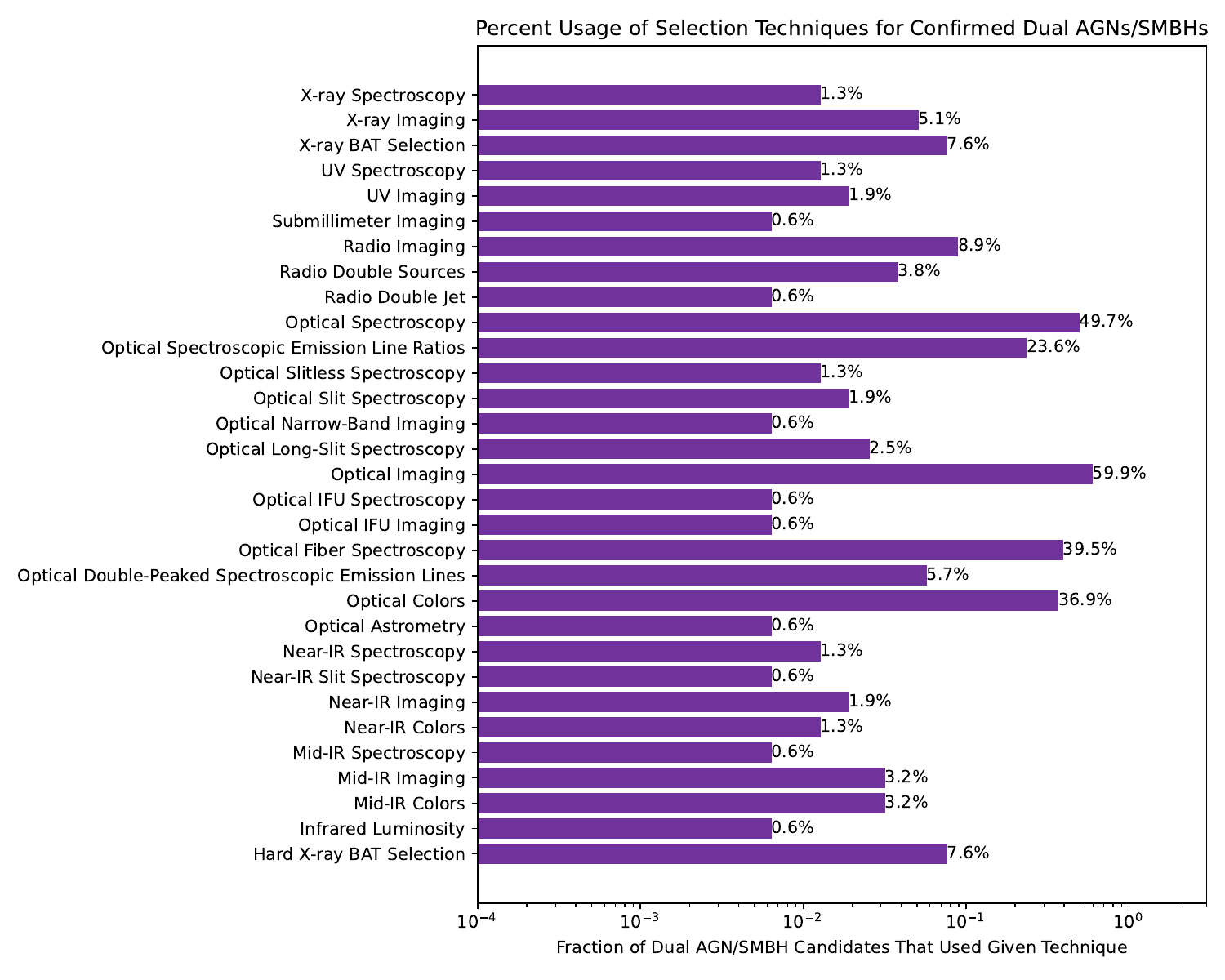}
    \caption{A breakdown of the selection methods used to identify multi-AGNs in the literature that are listed in this catalog as confirmed (confidence rank $=1$). This chart shows the aggregate percent usage of all multi-AGN selection techniques (normalized by the number of confirmed multi-AGNs in the catalog); note, these selection methods are not all necessarily mutually exclusive. %(for example: selection based on X-shaped radio sources necessarily requires radio imaging).
    }
    \label{fig:dual_confirm_selectmethods}
\end{figure*}

Figure~\ref{fig:dual_confirm_selectmethods} illustrates the methods used to select the confirmed dual AGNs from the literature (confidence rank $=1$). Optical spectroscopy was used during the selection process for about half (49.7\%) of all confirmed dual AGNs \citep[e.g.,][]{liu2011b,barrows2016,hennawi2006,more2016,inada2008,eftekharzadeh2017,bianchi2008,huang2014,ellison2017,stauffer1982,keel1985,bothun1989,piconcelli2010,smith2010,tadhunter2012,wang2009,liu2010b,inada2010,inada2012}; 39.5\% of confirmed duals were selected based specifically on fiber spectroscopy, while small portions relied upon other spectroscopic techniques for selection (slit: 1.9\%, long-slit: 2.5\%, slitless: 1.3\%, IFU spectroscopy: 0.6\%). 23.6\% of confirmed dual AGNs relied upon spectroscopic emission line ratios for selection \citep{liu2011b,wang2009,liu2010b,smith2010,huang2014,keel1985,keel2019,piconcelli2010}, while only 5.7\% of confirmed dual AGNs were selected via double-peaked spectroscopic emission lines \citep{wang2009,liu2010b,smith2010,ge2012}; this latter point is quite striking, given that double-peaked lines account for an overwhelming number of selected dual AGN candidates, but it is entirely in line with studies showing that less than 15\% of double-peaked emission line profiles are driven by dual AGNs \citep[e.g.,][]{fu2012,mullersanchez2015}. Optical spectroscopic techniques are the most frequently employed selection strategies for systems that later turn out to be confirmed dual AGNs, but this is simply due to the availability of archival optical spectra from surveys like the SDSS; this should not be construed as a higher success rate for finding bona fide dual AGNs when using optical spectroscopy, because this technique has also provided far larger samples of candidate (but not yet confirmed) dual AGNs \citep[e.g.,][]{liu2011b,wang2009,smith2010,liu2010b,ge2012,shi2014}, many of which turn out not to be dual AGNs \citep[e.g.,][]{rosario2010,rosario2011,fu2011b,fu2012,mullersanchez2015,mullersanchez2016}.  Optical imaging (59.9\%) and colors (36.9\%) are also commonly employed to select  dual AGNs that are later confirmed \citep[e.g.,][]{fu2015a,fu2018,satyapal2017,hennawi2006,eftekharzadeh2017,more2016,weedman1982,hagen1996,myers2007,hennawi2010,findlay2018,inada2008,djorgovski1984,huang2014,fu2018,tadhunter2012,koss2012}, where the frequent use of optical imaging naturally arises due to a need for morphological classification and color verification. Radio imaging \citep[8.9\%][]{fu2015a,owen1985,klamer2004,hewitt1987,mu1998,tadhunter2012,heckman1986,evans2008} and hard X-ray BAT preselection \citep[7.6\%][]{koss2011,koss2012} are used for the selection of smaller (but significant) fractions of dual AGNs that are later confirmed, where radio imaging includes selection based on the reported presence of double radio sources and double radio jets \citep[e.g.,][]{owen1985,klamer2004,fu2015a}, while BAT AGN preselection relies also upon the use of follow-up optical spectroscopy and soft X-ray imaging for selection. To a much lesser extent, mid-IR imaging/colors \citep[3.2\%,][]{satyapal2017,ellison2017,pfeifle2019a,schechter2017}, UV imaging and spectroscopy \cite[1.9\% and 1.3\%][]{hawkins1997a,junkkarinen2001,huang2014}, and X-ray imaging \cite[5.1\%][]{mu2001,komossa2003,ballo2004,guainazzi2005,bianchi2008,wong2008,eckert2017,piconcelli2010,iwasawa2011a} was used to select dual AGNs that were later confirmed. Other notable selection methods include near-IR imaging and colors \citep[1.9\% and 1.3\%, e.g.,][]{imanishi2014,imanishi2020} and infrared luminosity \cite[0.6\%][]{iwasawa2011a,iwasawa2011b,torres2018}. 

\begin{figure*}[t]
    \centering
    \includegraphics[width=1.0\linewidth]{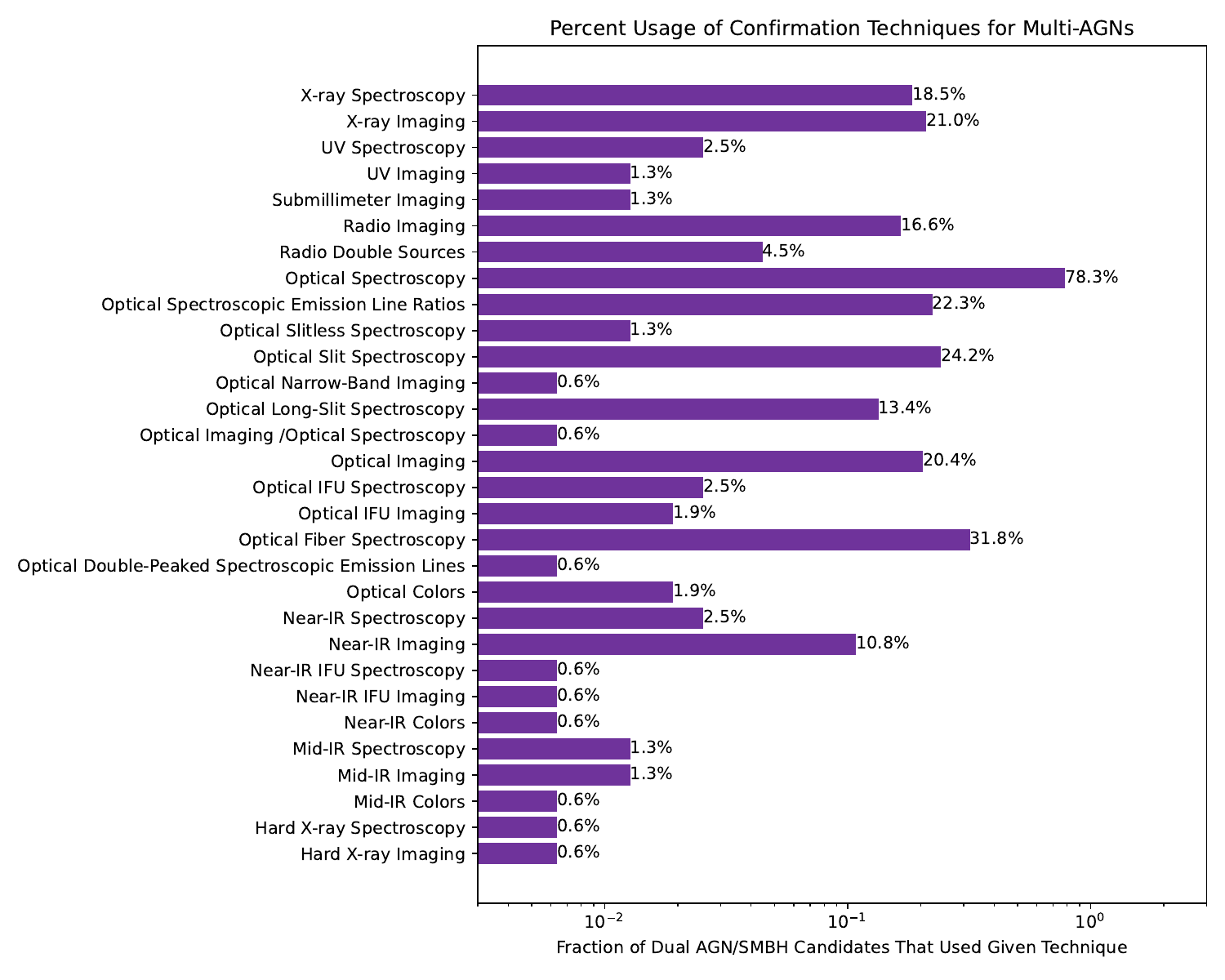}
    \caption{A breakdown of the confirmation methods used to identify dual AGNs in the literature. This chart shows the aggregate percent usage of all dual AGN confirmation techniques (normalized by the number of confirmed dual AGNs in the catalog); note, these confirmation methods are not all necessarily mutually exclusive. %(for example: selection based on X-shaped radio sources necessarily requires radio imaging).
    }
    \label{fig:dualconfirmmethods}
\end{figure*}

Figure~\ref{fig:dualconfirmmethods} illustrates the aggregate percent use of methods used for dual AGN confirmation, where the techniques shown are not all necessarily mutually exclusive; as in Sections~\ref{sec:aggselmethodduals}-\ref{sec:dualbreakradioimaging}, some techniques rely upon, or are derived from, other methods. Optical spectroscopy plays a common role in confirming dual AGNs (78.3\% of confirmed duals). Breaking this down further, optical fiber \citep[e.g.,][]{liu2011b,derosa2018,koss2012,hennawi2006,green2011,eftekharzadeh2017,inada2008,klamer2004}, slit \citep[e.g.,][]{hennawi2006,hennawi2010,eftekharzadeh2017,green2004,brotherton1999,morgan2000,popovic2010,hewitt1987,junkkarinen2001,impey2002}, and long-slit spectroscopy \citep[e.g.,][]{myers2008,green2011,nzhang2017,keel2019,tadhunter2012} are each used for confirmation in 31.8\%, 24.2\%, and 13.4\% of cases, while slitless (1.3\%) and IFU (2.5\%) spectroscopy \citep[e.g.,][]{ellison2017} are less common confirmation techniques. Optical spectroscopic emission line ratios are used in the confirmation of roughly a fifth (22.3\%) of all confirmed systems \citep[e.g.,][]{liu2011b,derosa2018,koss2012,huang2014,keel2019,ellison2017}. Optical imaging is used in about a fifth (20.4\%) of cases, too \citep[e.g.,][]{koss2011,tadhunter2012,evans2008,ellison2017}, where IFU imaging specifically was used in 1.9\% of confirmed systems. X-ray imaging was the next most prevalent technique used to confirm dual AGNs, with X-ray imaging and spectroscopy playing a role in the confirmation for 21.0\% and 18.5\% of confirmed systems, respectively \citep[][]{koss2012,liu2013,hou2019,hou2020,brassington2007,green2010,iwasawa2011b,wliu2019,guainazzi2005,turner2001,bianchi2008,wong2008,eckert2017,piconcelli2010,torres2018,evans2008,comerford2015}. Radio imaging follows these techniques as the next most common confirmation method \citep[16.6\% of all systems][]{fu2015a,fu2015b,owen1985,hewitt1987,mu1998,klamer2004,evans2008,tadhunter2012,mullersanchez2015,fu2011b,rubinur2019,green2004,brotherton1999,djorgovski1987b,djorgovski1987a}; specifically, 4.5\% of confirmed dual AGNs showed evidence for two AGN-driven radio sources \citep[e.g.][]{fu2015a,fu2015b,owen1985,klamer2004,hewitt1987,rubinur2019,fu2011b,mullersanchez2015}. Among common techniques, near-IR imaging is the final technique, used in  10.8\% of all confirmed systems \citep[e.g.][]{shen2011,gregg2002,imanishi2014,imanishi2020,mu2001,morgan2000,u2013}. Several other  waveband+technique combinations have also been employed in the literature, but for only small percentages of confirmed systems: near-IR, UV, and mid-IR spectroscopy are used in the confirmation of $<3\%$ each of all systems; sub-mm, mid-IR, and hard X-ray imaging are used in the confirmation of $<1.5\%$ each of all systems.

Only a single binary AGN has so far been confirmed in the literature: CSO 0402+379, which was serendipitously identified and confirmed in VLBA radio VLBI imaging \citep{maness2004,rodriguez2006}. VLBI imaging is currently the only way to identify and confirm binary AGNs due to the instrumental requirements for resolving mas-scale separated AGNs. To-date, no recoiling AGNs have been confirmed in the literature.

\subsection{Examining Pair Separation and Redshift Parameter Spaces}

The Big MAC DR1 quantifies the parameter space of multi-AGN redshifts and pair separations that have so far been probed in the literature (up to 2020). There is some incompleteness and ambiguity introduced to this analysis due to the fact that angular and physical separations are not available for large fractions of binary AGN candidates, as these quantities are not always derived for entire samples \citep[e.g., as in the case of ][]{graham2015b,charisi2015,eracleous2012,runnoe2017a,ju2013,xliu2014,shen2013}. The same can be said for some dual AGNs and candidates, but for those with resolvable nuclear separations we have supplemented our own angular projected pair separation measurements (as mentioned in Section~\ref{sec:catalog_construct}). In the case of binary AGNs and SMBHs without a constraint on separation \citep[such as those identified based on tentative periodicity or offset emission lines, e.g.,][]{graham2015b,eracleous2012,xliu2014}, we assume for now a (very generous) upper limit on the binary separation of $<1$\,pc. This upper limit is an order of magnitude or larger than the separation expected in the case of, for example, a binary AGN selected based upon optical periodicity ($<0.1$\,pc), but it is sufficient for the present work and can be adjusted in a future data release.

Figure~\ref{fig:proj_seps} shows the distribution of pair separations where the angular separation (in arcseconds) is shown on the x-axis and the physical, projected separation (in kpc) is shown on the y-axis. The data are color-coded by the logarithm of the system redshift (given on the auxiliary axis). Selection effects are immediately apparent within the distribution of multi-AGN systems: the distribution appears bimodal and redshift dependent, which is a direct result the different selection functions (which carry instrumental, survey, and wavelength-dependent biases) employed in low- and high-z multi-AGN searches.  Other features of note are the bands of triangle markers -- pointing diagonally down and to the left -- that stretch across the parameter space: these objects are selected based upon the presence of double-peaked optical spectroscopic emission lines in spectroscopic fibers. Separations are not available for the vast majority of these candidate multi-AGN systems, but the angular diameter of the spectroscopic fibers provides an upper limit on the separations of the candidates: if these double-peaked objects do harbor kpc-scale or sub-kpc-scale dual AGNs (as traced by the double-peaked emission observed within the 3'' or 2'' spectroscopic fibers in SDSS and/or LAMOST), then the candidate separation should be no larger than the angular size of the fiber. Note, however, that fewer than 15\% of such objects are expected to host multi-AGNs \citep[e.g.,][]{fu2012,mullersanchez2015}, and thus only a small fraction of these data points are expected to be associated with bona fide multi-AGNs.

This figure extends down to $\sim100$ parsec-scale separations, illustrating the handful of dual AGN and candidates that have sub-kpc separations \citep[e.g., ][]{komossa2003}, but does not illustrate the handful of dual AGN/binary candidates with sub-100\,pc separations \citep[e.g., ][]{gitti2006,gitti2013}, binary AGN candidates with pc-scale and sub-pc-scale separations, and the only known binary AGN, CSO 0402+379, which has a separation of 7.3\,pc \citep{rodriguez2006}. Due to observational and technological constraints, efforts in the literature have naturally been biased towards pair separations $>1$\,kpc; in fact, our samples are particularly weighted toward even larger separations ($>10$\,kpc) due to the ease of selecting large quantities of $>10$\,kpc dual AGN candidates in the optical via SDSS fiber spectroscopy \citep[e.g., ][]{liu2011b}. Greater efforts are certainly needed -- across all wavebands -- to identify larger samples of dual AGNs at $1-3$\,kpc and sub-kpc, particularly $\lesssim100$\,pc below which dual AGNs/SMBHs evolve into gravitationally-bound binary AGNs/SMBHs. This latter objective is largely limited by our technological abilities in the optical, infrared, and X-rays; radio imaging currently represents the only promising method by which larger samples of $<100$\,pc dual and binary AGNs can be identified thanks to the mas-scale resolutions achievable with VLBI. However, this does require both AGNs to be active in the radio \citep[e.g.,][]{rodriguez2006}, which is a less likely scenario \citep[e.g.,][]{burke2011,mullersanchez2015}. Though difficult to detect with current observational techniques, samples of dual AGNs with separations $100\,\rm{pc}<r_p<1000\,\rm{pc}$ can be identified and studied with great detail through synergistic multiwavelength observations \citep{koss2023}, and it is possible that large numbers of such dual AGNs lurk within optically-elusive mergers that masquerade as single galaxies \citep{koss2018}, awaiting discovery. On the other hand, dual AGN candidates at large pair separations are absolutely necessary for a complete picture of SMBH fueling across the merger sequence \cite{derosa2023,chen2023}, and further multiwavelength efforts are needed to identify such dual AGNs in earlier-stage mergers. 

\begin{figure}[t]
    \centering
    \includegraphics[width=\linewidth]{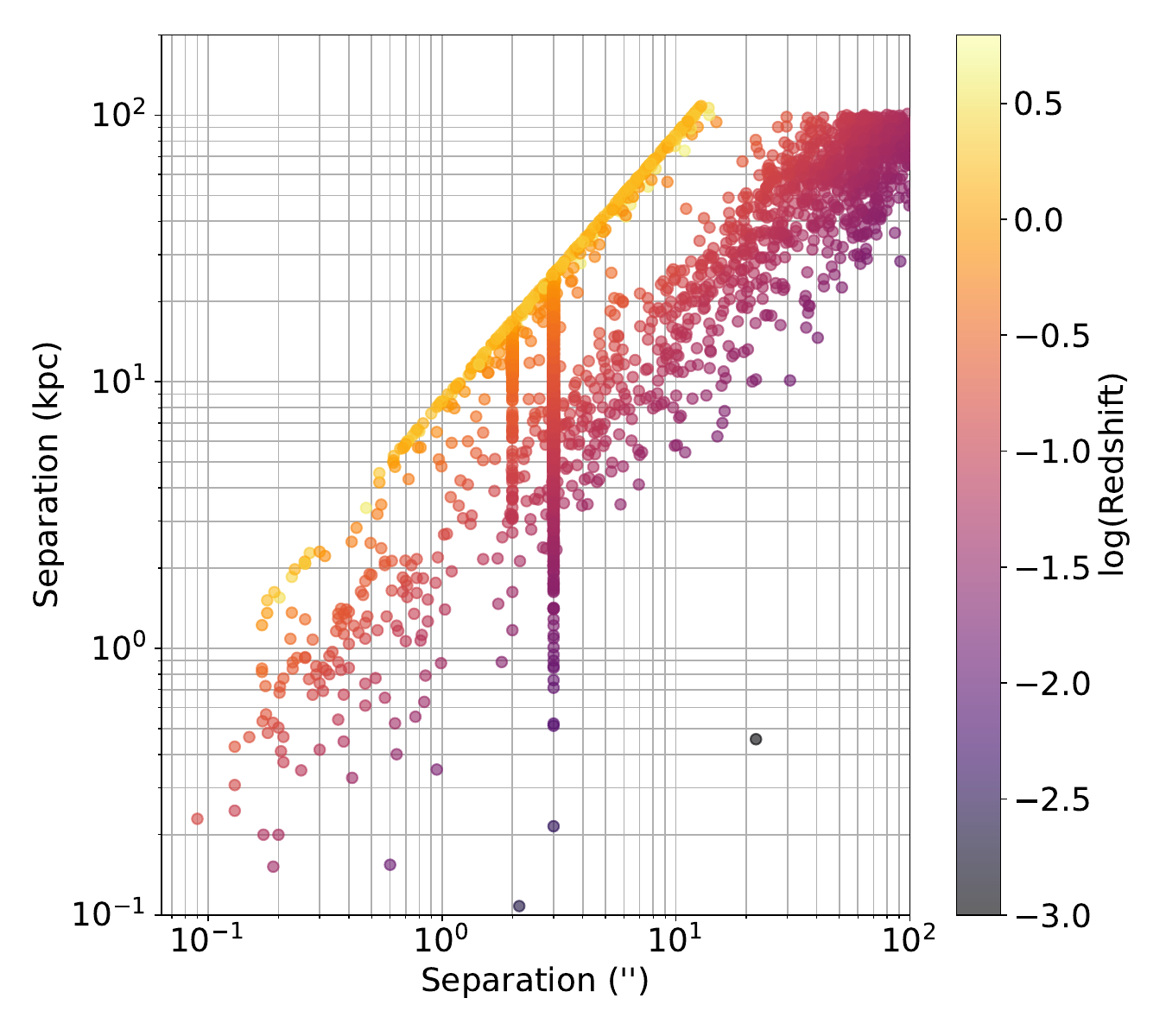}
    \caption{The distributions of angular separation, physical, projected separation ($r_p$, in kpc), and redshift for objects within the Big MAC DR1. Angular separation (in arcseconds) is given on the x-axis, and physical, projected separation (given in kpc) is given on the y-axis. The data are color-coded according to the logarithm of the redshift, given on the auxiliary color map.} %\textcolor{violet}{Possible adding some histograms/curves along the top axis and side axis to show the number sources distributed along each axis. This will really help to show the the biases.}}
    \label{fig:sepas_vs_seprp_z}
\end{figure}

We also illustrate the distribution of projected pair separations (in kpc) as a function of redshift for the DR1 multi-AGN candidates in Figure~\ref{fig:proj_seps}; this plot is analogous to Figure~1 in \citet{chen2022}. As in Figure~\ref{fig:sepas_vs_seprp_z}, the redshift distribution is fairly bimodal due to the selection functions used to draw the various samples of dual AGNs, and the vast majority of systems have been identified at local redshifts ($z<0.1-0.2$). With JWST and Euclid, along with ground-based high-resolution imaging soon to be provided by Vera C. Rubin, we are poised to expand both our local- and high-redshift samples of dual AGN candidates, particularly those with close angular separations. This will be particularly important, because high redshift dual AGNs are expected to be clear signposts of merger-enhanced SMBH accretion \citep{chen2023,foord2024} and exhibit higher accretion rates, higher obscuration, and higher host gas fractions \citep[e.g.,][]{blecha2018,chen2023} than local universe counterparts.

\section{Results}
\label{sec:results}
Big MAC DR1 comprises 5742 multi-AGN systems (confirmed or candidates) published in the literature between 1970-2020, and fall into the four categories of (1) dual AGN candidates (4336), (2) binary AGN/SMBH candidates (1369), (3) recoiling AGN/SMBH candidates (581), and (4) multi-AGNs or N-tuple AGN candidates (59); we show the breakdown of this sample into the specific classes in Figure~\ref{fig:class_breakdown}. Since each system may be cross listed across all three categories of dual, binary, and recoiling AGNs, we show the percentages of objects showing single system classes, two system classes, and three system classes in Figure~\ref{fig:class_breakdown}, and the associated bar charts illustrate the fractional contribution of the different classes/cross-listings to the objects with single, double, or triple system classes. For simplicity, we treat the N-tuple table as its own table separately from the duals, binaries, and recoil candidates.

Recall that dual and binary AGN candidates in this catalog satisfy or partially-satisfy (in cases of missing redshifts) the definitions laid out in Sections~\ref{sec:definingduals} and \ref{sec:definingbinaries}: dual AGNs in general should show projected physical separations $<110$\,kpc with host system velocity differences $<600$\,km\,s$^{-1}$; when broad lines are used for redshift determination, a cut of $|\Delta v| \lesssim 2000\,\rm{km}\,\rm{s}^{-1}$ is suggested \citep[as in][]{hennawi2006}, and such systems are typically referred to as candidates. Binary AGNs or binary SMBHs are required to be gravitationally-bound to one another, with physical separations at or below the `bound radius' ($r_b$, 10s of pc to $<1$\,pc depending on the mass of the binary). We show the sky distribution of the Big MAC DR1 in Figure~\ref{fig:sky_dist}, which is heavily weighted toward coverage by SDSS since a significant portion of the selected multi-AGN systems arise from systematic searches within the SDSS footprint.

\begin{figure}[t]
    \centering
    \includegraphics[width=\linewidth]{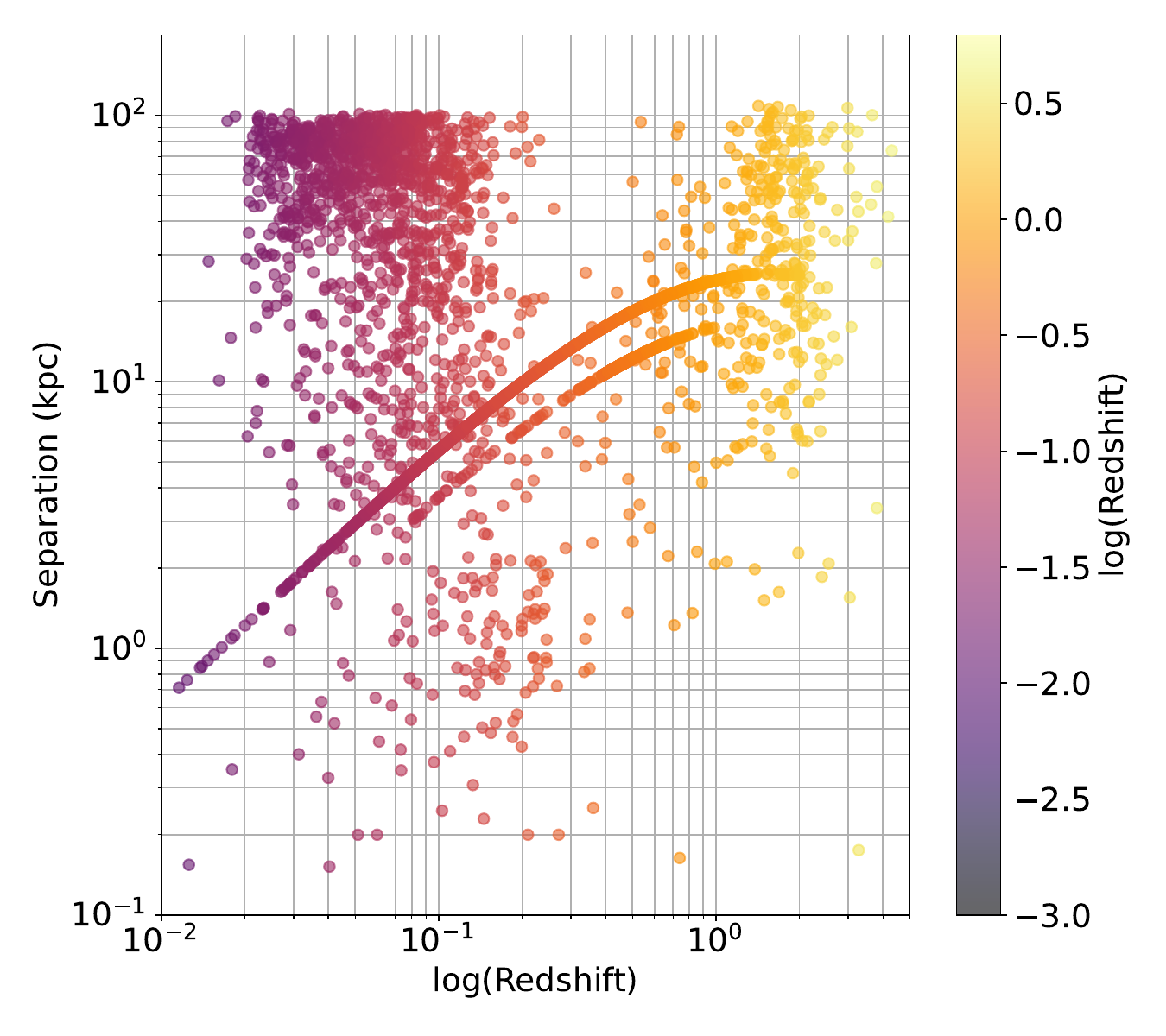}
    \caption{The distributions of redshift and projected, physical separations of objects in the Big MAC DR1. The logarithm of the redshift is given on the x-axis, and the projected, physical separation (in units of kpc)  is given on the y-axis. To aid the eye of the reader, the data are additionally color coded according to logarithm of the redshifts, with the color map provided on the auxiliary axis. This plot, while similar to the information provided in Figure~\ref{fig:sepas_vs_seprp_z}, is provided to complement a similar plot shown in \citet[][see their Figure 1]{chen2022}. %A series of black lines (solid: 5'', dashed: 1'',  dash-dotted: 0.5'', dotted: 0.1'') denote various angular separations (given in arcseconds). %\textcolor{violet}{Possible adding some histograms/curves along the top axis and side axis to show the number sources distributed along each axis. This will really help to show the the biases. Also need to add the various lines showing angular separation!}
    } %(solid black line denotes angular separations of 5''; the dashed black line denotes angular separations of 1''; the dash-dotted black line denotes angular separations of 0.5''; the dotted black line denotes angular separations of 0.1''.
    \label{fig:proj_seps}
\end{figure}

% technology biases also affect the bias. radio and xray imaging is biased toeard laregr pairs.
The DR1 redshift distribution extends from $0.001\lesssim z \lesssim 6.2$. The distribution of angular and physical, projected separations is shown in Figure~\ref{fig:proj_seps}, where we have limited the visible separations to $>100$\,pc due to the lack of sources below 100\,pc with firm separation constraints (rather than simply $<1$\,pc upper limits int he case of most binaries). The DR1 (and the current literature) is heavily biased toward systems with projected separations $>1-5$\,kpc. Many of the dual AGN candidates and binary AGN candidates have only upper limits for their projected separations; these upper limits were either derived in the literature or assigned during the construction of the catalog. As an example, for the double-peak selected dual AGN candidates from \citet{wang2009}, \citet{liu2010a}, and \citet{smith2010} that do not have separations available in the literature, we have assumed 3\arcsec{} angular separations as an upper limit since these objects are selected based upon their double-peaked spectroscopic emission lines observed via the 3\arcsec{} diameter SDSS fibers; double-peaked dual AGN candidates selected with BOSS 2'' diameter fibers are similarly assigned a 2'' separation upper limit. These were used to derive upper limits on the projected separations (in kpc) at the redshift of each source. Unless (angular/projected) positional offsets were observed for recoil candidates (representing the separation between the recoiling AGN and the nucleus of the host), the separations of recoil candidates are flagged as ``-99'' and are not included in this Figure~\ref{fig:proj_seps}. For binary AGN candidates that do not have a reported separation (or upper limit), we assign an upper limit of $<1$\,pc (and calculate an inferred angular separation upper limit given the redshift of the candidate(s)).

The vast majority of dual AGN candidates in the literature have been drawn from optical spectroscopic fiber surveys, with samples drawn from pairs of fibers covering distinct nuclei \citep[e.g.,][]{liu2011b,hennawi2006,hennawi2010} and selection based on double-peaked emission lines \citep[e.g.,][]{liu2010b,wang2009,smith2010,ge2012,shi2014,lyu2016}. Optical imaging has been used for the selection of a significant number of dual AGNs and candidates, particularly during searches for gravitational lenses and clustered quasars at higher redshift \citep[e.g.,][]{hennawi2006,myers2007,inada2008,inada2010,inada2012,eftekharzadeh2017,more2016,anguita2018,agnello2018,lemon2018} as well as searches for closely separated dual AGNs via varstrometry-selection \citep[e.g.,][]{hwang2020}. Smaller (but notable) samples of dual AGNs candidates ($\lesssim$5\% of the total candidate sample in Big MAC DR1) have been selected using hard X-ray preselection \citep{koss2011,koss2012}, mid-IR selection \citep[e.g.,][]{satyapal2017,pfeifle2019a,findlay2018}, radio imaging \citep[e.g.,][]{owen1985,klamer2004,fu2015a}, and near-IR colors \citep{imanishi2014,imanishi2020}. Among confirmed dual AGNs, optical spectroscopic selection is still a common technique \citep[optical spectroscopy was used to select $\sim50$\% of confirmed dual AGNs, e.g.,][]{hennawi2006,liu2010b,liu2011b,koss2012}, but it is not nearly as overrepresented relative to other selection methods. For example, optical imaging and colors were used in the selection of $\sim60$\% and $\sim37$\% of candidates that would later be confirmed as bona fide dual AGNs (whereas across the full sample of selected dual AGN candidates, optical imaging was only used in $\sim17$\% of cases). Similarly, hard X-ray BAT selection \citep{koss2011,koss2012}, soft X-ray selection \citep[e.g.,][]{komossa2003,bianchi2008,piconcelli2010,koss2011,koss2012,pfeifle2019a,liu2013,comerford2015}, mid-IR imaging \citep[e.g.,][]{satyapal2017,pfeifle2019a,findlay2018}, radio imaging \citep[e.g.,][]{fu2015a} played a role in selection of $\sim7.6$\%, $\sim3.2$\%, $\sim9$\% of candidates that would later be confirmed as bona fide dual AGNs (see Figure~\ref{fig:dual_confirm_selectmethods}. %a more significant role. Confirmation techniques are far more diverse, often requiring X-ray imaging, radio imaging, and follow-up optical spectroscopy for the confirmation of dual AGNs. 

In terms of confirmation methods for dual AGNs, optical spectroscopy again plays a large role ($\sim78$\%) as expected. However, soft X-ray imaging and spectroscopy played significant roles in confirming dual AGNs and were used in the confirmation of $\sim21$\% and $\sim18$\% of dual AGNs. This is in stark contrast to soft X-ray imaging or X-ray spectroscopic selection ($<5\%$ of confirmed duals relied upon soft X-ray selection), but this emphasizes a technological hurdle/bias: large-scale, high spatial resolution, soft X-ray imaging has not available, impeding efforts to select complete samples of dual AGNs via the soft X-rays, but targeted, follow-up X-ray observations have been very effective in confirming samples of dual AGNs \citep[e.g.,][]{komossa2003,bianchi2008,piconcelli2010,liu2013,comerford2015,koss2012,pfeifle2019a}. Optical imaging (used to identify galaxy mergers) and radio imaging also played key roles ($\sim20\%$ and $\sim17\%$, respectively) in the confirmation of dual AGNs.

Binary AGN candidates are overwhelmingly selected via optical spectroscopy, where fiber or long-slit spectra are searched for velocity offset emission lines or radial velocity shifts in emission lines over time \citep[e.g.,][]{eracleous1994,eracleous2012,shen2013,ju2013,xliu2014,runnoe2015,runnoe2017a,wang2017}, and binaries are also selected based on the presence of X-shaped radio sources observed in radio imaging\citep[e.g.,][]{dennett2002,cheung2007,yang2019}. Note: not all literature articles consider the binary hypothesis for X-shaped radio sources, and a variety of origins are viable for X-shaped radio sources. In this catalog release, however, all X-shaped radio sources are considered binary candidates. Optical imaging is another common selection strategy, used to select binary candidates that exhibit firm \citep{sillanpaa1988} or tentative periodicities \citep[e.g.,][]{graham2015b,charisi2016,tliu2016,tliu2018,tliu2019}. Despite extensive optical spectroscopic and periodicity searches, only a single binary AGN has so far been confirmed in the literature: CSO 0402+739, which was serendipitously identified \citep{maness2004} and confirmed \citep{rodriguez2006} in radio VLBI follow-up imaging. Several candidates exist within the literature that have also been selected serendipitously via radio VLBI imaging \citep[e.g.,][]{kharb2017,yang2017b}

Recoil candidates are most often selected based on the presence of emission lines with significant velocity offsets from the host restframe, as observed in optical fiber spectra \citep[e.g.,][]{komossa2008,steinhardt2012,kim2016}, X-shaped radio sources \citep[since jet precession following a SMBH merger could drive the formation of such structures, e.g.,][]{dennett2002,merritt2002,yang2019}, and to a lesser extent optical imaging to search for spatial offsets from the nuclei of galaxies \citep[e.g.,][]{batcheldor2010,lena2014,civano2010}. So far, no recoiling AGNs have been confirmed in the literature.

The statistical completeness of this catalog is limited by the statistical completeness of the literature; that is to say, this catalog is highly statistically incomplete because the selection methods used to identify multi-AGN systems are intrinsically incomplete. However, in terms of the known multi-AGN candidates and confirmed systems (along with their discovery articles or key analysis articles published in the literature between $\sim$1970-2020), the catalog should be considered overwhelmingly literature-complete. 
\section{Discussion}
\label{sec:discussion}

\subsection{The Need For New and/or Diverse Selection Approaches}
\label{sec:new_approaches}

In Section~\ref{sec:analysis}, we performed an exhaustive cross-examination of selection techniques employed in the multi-AGN literature to search for dual AGNs, binary AGNs, and recoiling AGNs. In this section, we focus on the need for new, diverse, and/or multiwavelength selection strategies specifically for dual AGNs. Our analysis in Section~\ref{sec:dualselectmethods} demonstrated that  dual AGN candidates selected via optical spectroscopic techniques are over-represented, specifically double-peaked emission lines \citep[e.g.][]{wang2009,liu2010b,smith2010,ge2012,shi2014,lyu2016} and/or spectroscopic emission line ratios \citep[e.g.][]{liu2011b}. Of course, there exists a variety of other selection techniques (including but not limited to mid-IR colors, near-IR colors, soft and hard X-ray imaging, and radio imaging) that have been used in concert with -- or independently of -- these optical spectroscopic techniques, but the percent usage of ``non-optical'' techniques is disproportionately low in comparison to optical spectroscopic selection. 

New and/or more diverse selection strategies and multi-wavelength follow-up observations must be the focus more frequently in the search for dual AGNs, and the community should not rely primarily on optical spectroscopy. Optical spectroscopic selection will undoubtedly remain one of the most common and popular selection techniques for identifying dual AGNs given the ever-increasing numbers of AGNs observed with the SDSS \citep[][which is now on its 18$^{\rm{th}}$ data release]{almeida2023}, LAMOST \citep{cui2012,zhao2012}, the large spectroscopic samples of AGNs soon to be available via the Dark Energy Survey Instrument \citep[DESI,][]{desi2016}, etc. However, dual AGNs have been notoriously difficult to confirm using a single waveband, particularly when it comes to the optical band: heavily obscured dual AGNs do not always manifest at optical wavelengths \citep[e.g.,][]{komossa2003,piconcelli2010,koss2012,satyapal2017,pfeifle2019a}, and optically elusive or optically-ambiguous dual AGNs \citep{komossa2003,koss2012}\footnote{as well as optically-elusive galaxy mergers \citep[e.g.,][]{koss2018}} necessitate selection or follow-up by shorter \citep[X-rays, e.g.][]{koss2012,hou2019,pfeifle2019a,pfeifle2019b,hou2020} or longer \citep[mid-IR, near-IR, or radio, e.g.,][]{fu2011b,shen2011,fu2015a,fu2015b,imanishi2014,satyapal2017,pfeifle2019a,imanishi2020} wavelengths for detection; longer wavelengths can be particularly useful for powerful and/or obscured systems, due to their insensitivity to the line-of-sight absorption. Furthermore, numerous dual AGNs have been confirmed via rigorous multiwavelength analyses \citep[e.g.,][]{koss2012,liu2013,comerford2015,hou2019,hou2020}. Using both single waveband/facility and multiwavelength/multi-facility selection techniques -- including techniques independent of optical spectroscopic selection -- is the key to uncovering the full population of dual AGNs.

Dual AGN surveys that are independent of optical spectroscopy can take place presently owing to the availability of all- or large-sky coverage surveys across the electromagnetic spectrum. For example, given the all-sky mid-IR coverage of the Wide-Field Infrared Survey Explorer \citep{wright2010} and large-scale radio coverage ($1.5''$ spatial resolution at $2-4$\,GHz for decl.$>-40\deg$) with the Very Large Array (VLA) All Sky Survey \citep[VLASS,][]{lacy2020}, it would be prudent to perform exhaustive searches for dual AGNs in the local universe where (a) each AGN exhibits red WISE colors \citep[][]{stern2012,assef2018} indicative of mid-IR AGNs, (b) each AGN exhibits a compact, flat-spectrum radio source \citep[like those identified in the Stripe82 survey by ][and who suggested such a radio-based search in VLASS]{fu2015a,fu2015b}, and (c) a complementary approach where one source is a mid-IR AGN and one is a radio AGN. Combining these techniques with optical spectroscopic fiber or long-slit measurements, one could also search for cases where the AGN pairs are comprised of one optical and one mid-IR AGN or one optical and one radio AGN. Recent efforts to select dual AGNs blindly with respect to optical spectroscopy have been made in the mid-IR: \citet{barrows2023} selected $\sim200$ new spatially resolved dual AGN candidates (based on photometric redshifts) to study the occupation fraction of single and dual mid-IR AGNs in galaxy mergers relative to isolated controls; they found increases in AGN fraction with increasing extinction, and AGN excesses at increasingly smaller merger separations (particularly for obscured AGNs). Similar efforts to uncover spatially-resolved mid-IR dual AGNs (nearly-)independently of optical spectroscopy are currently underway within both the BAT AGN Spectroscopic Survey sample (Pfeifle et al., in preparation) and within the DeCaLs imaging footprint (Pfeifle et al., in preparation). Techniques independent of optical spectroscopy necessarily depend upon the accessibility of large-scale sky coverage (with moderate to high spatial resolution) to identify clear signatures of tidal interactions and ongoing mergers, which is a current limitation particularly for higher redshift $\gtrsim0.1$ dual AGNs (Pfeifle et al., in preparation)\footnote{Though, even spectroscopic selection techniques like double-peaked optical emission lines \citep[e.g.]{wang2009,liu2010b,smith2010} almost ubiquitously require follow-up high spatial resolution imaging to search for multiple nuclei \citep[e.g.,][]{shen2011,fu2011a,fu2012,rosario2011}.}. When it begins science survey operations in 2025, the Vera C. Rubin observatory will aid in surmounting this limitation in the southern sky: Rubin will provide high spatial resolution and high cadence imaging that will play a key role in morphologically classifying dual AGNs selected ``blindly'' with respect to optical spectra in the southern sky \citep[e.g., WISE-selected samples like ][and Pfeifle et al., in preparation]{barrows2023}. Other avenues independent of spectroscopy  could include soft X-ray ($2-10$\,keV) selection of dual AGN in, for example, archival Chandra, Swift, XMM-Newton, and now eROSITA imaging \citep[$2-5\,\rm{keV}$,][]{predehl2021}. Although the PSF of eROSITA ($\sim30''$ in scanning mode) will not allow for the unambiguous identification of close ($\lesssim30''$) angular X-ray pairs, the large-scale sky coverage could be used to search for wider-separation (low redshift) X-ray dual AGNs or could be matched with other surveys (e.g., WISE, VLASS) to search for cases where positional offsets between the eROSITA X-ray centroid and multiwavelength AGN counterparts are due to dual AGNs manifesting in different wavebands. 
%relying upon mid-IR colors for AGN selection, optical imaging for merger confirmation, and spectroscopy for redshift determination;

Novel techniques, such as the varstrometry (based on astrometric variability in optical imaging of spectroscopically-selected quasars) \citep{shen2019,hwang2020} and GAIA multi-peak \citep{mannucci2022,ciurlo2023,mannucci2023} strategies have already been welcome advancements in the search for high redshift Type 1 multi-AGNs, and they have uncovered medium-sized samples of high redshift candidate dual AGNs and gravitational lenses \citep[e.g.,][]{mannucci2023,chen2024}, some of which have been confirmed via follow-up JWST spectroscopy and imaging \citep[.g.,][]{chen2024singleobj}. Though we remain biased against detecting high redshift obscured AGNs, the application of current as well as novel multiwavelength techniques are sure to circumvent this bias within this decade and provide a more statistically complete view of high redshift dual AGN activity. As a prime example, \citet{li2024} very recently employed a powerful multiwavelength selection approach to identify unobscured and obscured dual AGNs at high redshift: selecting high redshift AGNs first based on their (unresolved) X-ray emission \citep{civano2016,marchesi2016a}, \citet{li2024} used JWST near-IR and spatially-resolved broadband SED fitting to identify 28 new dual AGNs (and candidates) at $0<z<5$ in the COSMOS field. \textit{Euclid} will soon complement near- and mid-IR JWST-based dual AGN searches, bringing large-scale  ($\sim1/3$ of the sky) imaging and spectroscopic coverage in the near-IR with sensitivity and angular resolution comparable to HST. 

While we have discussed specific wavelength- and instrument-dependent selection techniques, it is also prudent to discuss the selection of dual AGNs as a function of projected separation ($r_p$) and merger stage. It is frequently quoted that dual AGNs should be preferentially found in late-stage mergers because the activity of the AGNs is predicted to be better correlated \citep[e.g.][]{vanwassenhove2012,blecha2018}. Many papers therefore focus specifically on dual AGNs with separations below 30\,kpc \citep[e.g.][]{stemo2021,comerford2018,li2024} or 10\,kpc \citep[e.g.,][]{fu2012,liu2013,comerford2015,satyapal2017,pfeifle2019a,hou2019}\footnote{Though this is also intrinsically tied to waveband selection method: dual AGN candidates selected via double-peaked emission lines are biased toward $<10$\,kpc separations \citep[e.g.,][]{wang2009,liu2010b,smith2010}.} and tend to ignore larger separation systems \citep[though with some exceptions, e.g.,][]{liu2011b,hennawi2006,derosa2015,derosa2023,koss2012,hou2020}. This is a serious deficiency: recent results in the ASTRID cosmological simulation showed that, when considering dual AGNs with separations $<30$\,kpc, $\sim50$\% of dual AGNs were found at separations $>10$\,kpc \citep{chen2023}. Understanding the full population of dual AGNs and how they (and AGN activity more generally) relate to the evolution of the galaxy merger process requires an understanding of how these systems ignite and evolve across the full merger sequence. The selection functions of dual AGN surveys should preferentially include larger separation dual AGNs in order to understand how they may be similar or different to their late-stage and/or close-separation counterparts. Furthermore, as outlined in Section~\ref{sec:definingmultiagns}, dual AGNs observed at close separations $<10$\,kpc may not yet actually reside in late-stage mergers due to the orbital dynamics of the host galaxies but instead may simply be observed during one of several pericenter passages. Samples comprised of close physical pairs of dual AGNs are likely to include both intermediate-stage dual AGNs (close to pericenter) as well as dual AGNs residing within truly late-stage mergers, blurring the line between earlier- and late-stage dual AGNs and biasing against a clear understanding how how these populations manifest and behave. Morphological and kinematic information is likely the key to deblending these distinct -- yet sometimes similar looking -- populations of dual AGNs and galaxy mergers, particularly through the use of Neural Networks trained on cosmological simulations \citep[e.g.,][]{ferreira2024}. It may turn out that dual AGNs in late-stage mergers are intrinsically more frequent, more luminous, and more obscured than their earlier-stage counterparts \citep[e.g.,][]{blecha2018,ricci2017mnras,vanwassenhove2012,koss2012}, but this has yet to be conclusively proven \citep[e.g.][]{guainazzi2021,derosa2023}.

\subsection{Known ICRF Targets Associated with Multi-AGN Systems}
\label{sec:ICRF}

The multiplicity of sources intrinsic to multi-AGN systems can lead to source confusion \citep[e.g.,][]{tadhunter2012} and multiwavelength positional offsets \citep[e.g.,][]{orosz2013,barrows2016} in these systems, making them potentially poor targets for celestial reference frame research \citep[CRF, e.g.,][]{dorland2020,fischer2021,sargent2024}. Thus, it is necessary to understand what overlap exists between the multi-AGNs in the Big MAC DR1 and the current catalog of ICRF AGNs \citep[ICRF3, ][]{charlot2020}. 

For each AGN/nucleus entry with distinct coordinates, we matched to the ICRF3 catalog \citep{charlot2020} to search for counterparts in the S/X, K, and X/Ka bands. Match tolerances of 1'' and 10'' were used, where the 10'' tolerance was specifically used to search for large-scale radio structure that originates from a multi-AGN candidate. We identified an overlap of 135, 38, and 32 objects at S/X, K, and X/Ka with a match tolerance of 1'' between Big MAC DR1 and ICRF3; an identical overlap was found for a match tolerance of 10'' at K and X/Ka band, but an additional two objects (BZQJ2156-2012, with an offset of $\sim1.05''$; PKS 1155+251, with an offset of $5.9''$) were recovered at S/X band. A subset of matching ICRF3 and Big MAC objects are listed in in Table~\ref{tab:icrf} for illustrative purposes, including the system classes, literature names, ICRF designations, coordinates (to AGN entry 1 in the Big MAC DR1), and the angular separation (in X/S, K, and X/Ka band) between the ICRF3 targets and the Big MAC entries. Among these overlapping targets, 11 are currently ICRF3 defining sources (and are flagged within the full version of the table); these targets should be removed as defining sources in the ICRF3 catalog, and all of the 135 objects shared between the ICRF3 and Big MAC DR1 should be avoided when conducting CRF research.

\begin{table}[t!]
\caption{A sample of ICRF3 Targets within Big MAC DR1}
\begin{center}
\hspace{-2.2cm}
\begin{tabular}{cccccc}
\hline
\hline
\noalign{\smallskip}
AGN1 Name & ICRF3 Designation & $\Delta\theta_{\rm{S/X}} ('')$\\%, $\Delta\theta_{\rm{K}}$, $\Delta\theta_{\rm{X/Ka}}$ ('')\\
\noalign{\smallskip}
\hline
\noalign{\smallskip}
%Binary Quasar & SDSSJ0812+2520A & 105.8 & ICRF J081247.7+252242 \\
%Binary Quasar & J13488.7+28407.0 & 59.09 & ICRF J134804.3+284025 \\
%Recoil Candidate & J1357+4807 &  0.0 & ICRF J174614.0+622654 \\
%Recoil Candidate & J1424+2637 & 0.0 & ICRF J142440.5+263730 \\
%Dual AGN / Quasar Candidate & PKS 1614+051 & 6.55 & ICRF J161637.5+045932 \\
%Binary Quasar & PKS 1145-071 & 4.2 & ICRF J114751.5-072441 \\
%Binary SMBH Candidate & Arp 102B & 0.4104 & ICRF J171914.4+485849 \\
%Binary SMBH Candidate & OJ 287 & 2.2146E-5 & ICRF J085448.8+200630 \\
%Binary SMBH Candidate & 3C 120 & 3.0379E-5 & ICRF J043311.0+052115 \\
%Binary SMBH Candidate & OX 169 & 0.05835 &  ICRF J214335.5+174348 \\
%Binary SMBH Candidate & 3C 390.3 & 0.1725 & ICRF J184208.9+794617 \\
%Binary SMBH Candidate & 1928+738 & 3.0E-6 & ICRF J192748.4+735801 \\
%Binary SMBH Candidate & 3C 66B &  0.4598 &  ICRF J022311.4+425931 \\
%Binary Quasar Candidate & PKS 0537-441 &  3.9 & ICRF J053850.3-440508 \\
%Binary SMBH & CSO 0402+379 & 0.0068 & ICRF J040549.2+380332 \\
%Binary SMBH Candidate & 3C 345 & 3.0084E-6 & ICRF J164258.8+394836 \\
%Binary SMBH Candidate & SDSS J1048+0055 & 0.0025 & ICRF J104807.7+005543 \\
%Binary Quasar & SDSSJ1310+0044A  & 65.5 & ICRF J131028.5+004408 \\
BZQJ0842+4525 & ICRFJ084215.3+452544 & 0.9835 \\%, , \\
3C 298.0 & ICRFJ141908.1+062834 & 0.4189 \\%, , \\
BZQJ2156-2012 & ICRFJ215633.7-201230 & 1.0486 \\%, , \\
TEX 1428+370$\dagger$ & ICRFJ143040.5+364903$\dagger$ & 0.04575 \\%, , \\
PKS 2203-215 & ICRFJ220641.3-211940 & 0.5709 \\%, , \\
%OQ 208 / Mrk 668 / 140700.40+282714.6 & ICRFJ140700.3+282714 & 0.1163 \\%, , \\
%151443.07+365050.4 & ICRFJ151443.0+365050 & 0.04801 \\%, , \\
J1357+4807$\dagger$ & ICRFJ174614.0+622654$\dagger$ & 7.245e-5 \\%, , \\
J1424+2637 & ICRFJ142440.5+263730 & 0.0357 \\%, , \\
J0038-2120 & ICRFJ003829.9-212004 & 0.0040 \\%, , \\
J0041+1339 & ICRFJ004117.2+133927 & 6.8792e-4 \\%, , \\
J0106+3402 & ICRFJ010600.2+340202 & 0.00212 \\%, , \\
J0106+2539 & ICRFJ010610.9+253930 & 5.0982e-4 \\%, , \\
J0106-0315 & ICRFJ010643.2-031536 & 0.001 \\%, , \\
J0146+2110 & ICRFJ014658.7+211024 & 0.0039 \\%, , \\
J0216-0118 & ICRFJ021605.6-011803 & 6.0575e-4 \\%, , \\
J0216-0105 & ICRFJ021612.2-010518 & 0.0027 \\%, , \\
J0334+0800 & ICRFJ033453.3+080014 & 3.7039e-4 \\%, , \\
J0335-0709 & ICRFJ033557.0-070955 & 0.0022 \\%, , \\
J0431+2037 & ICRFJ043103.7+203734 & 2.475e-4 \\%, , \\
J0435+2532 & ICRFJ043534.5+253259 & 0.0017 \\%, , \\
J0523+6007 & ICRFJ052311.0+600745 & 0.0043 \\%, , \\
J0552+0313 & ICRFJ055250.1+031327 & 0.0013 \\%, , \\
J0644+2911 & ICRFJ064444.8+291104 & 0.0017 \\%, , \\
J0729-1320 & ICRFJ072917.8-132002 & 5.7656 \\%, , \\
J0736+2954 & ICRFJ073613.6+295422 & 0.0010 \\%, , \\
J0817+3227 & ICRFJ081728.5+322702 & 0.0014 \\%, , \\
J0843+4537 & ICRFJ084307.0+453742 & 0.0066 \\%, , \\
J0854+6218 & ICRFJ085450.5+621850 & 0.0017 \\%, , \\
J0902+4310 & ICRFJ090230.9+431014 & 4.18e-4 \\%, , \\
J1022+4239 & ICRFJ102213.1+423925 & 3.79e-4 \\%, , \\
J1033+3935 & ICRFJ103322.0+393551 & 7.61e-5 \\%, , \\
J1150+4332 & ICRFJ115016.6+433205 & 0.0011 \\%, , \\
J1254+0859 & ICRFJ125458.9+085947 & 0.0030 \\%, , \\
J1301+4634 & ICRFJ130132.6+463402 & 4.20e-4 \\%, , \\
J1312+2531 & ICRFJ131214.2+253113 & 0.0010\\%, , \\
\dots & \dots & \dots\\
\noalign{\smallskip}
\hline
\end{tabular}
\end{center}
\tablecomments{A sample of ICRF3 targets identified within the Big MAC DR1. Column 1: AGN1 designation within Big MAC DR1. Column 2: ICRF3 designation. Column 3: angular separation between the Big MAC AGN entry and the ICRF target (in arcseconds) in the S/X bands (when available). $\dagger$ICRF3 defining sources. An abridged version of this table is provided here; a complete table is available in the Big MAC repository.} %, K, and X/Ka
\label{tab:icrf}
\end{table}

\subsection{Recommended Best Practices in Multi-AGN Science}
\label{sec:best_practices}
Over the course of the catalog creation, a number of issues arose related to the uniformity (or lack thereof) with which relevant information (specifically, coordinates and redshifts) is accessible in literature articles and the lack of uniformity in multi-AGN naming schemes and definitions. Here in this section we outline several major issues that are commonplace in the multi-AGN literature and emphasize best practices that can help to resolve these issues moving forward.

\subsubsection{The Prevalence of Imprecise and Missing Coordinates}
\label{sec:missing_coords}

There are excellent examples of works throughout the literature that provide precise and distinct coordinates for each AGN component in a given multi-AGN system \citep[including but not limited ][]{bondi2010,koopmans2000,comerford2015,findlay2018,hennawi2006,hennawi2010,myers2007,myers2008,liu2011b,liu2011a,liu2013,hewitt1987,gregg2002,crampton1988,weedman1982,fu2015a,fu2015b,fu2018,fu2011b,frey2012,comerford2015,liu2013,owen1985,klamer2004,herzog2015}. However, a large number of articles lack coordinate information for the targets in question, ranging from the provision of (1) imprecise coordinates and/or (2) only a partial amount of coordinate information (i.e., coordinates are only listed for one object in a system, or only relative coordinates from one source to another are provided), or (3) an overall absence of this information (i.e., no objects in a given system have coordinates listed). This is not isolated to any one object type in the catalog; in fact, the omission of coordinate information is prevalent among both confirmed and candidate dual AGNs, binary AGNs, and recoils, and it is a problem that extends even back to some of the earliest papers. In the following, we discuss examples for each of these cases of coordinate information omission.

In a significant number of papers \citep[a few examples include][]{liu2010b,smith2010,liu2011b}, target designations are published in place of (or as a proxy for) coordinates, for example: J123456.78$\pm$123456.78; generally speaking, this can offer an efficient way to store both the target designation and coordinate information of the object(s). However, a significant level of source confusion and/or positional error is introduced when these designations lack decimal precision (falling under cases [1] and [2] outlined above). For example, \citet{wang2009} used the format J123456$\pm$123456, leaving off the decimal precision associated with the SDSS positions; when the \citet{wang2009} catalog was matched to the \citet{liu2010a} and \citet{smith2010} catalogs, match tolerances of $10''-15''$ (or larger) were required in some cases ($\sim10\%$) to properly match all sources together. The double-peaked emission line AGNs in all three catalogs were originally selected from 3'' diameter fibers, so naively it was expected that a match tolerance of only 3'' would be required to match and combine these tables while ensuring no duplicate objects remained after the matching process. As evidenced during our own work here, this lack of coordinate precision introduces unnecessary source confusion and directly impedes the matching of sources across different sample catalogs (potentially leading to the erroneous inclusion of multiple instances of the same object in a single table), impedes the matching of multiwavelength source counterparts, and increases the risk for error during target acquisition and follow-up observations. We strongly recommend every work publishing on multi-AGN systems provide the most precise coordinates possible to the object(s) in question. 

Another (common) practice in the multi-AGN literature involves including only a partial amount of coordinate information (falling under case [2] outlined above) for a given multi-AGN system. One example of this tendency is the inclusion of coordinates for one object in a multi-AGN system, but omitting the coordinates to any other object identified in the same system. In some cases, such as binary AGN candidates with unresolvable separations \citep[e.g.][]{tsalmantza2011,eracleous2012,ju2013,shen2013,xliu2014,graham2015b,charisi2016,wang2017} or kpc-scale dual AGN candidates selected based on double-peaked or velocity offset emission lines observed in the fiber spectrum of an unresolved host galaxy \citep[e.g.][]{wang2009,comerford2009a,liu2010b,smith2010,ge2012,comerford2013,comerford2014,shi2014,lyu2016}, it is natural to expect only a single set of coordinates to be reported. There are, however, a substantial number of follow-up works that identify distinct galaxy nuclei and/or AGN components in, for example, double-peak selected dual AGN candidates \citep[e.g.,][]{liu2010a,rosario2011,shen2011,fu2011a,fu2012,comerford2012,mullersanchez2015,mcgurk2015,xliu2018b} using optical imaging, near-IR imaging, and/or optical spectra but do not report the associated coordinates. This is also common among large-scale systematic searches for gravitational lenses \citep[which also turn up high redshift dual AGN candidates,][]{inada2008,inada2010,inada2012,more2016,schechter2017,lemon2018,spiniello2018,agnello2018,anguita2018,lemon2019,rusu2019,lemon2020}, though some lens-based works do provide distinct coordinates for a portion of their samples \citep[e.g.][]{inada2008,inada2010,inada2012}. Moreover, there are a substantial number of initial discovery and/or confirmation papers (for small numbers of objects) that do not provide the coordinates to distinct nuclear components (in at least a single waveband band) available through their imaging or spectroscopic observations \citep[e.g.][]{gerke2007,mcgurk2011,comerford2009b,hewett1998,morgan2000,pindor2006,tinney1995,djorgovski1987b,pfeifle2019a, satyapal2017}. This can be particularly problematic when high resolution imaging is used that can reach far superior angular resolutions than the optical and non-AO near-IR imaging: for example, coordinates are often reported for one radio source in radio VLBI imaging, but the coordinates to any secondary radio components -- argued to be AGNs -- are often omitted altogether \citep{rodriguez2006,kharb2014,kharb2017,kharb2019,gitti2013}, making it virtually impossible to independently compute projected pair separations and reducing the reproducibility of the work. A related problem is the inclusion of precise coordinates for the ``primary'' source, but the coordinates for any secondary object are listed as relative to the primary source \citep[e.g.][and sometimes even precise coordinates for the primary AGN component are not included]{mu1998,mu2001,faure2003,hewett1994}; like the lack of precision discussed above, relative coordinates introduce unnecessary risk for error when calculating the true coordinates of the secondary object(s) and impedes the matching of objects and reproducibility of the work. Precise and distinct equatorial coordinates -- not relative coordinates -- to each object in a given multi-AGN system should always be clear and present when distinct/spatially-resolved sources are reported. 

One final disadvantageous practice is the complete omission of coordinates for all objects within a multi-AGN system, and here we provide clear examples -- including well known dual AGN systems -- in the literature. The poster-child for dual AGNs, NGC 6240, was discovered to host a dual AGN based on the identification of two hard X-ray point sources within the late-stage merging system, but nowhere in the discovery paper were the specific coordinates listed for each AGN\citep{komossa2003}; specific coordinates were reported in several follow-up works \citep[e.g.,][]{max2005,hagiwara2011,treister2020}. Mrk 463 was proven to host a dual AGN based on the presence of two hard X-ray sources detected in Chandra X-ray imaging \citep{bianchi2008}, but the discovery paper did not report coordinates to the specific X-ray AGNs; later works did include specific coordinates \citep[e.g.,][]{treister2018,pfeifle2019a}. 3C 273 has long been studied in samples of LIRGs and ULIRGs, with the discovery that it hosts a dual AGN/dual SMBH system arriving through Chandra X-ray imaging and Keck near-IR IFU imaging \citep{iwasawa2011a,u2013,wliu2019}; despite the difficulty in identifying the two AGNs in 3C 273, precise coordinates to the active nuclei in this system have not been published in any of the papers reviewed for this catalog. 3C 321, originally identified by \citet{heckman1986} as a radio AGN coincident with a galaxy merger, represents the unique case of a dual AGN wherein the radio jet launched from one nucleus is plowing through the nucleus of the active companion. The unambiguous dual AGN nature of 3C 321 was finally revealed by \citet{evans2008} using Chandra X-ray imaging in concert with radio imaging and the archival optical spectra \citep{filippenko1987,draper1993,robinson2000}, but coordinates to the individual nuclei were not provided in the discovery paper. \citet{tadhunter2012} serendipitously identified a dual AGN comprised of a radio-loud AGN and a separate Seyfert 1 AGN in PKS 0347+05; previous works had erroneously identified the radio AGN as coinciding with the Seyfert 1 AGN \citep{diseregoalighieri1994,allingtonsmith1991}, but \citet{tadhunter2012} used optical imaging and spectroscopy -- along with archival radio imaging -- to finally resolve the source confusion and reveal the dual AGN nature of this interacting system. Despite resolving the prior issue of source confusion, \citet{tadhunter2012} did not include coordinates to either of the objects. These few examples are representative of the large number of articles that do not publish distinct coordinates for either/any of the candidate AGNs in a given multi-AGN system \citep[including but not limited to][]{hewett1989,maness2004,comerford2011,koss2012,imanishi2014,imanishi2020,gattano2014}. 

The inclusion of coordinate information -- right ascension, declination, equinox, coordinate waveband and coordinate source, and (when possible) coordinate epoch -- should be common practice within this community. Precise coordinates are not only imperative for determining both the angular and projected, physical separation of a multi-AGN system, but also for targeted follow-up observations and for cross-matching distinct AGNs to archival multiwavelength catalogs (particularly for systems with stale or uncommon naming conventions). Every single article should provide precise and distinct coordinates, whenever possible. Even in cases where the host galaxies are well known (for example, NGC 6240), the specific coordinates to the AGNs should always be listed or referenced \citep[see][as an example]{treister2020} for the reader.

\begin{table*}[t!]
\caption{Recommended Best Practices when Publishing Multi-AGN Science}
\begin{center}
\hspace{-2.2cm}
\begin{tabular}{cc}
\hline
\hline
\noalign{\smallskip}
Topic & Guidance \\
\noalign{\smallskip}
\hline
\noalign{\smallskip}
Coordinates & Provide coordinates for both nuclei and/or detected sources in the analyzed waveband(s)\\
& If multiple wavebands studied, best to include coordinates in each waveband.\\
& Include decimal level of precision, i.e. $12345.67\pm12345.67$.\\
& Do not list coordinates relative to one another or relative to another point.\\
\noalign{\smallskip}
\noalign{\smallskip}
Redshifts & Include (whenever possible) redshifts for both nuclei/AGN components.\\
& Always indicate which redshift corresponds to which AGN component. \\
& Velocity differences are useful, but are not sufficient substitutes for redshifts. \\
& If only one redshift is available, indicate associated nucleus.\\
& Spectroscopic redshifts are preferred over photometric redshifts.\\
& $
        \begin{array}{c}
                \textrm{Without clear merger morphological evidence, systems with only photometric}\\
                \textrm{redshifts should only be considered candidates.} 
    	\end{array}
    $\\
\noalign{\smallskip}
\noalign{\smallskip}
$\begin{array}{c} \textrm{ Nomenclature and} \\ \textrm{Definitions} \end{array}$ & $\begin{array}{c} \textrm{Definitions are subject} \\ \textrm{to change over time, but:} \end{array}$ $\left\{
        \begin{array}{c}
                \textrm{Consider adopting the uniform nomenclature provided in Section~\ref{sec:definingmultiagns}.}\\
                \textrm{Consider adopting the uniform definitions provided in Section~\ref{sec:definingmultiagns}.} 
    	\end{array}
    \right.$\\
%& Consider adopting the uniform definitions provided in Section~\ref{sec:definingmultiagns}.\\
%& For dual AGNs: consider adopting the new physical definition provided in Section~\ref{sec:definingduals}.\\
\noalign{\smallskip}
\noalign{\smallskip}
Instrument Setup & Include telescope and instrument name, and complete instrument configuration. \\
& Ex: for spectrograph, include wavelength range, slit width and slit length, etc. \\
& $\begin{array}{c} \textrm{To gauge chance for spillover} \\ \textrm{for closely separated sources:} \end{array}$ $\left\{
        \begin{array}{c}
                \textrm{Consider including seeing conditions, and/or}\\
                \textrm{consider including 2D spectra.} 
    	\end{array}
    \right.$\\
%include seeing conditions and/or 2D spectra to gauge chance for spillover. \\
\noalign{\smallskip}
\hline
\end{tabular}
\end{center}
\tablecomments{A summary of guidance on best practices in multi-AGN science as discussed in Section~\ref{sec:best_practices}. }
\label{tab:bestpractices}
\end{table*}

\subsubsection{The Prevalence of Missing Redshifts}
\label{sec:missing_redshifts}
 
There are many works that serve as excellent examples for redshift reporting, providing distinct redshift measurements for both/all components in a multi-AGN system when possible \citep[e.g.,][]{klamer2004,liu2011a,liu2011b,liu2013,fu2015a,fu2015b,more2016,eftekharzadeh2017}. However, a substantial number of literature articles omit redshift information from tables and the main text. This omission typically occurs in one of two ways: (1) only one redshift is provided, and it is not always clear with which AGN component this redshift is associated, or (2) no redshifts are provided at all. In the case of (1), this commonplace practice can naturally lead to source confusion (the redshift of one object may be erroneously associated with the wrong AGN component) and -- in the absence of clear, unambiguous morphological evidence of an ongoing interaction or merger of galaxies -- can introduce uncertainty as to whether the two components in a candidate multi-AGN system are truly physically associated. Both of these issues unnecessarily impede the clarity of the results contained within the article and impede follow-up spectroscopic investigations. Even in cases where the AGNs reside in host galaxies that are clearly interacting (and thus, are physically associated), the provision of distinct redshifts is important for placing constraints on the velocity difference between the two AGNs. Taken across a full sample of objects, such velocity differences and morphological evidence could be compared to simulated expectations of galaxy mergers and AGNs triggered in mergers \citep{byrne-mamahit2024,patton2024,chen2023}. In the case of (2), this is as serious of a problem as the omission of coordinates (see Section~\ref{sec:missing_coords}); the omission of redshifts not only obfuscates the redshift space of the targets, it also makes it more difficult -- if not impossible -- to independently determine the physical projected size of a multi-AGN system without a rigorous follow-up analysis. In the following paragraphs, we discuss clear examples for each of these cases of redshift omission.

Large-scale systematic searches for clustered quasars at high redshift (often referred to as ``binary quasars'' and which often include high redshift dual AGNs) that select quasar pairs based on distinct SDSS photometric or spectroscopic fiber positions offer excellent examples of redshift reporting in the literature. These systematic searches typically provide both coordinates and redshifts for distinct AGN components in each pair \citep[e.g., ][]{hennawi2006,hennawi2010,myers2007,myers2008,eftekharzadeh2017,findlay2018} even as separations approach $\lesssim 3''$, and this is generally true also for articles focusing on serendipitously identified binary quasars \citep[e.g.,][]{husemann2018a}. Articles focusing on the search for gravitational lenses vary more considerably when it comes to reporting distinct redshifts for high redshift dual AGN candidates: some articles provide both redshifts and coordinates for distinct AGN components (\citealp[see, e.g., Table~3 in][]{inada2008,inada2010}, and \citealp[Table~4 in][]{inada2012}), some provide only a single set of coordinates but distinct redshifts for some portion of objects \citep[e.g.,][]{more2016}, while others provide only a single redshift \citep[and sometimes only a single set of coordinates, e.g.,][]{agnello2018,anguita2018,lemon2018,lemon2019,lemon2020,more2016}. Redshift reporting in lens articles can also change from table to table within a single article, with some tables reporting distinct redshifts and coordinates for each component AGN, while other tables present only a single set of redshifts and/or coordinates, or a mixture of these two formats within even a single table \citep[e.g.,][]{inada2008,inada2010,inada2012,anguita2018,more2016}. An additional layer of ambiguity is introduced when only one set of coordinates is included because the correct association between the AGN components and the provided redshift(s) is not clear (nor is it clear whether the provided coordinates are for the system as a whole or for a specific component AGN within the system). 

There are instances where the omission of the redshift for a secondary component is seemingly justified based upon the interpretation of the object in question. Articles that identify so-called ``nearly-identical quasars'' (NIQs) while searching for  gravitational lenses \citep[e.g.,][]{schechter2017,lemon2018,lemon2019,lemon2020,anguita2018} are illustrative examples. Given the striking similarities in the spectra (and redshifts) of these objects, it is difficult to discern whether NIQs are truly lensed images of the same quasar or if these are dual AGNs at high redshift; lens-focused articles often quote only a single redshift when discussing these systems \citep[e.g.,][]{lemon2018,lemon2019,lemon2020}. Under the assumption that these objects constitute lensed quasars, there would be no need to include the component redshifts because they should be identical. However, these objects are commonly found at close angular scales; in sub-optimal seeing conditions, fiber or slit spillover becomes a serious risk and could affect the results of spectroscopic observations used to determine the redshift(s) and/or emission line properties \citep{husemann2020,pfeifle2023b}. Not only is careful interpretation required for NIQs to ensure they are properly classified as either dual AGNs or lenses, the inclusion of precisely measured redshifts (and seeing conditions) could be beneficial to future follow-up observations or future articles that publish congruent or contrasting redshift results. 

Another case where the omission of individual redshifts may be initially justified is when the host is unresolved and it is not clear how many nuclear components exist. For example, dual AGN candidates selected based on double-peaked emission lines in SDSS fiber spectra have almost always been reported prior to examining high-resolution imaging to search for distinct nuclei \citep[and $<3''$ companions could not be discerned from SDSS alone;][]{liu2010b,smith2010,wang2009,ge2012}. Even in cases where high-resolution imaging was later examined and multiple nuclei were detected within the diameter of the fiber position \citep[e.g.,][]{fu2011a,shen2011,rosario2011}, it was not clear with which of the nuclei the observed redshift was associated (while one could derive distinct redshifts based on the blue and red peaks of the emission lines, no spatial information could be gleaned from the fiber spectra alone). There are a few excellent examples where distinct redshifts and coordinates were reported based on the follow-up long-slit or IFU spectroscopic observations \citep[e.g.,][]{liu2013}. Unfortunately, even after distinct nuclei have been identified and spatially-distinct spectra have been obtained, it is still quite common for articles to omit the distinct redshifts (and coordinates) for the individual AGNs/nuclei and list only a single systemic redshift and the velocity difference between the nuclei \citep[e.g.,][]{liu2010a,comerford2012,fu2012,mcgurk2011,mcgurk2015,comerford2015}. These examples from the double-peak literature (and from the lens literature, discussed above) are representative of a larger number of works that reported/included only a single redshift for a given multi-AGN system even when multiple redshifts were available. 

Of course, spectroscopic redshifts are not always available for each individual AGN (and candidate) in a given system, and the omission of redshift information for one component is natural and expected in such cases. \citet{eftekharzadeh2017} is a good example of an article that clearly denoted which AGN components had and did not have spectroscopic observations and available redshift information (see their Table~5), and all available spectroscopic redshifts were included along with the coordinates to the component AGNs. Another good example is \citet{myers2007,myers2008}, who provided photometric redshifts when spectroscopic redshifts were not available (and clearly differentiated between the two). As was done by \citet{myers2007,myers2008} and \citet{eftekharzadeh2017}, when redshift information is available for a given system/pair of AGNs, it should be listed in the article and it should be stated clearly which component in the system is associated with the known redshift(s). In a minority of cases, redshifts may not be available for any of the component AGNs in a given multi-AGN system \citep[e.g.,][]{spiniello2018,rusu2019}; such multi-AGN candidates should be priority targets for spectroscopic follow-up. The most detrimental scenario is when neither AGN component has a reported redshift, despite (in some cases) the use of the omitted redshift(s) in the article.

Inclusion of redshifts makes it possible to determine the redshift distribution for the sample/sources, calculate velocity differences, and measure the projected physical sizes of multi-AGNs, the latter two of which represent critically important criteria necessary for determining whether the objects in question represent a physically-associated pair of AGNs. Partial or complete omission of redshifts for any components in a multi-AGN candidate directly impedes follow-up spectroscopic observations because the instrumental configuration cannot be optimized (in terms of filter choice, grating choice, central wavelength, etc.) for the redshift of the source. Every paper should provide redshift information when it is available - whether it be in tables or within the text itself - for the distinct AGN (or candidate) components.

\subsubsection{Lack of Uniformity in Nomenclature in the Literature}
\label{sec:uniform_names}
As discussed in Section~\ref{sec:catalog_construct}, one of the first hurdles we faced in constructing this literature library and multi-AGN catalog is the variety of nomenclature used in the multi-AGN field (see Table~\ref{table:aliases}), and the breadth of physical systems these nomenclatures refer to. Here we discuss the evolution of nomenclature in the field and the challenges that arise as a result of nomenclature diversity. We conclude by emphasizing the utility of using the more uniform, physically-motivated nomenclature developed in Section~\ref{sec:definingmultiagns}. 

As with every scientific field, nomenclature is introduced to describe particular phenomenon, and this nomenclature (and any associated definitions) naturally evolves over time as the field of study matures. This organic evolution introduces a variety of nomenclature to the literature that must be carefully tracked to avoid confusion during literature searches; often times different (or simply more general) terminology is used interchangeably, which can hinder searches in the literature. The term ``double quasar'' and ``double QSO'' were coined seemingly by \citet{stockton1972} and \citet{wampler1973} -- well before\footnote{``Double radio sources'' were already known by this time and referred to as such.} the discovery of physically-associated high redshift quasars pairs or gravitational lenses -- to describe the discovery of a close angular pair of quasars with discordant redshifts. A broad term like ``double quasar'' was reasonable at the time, particularly because so few objects falling under this umbrella term were known. When the first gravitational lens was discovered \citep[0957+561A,B][]{walsh1979}, it was referred to as a ``twin quasi-stellar object'' and a ``gravitational lens''; naturally, the general terms ``double quasar'' and ``double QSO'' were then used to refer to this system in follow-up works \citep[e.g.][]{porcas1979,lebofsky1980} as well as later lens discoveries \citep[e.g.,][]{subramanian1984,weir1991,hagen1996}. Additional lens discoveries shortly thereafter would also consistently use ``quasar pair'' and ``gravitational lens'' \citep[e.g,][]{weedman1982}. The appearance of ``binary quasar'' appears to trace back to \citet{chaffee1980}\footnote{But the term ``binary'' in the general context of quasars goes back much further; see \citet{komberg1968}.}, where it was used to describe the alternate hypothesis to the lens scenario for 0957+561 A,B, namely that the two quasar images were distinct, physically associated quasars. ``Binary quasar'' then became the more precise nomenclature used for physically associated quasar pairs on scales of kpc and tens of kpc \citep[e.g.,][]{djorgovski1987b,mu1998} and discussed in the lens literature as the alternative scenario whenever a lens candidate was discovered \citep[e.g.,][]{kochanek1997,hawkins1997a,inada2008,inada2010,inada2012,more2016}. The more general terminology of ``quasar pairs'' and ``double QSOs'' remained present, however, and was used interchangeably with ``binary quasar''. %ryan: citations?

In the 1980s through the 2000s, the ``binary quasar'' nomenclature evolved further: in the early 1980s, \citet{begelman1980} outlined the theoretical framework for the presence of gravitational-bound pairs of SMBHs within AGNs, terming these SMBH pairs as ``binaries,'' and  nomenclature such as ``binary black hole'' and ``binary supermassive black hole'' was then adopted in later observational and theoretical works in the 1980s \citet[e.g.,][]{whitmire1981,sillanpaa1988} and 1990s \citep[e.g.,][]{roos1993,eracleous1994} to describe these unresolved, gravitationally bound binary SMBHs/AGNs. In the 2000s, the term binary quasar was used to refer to not only kpc-scale, higher redshift quasar pairs \citep[e.g.,][]{hennawi2006,myers2007,myers2008}, but also pairs of quasars with separations up to $\sim1$\,Mpc \citep[e.g.,][]{hennawi2006} that are unlikely to be associated with a physically interacting system of galaxies. At the same time, ``binary AGN'' or ``binary active galactic nucleus'' became the terminology used to refer to kpc-scale pairs of AGNs in the local universe \citep[e.g.,][]{komossa2003,komossa2003review,guainazzi2005,jim2007,liu2010b}. Thus, in addition to the plethora of nomenclature used within the literature by the early 2000s, incredibly similar or identical nomenclature was used to describe multi-AGN phenomenon \citep[and clustered quasars, e.g.,][]{hennawi2006} covering effectively $>6$ orders of magnitude in projected separation space, from  hundreds of kpc down to $<0.01$\,pc.

The term ``dual AGN'' was coined  by \citet{gerke2007} to describe kpc-scale AGN pairs in ongoing galaxy mergers, and this was adopted as another nomenclature choice in the literature \citep[e.g.,][]{wang2009,piconcelli2010,comerford2011,peng2011}. Over the last two decades, observational and theoretical works have slowly but progressively adopted the ``dual AGN'' terminology to describe (physically-associated) kpc-scale and sub-kpc-scale AGN pairs in interacting/merging galaxies \citep[where the SMBHs are not gravitationally-bound to one another][]{blecha2018,chen2023,volonteri2022,saeedzadeh2024} while the ``binary'' terminology has been more reserved for high redshift AGN pairs \citep[``binary quasars'', e.g.,][]{hennawi2010,inada2012,eftekharzadeh2017} and gravitationally-bound AGN pairs \citep[``binary AGNs'' and ``binary SMBHs'' at pc- and sub-pc scales, e.g.,][]{eracleous2012,shen2013,xliu2014}.

Despite this, the organic expansion of nomenclature over the last five decades has unintentionally: (1) made it difficult to track down and identify relevant multi-AGN articles because so many multi-AGN aliases are required during literature searches \citep[resulting in lower citation counts and by extension the loss of relevant articles][]{parma1991,hawkins1997a,klamer2004,jim2007}; (2) led to the near-simultaneous introduction of terminologies that are permutations of one another (making it further difficult to find and track articles); (3) led to a slow, uneven transition from one adopted set of nomenclature to the next, with many articles still adopting stale nomenclature choices; and (4) led to a variety of distinct definitions for these different object classes, as described in Section~\ref{sec:definingmultiagns}. For example (related to [3]), despite the overall transition to the ``dual AGN'' terminology for kpc-scale and sub-kpc scale AGN pairs in ongoing galaxy mergers \citep{comerford2015,xliu2018a,xliui2018b,blecha2018,pfeifle2019a,kim2020,chen2023}, some recent articles still use ``binary AGN'' to describe this same phenomenon, effectively prolonging the period of time these two terminologies will be viewed as interchangeable. As another example (related to [2]), several systematic searches for binary SMBHs in the years 2010-2013 \citep[e.g.,][]{eracleous2012,ju2013} were not found during our initial literature searches because those studies used different permutations of the alias ``binary supermassive black hole'' not included in our early ADS keyword search.

The issues stemming from the evolution and non-uniformity of multi-AGN nomenclature only provided further motivation for the exhaustive retrieval of literature in the present work and catalog, and it provided direct motivation for the derivation and adoption of more formal nomenclature and definitions for dual AGNs, binary AGNs, and recoiling AGNs in the Big MAC DR1 (as described in Section~\ref{sec:definingmultiagns}). This catalog adopts specific terminology for each of these system classes (coined in the literature over the last couple of decades but used non-uniformly) and emphasizes the use of these distinct terms when referring to different classes of multi-AGN systems. For dual AGNs in particular, this is the first time this nomenclature has been defined in the literature using physically-motivated criteria drawn from realistic cosmological simulations \citep[e.g.,][]{patton2024}, providing a more solid grounding for dual AGN sample selection in the future.

It is possible (and, in fact, likely) that despite the variety of multi-AGN aliases used to perform the ADS searches for this literature archive, the Big MAC DR1 is likely still not 100\% complete (up to 2020); the use of so many similar but distinguishable aliases in the literature makes an exhaustive multi-AGN literature retrieval an arduous task, but this living library will continue to be updated as additional articles are recovered. While specific nomenclature (or permutations of nomenclature) choice in a given article is up to the discretion of individual authors, it would be beneficial to adopt a more uniform approach to multi-AGN nomenclature and definitions; we recommend that authors rely upon the definitions and nomenclature provided in Section~\ref{sec:definingmultiagns} when determining multi-AGN system system classes and to use the nomenclature associated with these definitions at least once within the text of a given article so that it can be recovered in a single ADS search by members of the community.

\subsection{Community Input and Contributions via GitHub}
\label{sec:community}

The Big MAC serves as a new tool for the community: it provides a strong foundational archive for follow-up studies of known multi-AGNs and candidates, and it strongly motivates new observational techniques that build upon (circumvent) current selection functions ( selection biases) in the literature. To enhance the utility of this catalog and library, the Big MAC is housed in a public GitHub repository\footnote{https://github.com/thatastroguy/thebigmac} and is complemented by a public landing page\footnote{https://thatastroguy.github.io/thebigmac/} that describes the library, catalog, and repository. Within the repository, community members will find not only the Big MAC itself (via machine-readable comma-separated value tables) but also all Jupyter Notebooks and table files used to construct the Big MAC from the ground-up. The benefit of the repository is manifold: (1) this living library will undoubtedly require small revisions and adjustments somewhat regularly, which can be pushed to the repository easily along with commit notes, and these adjustments can be easily version-controlled; (2) researchers in and outside of the field of multi-AGN science can follow the repository and can opt-in to notifications alerting them to changes and additions to the repository; (3) new multi-AGN discoveries can be contributed by community members through the use of the ``raise issue'' feature on GitHub. This point is particularly enticing, as this allows the community to make direct contributions to the catalog (pending review, potential modifications, and acceptance of the additions/changes) and ensure the catalog remains up-to-date with the latest additions to the literature. An ultimate goal of the Big MAC is to foster and enable large-scale collaboration across the community, and one way we can embark on this path is through community contributions and feedback. (4) Any issues with the library and catalog in its current form can be flagged by individual members of the community by raising an issue, and these corrections can be promptly addressed and new versions of the catalog can be quickly disseminated to the community (see point 2 above). The GitHub repository will also house the follow-up data releases as we move toward a complete multiwavelength picture for these objects.

\section{Conclusions and the Road Forward}
\label{sec:conclusion}

Multi-AGN systems -- dual AGNs, binary AGNs, and recoiling AGNs -- are predicted to be important tracers of hierarchical SMBH growth and evolution across cosmic time \citep[e.g.,][]{derosa2019,foord2024}. In this work, we have motivated and described the creation of the first literature complete catalog of multi-AGN systems, assembled from approximately 600 literature articles and comprised of confirmed and candidate dual AGN, binary AGN/SMBH, recoiling AGNs, and N-tuple AGNs. This catalog is designed to bring uniformity, structure, and clarity to the literature and it provides  a strong foundation for targeted multiwavelength follow-up investigations of these multi-AGN systems using archival as well as new ground- and space-based observations. We summarize the content and results of our work as follows:

\begin{itemize}
    \item The Big MAC DR1 includes: 156 confirmed dual AGNs and 4180 dual AGN candidates; 1 confirmed gravitationally-bound binary AGN and 1368 binary AGN/SMBH candidates; and 581 recoiling AGN/SMBH candidates. 
    \item The overall redshift distribution for the catalog is $0.001<z<6.2$. The observed redshift distribution for dual AGNs and candidates is $0.001<z<6.2$, for binary AGNs and candidates it is $0.002<z<3.2$, and for recoiling candidates it is $0.002<z<1.4$. Among confirmed dual AGNs (confidence = 1), the redshift distribution is $0.001<z<4.2$; the only known binary AGN (CSO 0402+739) resides at $z=0.05$.%with separations X<r<X % (X<r<X)
    \item We have outlined formal, physically-motivated definitions for dual AGNs ($0.03\lesssim r_p \lesssim 110$\,kpc) and binary AGNs ($r_p\lesssim30$\,pc), where the dividing line between dual and binary AGNs depends upon the mass of the SMBHs and the host stellar velocity dispersion. Our definitions for dual AGNs are informed by the reconstructed orbits of merging galaxy pairs in Illustris-TNG100 \citep{patton2024}. We recommend the adoption of these definitions in the literature going forward. We also discuss and reinforce general definitions for recoiling AGN candidates.
    \item Dual AGNs have been predominantly selected via optical spectroscopic techniques (emission line ratios, double peaked emission lines), but notable samples have also been assembled via hard X-ray \citep[e.g.,][]{koss2012}, mid-IR \citep[e.g.,][]{satyapal2017}, radio \citep[e.g.,][]{fu2015a}, and optical imaging selection \citep[e.g.,][]{myers2007,eftekharzadeh2017}. Binary AGNs are predominantly selected via optical spectroscopic techniques \citep[velocity offset emission lines and emission line radial velocity shifts, e.g.,][]{eracleous2012,xliu2014}, and X-shaped radio sources \citep[e.g.,][]{merritt2002,lal2007}. Recoiling AGNs are predominantly selected via velocity offset emission lines \citep[e.g.,][]{kim2016}, X-shaped radio sources \citep[e.g.,][]{cheung2007}, and spatial offsets between the AGNs and the host galaxies \citep[e.g.,][]{lena2014}.
    \item Among confirmation techniques, optical spectroscopic techniques have played a role in most of the confirmed dual AGN population ($\sim78\%$), though soft X-ray imaging, optical imaging, and radio imaging have each played significant roles in the confirmation of dual AGNs (used in the confirmation 21\%, 20\%, and 17\% of known duals, respectively).  The only confirmed binary AGN, CSO 0402+739, relied upon radio VLBI imaging for (serendipitous) selection and later confirmation. No recoiling AGNs have so far been confirmed.
\end{itemize}

Here we also summarize our recommendations for the publishing of multi-AGN articles in the field:
\begin{itemize}
    \item The physically-motivated nomenclature presented in Section~\ref{sec:definingmultiagns} of this work provides a clear and uniform approach for classifying multi-AGN systems in future literature articles. Adoption and use of these terms and definitions by future works focused on multi-AGN systems will ensure uniformity and accessibility within the field, particularly important in an age where publication rates far outpace those of previous decades. These definitions will be presently submitted to the Unified Astronomy Thesaurus (UAT) overseen by AAS publishing as well as the International Astronomical Union for formal adoption.
    \item Thorough in-text citations play a critical role in the documentation of the field: articles should attempt to be as thorough (and/or diverse in citation choice) as reasonably possible while adhering to journal guidelines and maintaining a clear, logical connection to the work presented. Excellent examples include \citet{eracleous2012}, \citet{liu2011b}, \citet{shen2013}, and \citet{xliu2014}. 
    \item Distinct coordinates for each component AGN/SMBH in a multi-AGN system should always be listed within the tables or text of an article. These coordinates should be a precise as possible (decimal precision) and should provide the distinct, standalone positions of the AGN components; relative coordinates should not be listed.
    \item Redshifts -- and specifically distinct redshifts -- should be listed wherever possible. Even if a target is well known, literature articles should still provide a measurement of distance to the system in question. 
    \item Instrument configurations -- including camera, spectrograph, slit sizes, and other configurations -- should always be clearly listed for each target. These configurations are relevant specifically for understanding the selection methods of dual AGN samples. The inclusion of the associated atmospheric seeing conditions for a given set of observations is also preferable if the observations were focusing on closely-separated sources, as this information will help to gauge the likelihood of fiber/slit spillover effects. 

\end{itemize}

Providing adequate information to reproduce, follow-up, and confirm multi-AGNs will ensure the reliability and robustness of our studies, is essential to producing sound scientific results, and will ultimately aid in our efforts to determine the relative importance of multi-AGN systems in the fueling and evolution of SMBHs and their hosts. 

The next major milestone for the Big MAC is the inclusion of all literature articles from December 2020 - May 2024 (data release 2, approximately another $\gtrsim100$ articles), which will be followed by the Big MAC mid-IR and near-IR counterpart catalog. In particular, this IR counterpart catalog will provide the first large-scale study on the mid-IR AGN fraction across all samples of dual AGNs discovered to-date, as well as the relative fraction of mid-IR - mid-IR dual AGNs (where each AGN is a mid-IR WISE AGN) across these samples. The mid-IR counterpart analysis will dovetail with the first two investigations focused on spatially-resolved mid-IR dual AGNs (\citealp{barrows2023}; Pfeifle et al., in preparation). 
\acknowledgements
R. W. P. gratefully acknowledges support through an appointment to the NASA Postdoctoral Program at Goddard Space Flight Center, administered by ORAU through a contract with NASA. We thank Joan Wrobel for helpful discussions on manuscript formatting. 

This research has made extensive use of NASA’s Astrophysics Data System. This research made use of ChatGPT, an OpenAI software, to assist with python code development and refinement, however OpenAI software was not used to write content for this work or read and summarize journal articles. All review and synthesis of journal articles was undertaken by the corresponding author of this work. Articles written using ChatGPT, or other AI software, will not be included in future releases.

This work encompasses approximately 55 years of AGN and galaxy merger science from a large variety of international research groups and individuals; we thank the community for their long-standing and continued efforts that have made the creation of this catalog possible. We acknowledge that many discoveries and analyses of confirmed and candidate multi-AGNs have relied upon telescopes and facilities built upon lands considered sacred or culturally important to historically marginalized, underrepresented, and disregarded communities; we urge the scientific community to not only acknowledge these communities in future works, but to specifically bear in mind and engage with these communities when designing and planning future observatories and facilities. %Astronomy must 

This work is the result of approximately four years of efforts largely undertaken by a single person (R. W. P.). R. W. P., from the bottom of his heart, thanks his family, his friends, he and his wife's Corgi (O. L. P.), and most of all his wife, N. M. L., without whom this work could never have been accomplished. %, often at the expense of personal affairs and mental and physical health

\software{APLpy \citep{robitaille2012}, pandas \citep{mckinney2010}, NumPy \citep{oliphant2006,walt2011,harris2020array},  \textsc{ds9} \citep{joyce2003}, \textsc{Astropy} \citep{2013A&A...558A..33A,2018AJ....156..123A}}%SciPy \citep{virtanen2020}, \textsc{matplotlib} \citep{hunter2007}, \textsc{heasoft} \citep{heasoft}, \textsc{xspec} \citep{arnaud1996}, \textsc{ciao} \citep{fruscione2006}, \textsc{sas} \citep{2004ASPC..314..759G}, \textsc{nustardas}, , \textsc{topcat} \citep{2005ASPC..347...29T} 

\bibliography{references}{}
\bibliographystyle{aasjournal}

\end{document}